\documentclass[12pt]{article}
\usepackage{empheq}
\usepackage[most]{tcolorbox}
\usepackage{braket}
\usepackage{multirow}
\usepackage{jheppub}
\usepackage{ifpdf,tensor}
\usepackage{booktabs}
\usepackage{array}

\usepackage{color}
\usepackage{eufrak}
\usepackage{graphicx,psfrag, subfigure}
\usepackage{amsmath}
\usepackage{upgreek}
\usepackage{stmaryrd}
\usepackage{blkarray}
\usepackage{dsfont}
\usepackage{yfonts}
\usepackage{setspace}
\usepackage{soul}
\definecolor{dm}{cmyk}{.20, 0, .30, 0}
\sethlcolor{dm}
\usepackage{mathrsfs}
\numberwithin{equation}{section}
 \usepackage{relsize}
\usepackage{hyperref}
\usepackage{tabularx}
\usepackage{amssymb}
\usepackage[mathscr]{eucal}
\newcommand{\bs}{\boldsymbol}

\newcommand{\diag}{\text{diag}}

\DeclareMathAlphabet\mathbfcal{OMS}{cmsy}{b}{n}
\DeclareSymbolFont{bbold}{U}{bbold}{m}{n}

\usepackage{titlesec}
\titleclass{\subsubsubsection}{straight}[\subsection]

\usepackage{amsthm}
\theoremstyle{plain}
\newtheorem*{lemma*}{Lemma}

\newcounter{subsubsubsection}[subsubsection]
\renewcommand\thesubsubsubsection{\thesubsubsection.\arabic{subsubsubsection}}

\titleformat{\subsubsubsection}
  {\normalfont\normalsize\bfseries}{\thesubsubsubsection}{1em}{}
\titlespacing*{\subsubsubsection}
{0pt}{3.25ex plus 1ex minus .2ex}{1.5ex plus .2ex}

\makeatletter
\renewcommand\paragraph{\@startsection{paragraph}{5}{\z@}
  {3.25ex \@plus1ex \@minus.2ex}
  {-1em}
  {\normalfont\normalsize\bfseries}}
\renewcommand\subparagraph{\@startsection{subparagraph}{6}{\parindent}
  {3.25ex \@plus1ex \@minus .2ex}
  {-1em}
  {\normalfont\normalsize\bfseries}}
\def\toclevel@subsubsubsection{4}
\def\toclevel@paragraph{5}
\def\toclevel@paragraph{6}
\def\l@subsubsubsection{\@dottedtocline{4}{7em}{4em}}
\def\l@paragraph{\@dottedtocline{5}{10em}{5em}}
\def\l@subparagraph{\@dottedtocline{6}{14em}{6em}}
\makeatother

\DeclareSymbolFontAlphabet{\mathbbold}{bbold}

\makeatletter
\newcommand{\fixed@sra}{$\vrule height 2\fontdimen22\textfont2 width 0pt\shortrightarrow$}
\newcommand{\shortarrow}[1]{
  \mathrel{\text{\rotatebox[origin=c]{\numexpr#1*45}{\fixed@sra}}}
}

\newcommand{\trafo}[2]{\,\bs T_{{{#1 #2}}}\,}

\newcommand{\Po}{\bs P}
\newcommand{\Pob}{\bs  P^{\perp}}
\def\simleq{\; \raise0.3ex\hbox{$<$\kern-0.75em
      \raise-1.1ex\hbox{$\sim$}}\; }

\def\simgeq{\; \raise0.3ex\hbox{$>$\kern-0.75em
      \raise-1.1ex\hbox{$\sim$}}\; }

\def\be{\begin{equation}}
\def\ee{\end{equation}}
\def\bea{\begin{eqnarray}}
\def\eea{\end{eqnarray}}

\def\be{\begin{equation}}
\def\ee{\end{equation}}
\def\bea{\begin{eqnarray}}
\def\eea{\end{eqnarray}}

\definecolor{mkcolor}{rgb}{1, .83,.83}
\definecolor{tbcolor}{rgb}{.83, .9,.83}
\definecolor{ojcolor}{rgb}{.83,.83, 1}
\definecolor{kecolor}{rgb}{1, 1, .5}

\usepackage{marginnote}

\begin{document}

\begin{titlepage}

\setcounter{page}{1} \baselineskip=15.5pt \thispagestyle{empty}

\bigskip\

\begin{center}
{\Large \bf Systematics of Aligned Axions}

\vspace{2cm}

{Thomas C. Bachlechner$^\star$, Kate Eckerle$^{\star,\sharp}$, Oliver Janssen$^\dagger$ and Matthew Kleban$^\dagger$}

\vspace{1cm}

\textsl{$^\star$Department of Physics, Columbia University, New York, USA} \\ \vspace{0.2cm}
\textsl{$^\sharp$Dipartimento di Fisica, Universit\`a di Milano-Bicocca, Milano, Italy} \\ \vspace{0.2cm}
\textsl{$^\dagger$Center for Cosmology and Particle Physics, New York University, New York, USA}
\end{center}

\vspace{1cm}

We describe a novel technique that renders theories of $N$  axions tractable, and more generally can be used to efficiently analyze a large class of periodic potentials of arbitrary dimension. Such potentials are complex energy landscapes  with a number of local minima that scales as $\sqrt{N!}\,$, and so for large $N$ appear to be analytically and numerically intractable. Our method is based on uncovering a set of approximate symmetries that exist in addition to the $N$ periods. These approximate symmetries, which are exponentially close to exact, allow us to locate the minima very efficiently and accurately and to analyze other characteristics of the potential.  We apply our framework to evaluate the diameters of  flat regions suitable for slow-roll inflation, which unifies, corrects and extends several forms of ``axion alignment'' previously observed in the literature.  We find that in a broad class of random theories, the potential is smooth over diameters enhanced by $N^{3/2}$ compared to the typical scale of the potential. A Mathematica implementation of our framework is available online.

\vfil

\begin{flushleft}
\small \today
\end{flushleft}
\end{titlepage}
\tableofcontents
\newpage

\section{Introduction}\label{intro}
Fields protected by exact or approximate shift symmetries play an integral role in a wide variety of physical phenomena, ranging from magnetic fluctuations of topological insulators in condensed matter physics to inflation in early universe cosmology. Axion fields are a particularly interesting example: to all orders in perturbation theory they have a continuous shift symmetry which is broken to a discrete one by non-perturbative effects. Axions were originally proposed to solve the strong CP problem of QCD \cite{Peccei:1977hh}, and many ($\sim 100$) axions often arise in compactifications of string theory (see e.g. \cite{Wen:1985jz,Dine:1986vd,Dine:1986zy,Denef:2004cf,Svrcek:2006yi,Denef:2007pq,Denef:2008wq,Baumann:2014nda,Long:2016jvd,Halverson:2017deq,Halverson:2017ffz}). Axions can be dark matter \cite{Hu:2000ke, Arvanitaki:2009fg, Hui:2016ltb,Marsh:2015xka,Diez-Tejedor:2017ivd}, drive inflation \cite{Natural,Dimopoulos:2005ac,Kim:2004rp,SW,McAllister:2008hb,Kaloper:2008fb,Kaloper:2011jz} and (similar to quantized fluxes \cite{Bousso:2000xa}) can account for the observed vacuum energy \cite{Bachlechner:2015gwa}. In order to analyze the interplay between these distinct phenomena, one requires a comprehensive framework for multi-axion theories. The purpose of this paper is to present such a framework, which can be employed to unify the cosmological mechanisms mentioned above.

Theories of $N \gg 1$ axion fields constitute an extremely complex ``landscape" -- that is, they have  an exponentially large number of minima with different energies and a large diversity of regions of the potential. We will study general multi-axion theories, providing a systematic approach that renders even complex theories analytically tractable.\footnote{While we focus on axion field theories, our techniques carry over to the analysis of more general theories where the potential is a sum of terms with discrete shift symmetries.} In this paper we focus on properties of the axion potential. We identify all exact and approximate shift symmetries, provide the location of local minima and characterize features of the potential through a natural partition of the axion field space. In a companion paper \cite{bejk2} we study the dynamics of these theories in the context of cosmology.  A brief summary of our results can be found in \cite{Bachlechner:2017zpb}. In this paper and its companions we find that generic theories of $N \sim 100$ axions with a single energy scale close to the fundamental scale and with ${\cal O}(1)$ random coefficients can simultaneously account for the Big Bang (tunneling from a parent minimum), inflation (because such potentials generically have light directions with enhanced field ranges), ``fuzzy" dark matter \cite{Hu:2000ke} with roughly the correct abundance \cite{Arvanitaki:2009fg,Hui:2016ltb}, and provide many minima with energies consistent with observation that can solve the cosmological constant problem anthropically \cite{1981RSPSA.377..147D,Sakharov:1984ir,Banks:1984cw,Linde:1986dq,Weinberg:1987dv}.\footnote{See also \cite{Linde:2015edk} and references therein.}

This paper is divided into three parts. In \S\ref{symm} we identify the exact and approximate discrete shift symmetries of multi-axion theories by introducing a $P$-dimensional auxiliary field space. We discuss how in many cases the approximate shift symmetries can be used to eliminate all phases from the potential to good accuracy. Following the discussion of symmetries, the two subsequent sections can be read independently. In \S\ref{seccminima} we apply the framework of symmetries to locate the critical points of the potential. We provide an algorithm that finds all minima in exponential time, while a representative sample of all minima can be obtained in polynomial time. More specifically we demonstrate how an exponentially large number of minima can be located analytically via a polynomial in $N$ algorithm. We estimate the magnitude of the remaining phases, as well as the number of minima in certain ensembles of random axion theories. In \S\ref{seccdiameters} we provide a general discussion of the geometry of the approximately quadratic regions of the potential surrounding minima. This discussion generalizes and corrects misleading prior results in the literature (including those of one of the authors \cite{Bachlechner:2014rqa}).

A Mathematica demonstration of our framework for multi-axion theories is available online \cite{mathematicademo}.
\begin{figure}
  \centering
  \includegraphics[width=1\textwidth]{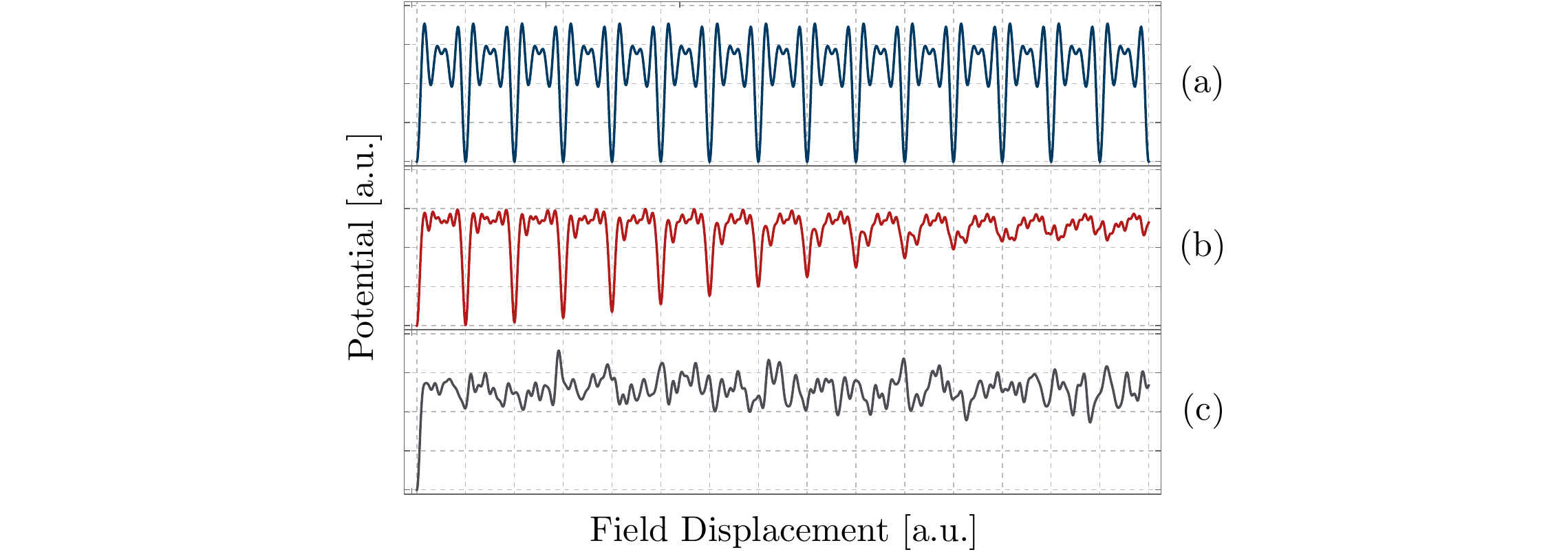}
  \caption{\small The potential plotted along three different rays through field space (starting at the global minimum), for an example of the potential in \eqref{lagrtheta} with $N=23, P=40$. \underline{Top pane}: a line oriented along an exact symmetry direction. \underline{Middle pane}: a line oriented along an approximate symmetry direction. \underline{Bottom pane}: a random direction. 
  \label{coordinateexamplephi1}}
\end{figure}

\subsection{Systematic framework}
Consider the general two-derivative theory of $N$ axions $\theta^i$.   The $N$ continuous shift symmetries of the free theory are broken to discrete ones by $P$ leading non-perturbative effects.\footnote{In some cases the shift symmetries are entirely broken for some linear combinations of fields by couplings to classical objects, such as sources or fluxes. In this work we restrict our attention to fields that retain discrete shift symmetries. Appendix \ref{app:classical} discusses the appropriate coordinate transformations that eliminate axions that couple to classical sources from the  theory.}
The Lagrangian takes the form
\be\label{lagrtheta}
{\mathcal L}={1\over 2} \partial\bs\theta^\top \bs K \partial\bs\theta-\sum_{I=1}^P \Lambda_I^4  \left[ 1-\cos\left({\mathbfcal Q}\bs\theta +\bs \delta \right)^I \right] + \dots\,,
\ee
where $\bs K$ is the metric on field space, ${\mathbfcal Q}$ is the $P\times N$ integer  matrix where the $I$th row contains the charge associated with the axions' coupling to the $I$th non-perturbative effect, and $\Lambda_I$ is the energy scale of this effect.  The dots denote subleading terms in the potential that we will generally neglect (but see \S\ref{bandstructuresec}), as well as a possible additive constant (a bare cosmological constant) that will be irrelevant in this paper, as we do not consider coupling to gravity here (but see \cite{bejk2}). We denote matrices and vectors by bold font, with upper and lower indices identifying rows and columns, respectively. The lower case indices $i,j$ run from $1$ to $N$, $a$ runs from $1$ to $P-N$, and $I$ runs from $1$ to $P$.\footnote{Without loss of generality we may assume that $P \geq N$ and ${\mathbfcal Q}$ has maximal rank $N$. See appendix \ref{app:massless} for a discussion.} Throughout this work we assume that $\bs K$ is independent of the axions $\bs \theta$.

Potentials of the form  appearing in \eqref{lagrtheta} are very complex when $N, P \gg 1$. However, because the cosine arguments consist of integer linear combinations of the axions, the potential is manifestly invariant under the $N$ discrete shifts $\theta^i\rightarrow \theta^i+2\pi$.\footnote{In general these shifts are linear combinations of ``minimal" discrete symmetries of the potential, in a sense which we will clarify.} The existence of  exact discrete symmetries is a fundamental characteristic of axion theories, and it allows us to restrict our attention to a finite region in field space -- a single periodic domain defined by these symmetries. What is not so obvious is that the potential in \eqref{lagrtheta} additionally possesses as many as $P-N$ \emph{approximate} discrete shift symmetries, which can be used to eliminate all phases $\delta^I$ to good accuracy. We develop a framework that identifies these symmetries, and provides a natural division of the field space into domains over which none of the individual terms in the potential exceed their period. As we shall see, the identification of approximate symmetries consists of finding short lattice vectors of a $P$-dimensional rank $P-N$ lattice, which at fixed $P-N$ requires a number of evaluations that scales polynomially in $N$.  The (approximate) symmetries then allow us to identify regions that are very similar. Furthermore, our framework allows us to identify a vast number of distinct minima by considering the approximate symmetry transformations away from a given minimum. 

Axions are protected from perturbative corrections of the potential and therefore constitute prime candidates in the constructions of models for large field inflation and tests of quantum gravity more generally \cite{Vafa:2005ui,ArkaniHamed:2006dz,Rudelius:2014wla,Cheung:2014vva,Heidenreich:2015wga,Bachlechner:2015qja,Rudelius:2015xta,Hebecker:2015rya,Ibanez:2015fcv,Heidenreich:2016aqi}. Large field inflation requires very flat potentials, therefore there has been much interest in the invariant distances over which axion potentials remain featureless \cite{Bachlechner:2014hsa,Choi:2014rja,Bachlechner:2014gfa,HT1,HT2,Shiu:2015uva}. The potential certainly is featureless in a field space region within which none of the cosine terms traverses more than its period. The invariant size of these regions depends both on the kinetic matrix $\bs K$ and the charge matrix $\mathbfcal Q$. Historically, two mechanisms have been proposed to construct theories with potentials that remain featureless over large invariant distances: lattice alignment \cite{Kim:2004rp}, which relies upon an almost exact degeneracy between the axion charges, and kinetic alignment \cite{Bachlechner:2014hsa}, which relies upon the delocalization of eigenvectors of the kinetic matrix. In this work we clarify the relation between these mechanisms and demonstrate that the diameter of featureless regions is bounded from above by
\begin{equation} \label{D0boundsss}
	{\cal D}\le 2 \pi\sqrt{P}{1\over \lambda_\text{min} \left( |{\mathbfcal Q} \bs K^{-1/2}| \right)}  \,,
\end{equation}
where ${\lambda_\text{min}}( \, \cdot \,)$ returns the smallest eigenvalue and we defined $| \mathbfcal Q | \equiv \sqrt{\mathbfcal Q^\top \mathbfcal Q}$. The bound (\ref{D0boundsss}) is approximately saturated in large classes of axion theories. Note that the diameter, perhaps surprisingly, scales with $\sqrt{P}\ge \sqrt{N}$.

\subsection{Results for random ensembles}
To illustrate our framework we apply it to random ensembles of axion theories defined by a collection of integer charge matrices $\mathbfcal Q$, energy scales $\Lambda_I^4 $ and axion-independent field space metrics $\bs K$ that are  loosely motivated by  flux compactifications of string theory. While our techniques apply to all $N,P$, they are most powerful in the regime $N \gg 1$ and $P < 2N$.  We will take $\mathbfcal Q$ to be a  $P \times N$ matrix of independent, identically distributed (i.i.d.) random integer entries with mean zero and standard deviation $\sigma_{\mathcal Q}$. As long as at least a small fraction $\gtrsim 3/N$ of the entries is non-vanishing -- which at large $N$ includes very sparse matrices -- the universality of random matrix theory takes over and yields simple analytic results.\footnote{Note that $P \times N$ matrices with a fraction of non-zero entries fewer than $1/P$ cannot be full rank, and can be dealt with using the techniques of appendix \ref{app:massless}. Hence our methods apply to all matrices except those with a fraction of non-zero entries between $1/P$ and $3/N$.}  As it turns out, in this regime the approximate shift symmetries become  exponentially close to exact.

Even for the simplest case $P =N+1$ we will find that the number of minima scales factorially with the number of terms in the potential (see also \cite{Bachlechner:2015gwa}),
\be
{\cal N}_\text{minima}\propto  \sigma_{{\mathcal Q}}^P\sqrt{P!} \,,
\ee
with a simple generalization to larger $P-N$.  In these potentials there is a natural definition of neighboring minimum: those that are separated by no more than one traversal of the maxima of each cosine term of the potential. We will find that even when $P=N+1$ the neighboring minima realize a wide variety of  energy densities so long as $N \gg 1$. In other words these theories have extremely complex potentials that look random in the vicinity of any point or along a randomly chosen direction.  However, they also possess nearly exact symmetries that make their analysis tractable.  In particular, we can use the symmetries to  identify minima with  energy  close to any desired value to exponential accuracy in polynomial time \cite{Bachlechner:2017zpb}. This kind of tractability in complex landscapes was recently discussed in \cite{Bao:2017thx}.  

We will consider both specific examples and ensembles of isotropic, positive definite kinetic matrices and parametrize the resulting field space diameters in terms of the largest eigenvalue $f_\text{max}^2$ of $\bs K$. Both the field space diameters and the distribution of energies in minima have previously been studied in such random axion theories. In this work we unify and generalize many of those results. We find that the  field space distance suitable for inflation is typically as large as (see also \cite{Bachlechner:2014gfa})
\be
{\cal D} \lesssim \begin{cases} N^{3/2} f_\text{max} \,,~~~\text{for}~~P-N=\text{constant}\,,\\ N^{1/2} f_\text{max} \,,~~~\text{for}~~P-N\propto N\,. \end{cases}
\ee
This result is robust even when large hierarchies are present between the energy scales $\Lambda_I^4$.

The approximate shift symmetries are lost in the limit $P\gg N$. In this case the potential ceases to be analytically tractable and approaches a Gaussian random field instead. In appendix \ref{gaussianfield} we discuss a connection between multi-axion theories in this limit and Gaussian random fields with a Gaussian power spectrum.

\section{Exact and approximate axion symmetries} \label{symm}
The leading non-perturbative  potential for the $N$ axions $\bs\theta$ in (\ref{lagrtheta}) is
\be \label{potentialnophases}
V(\bs \theta) =\sum_{I=1}^P \Lambda_I^4 \left[ 1-\cos\left({\mathbfcal Q} \bs\theta \right)^I \right] \,,
\ee
where we postpone the discussion of non-vanishing phases $ \delta^I$ to \S{\ref{phases}}. This potential depends  on the $P$ energy scales $\Lambda_{I}$ and the $P   N$  integers in the charge matrix ${\mathbfcal Q}$.  At large $N$ the field space volume becomes  large, but given the limited number of parameters and the periodicity of the cosines, one might suspect that the entire structure of the potential is analytically tractable, at least so long as $P$ is not too great. In the following we will make this expectation precise.

\subsection{Auxiliary fields and a geometric picture}\label{sec:geompic}
\begin{figure}
\centering
\includegraphics[width=1\textwidth]{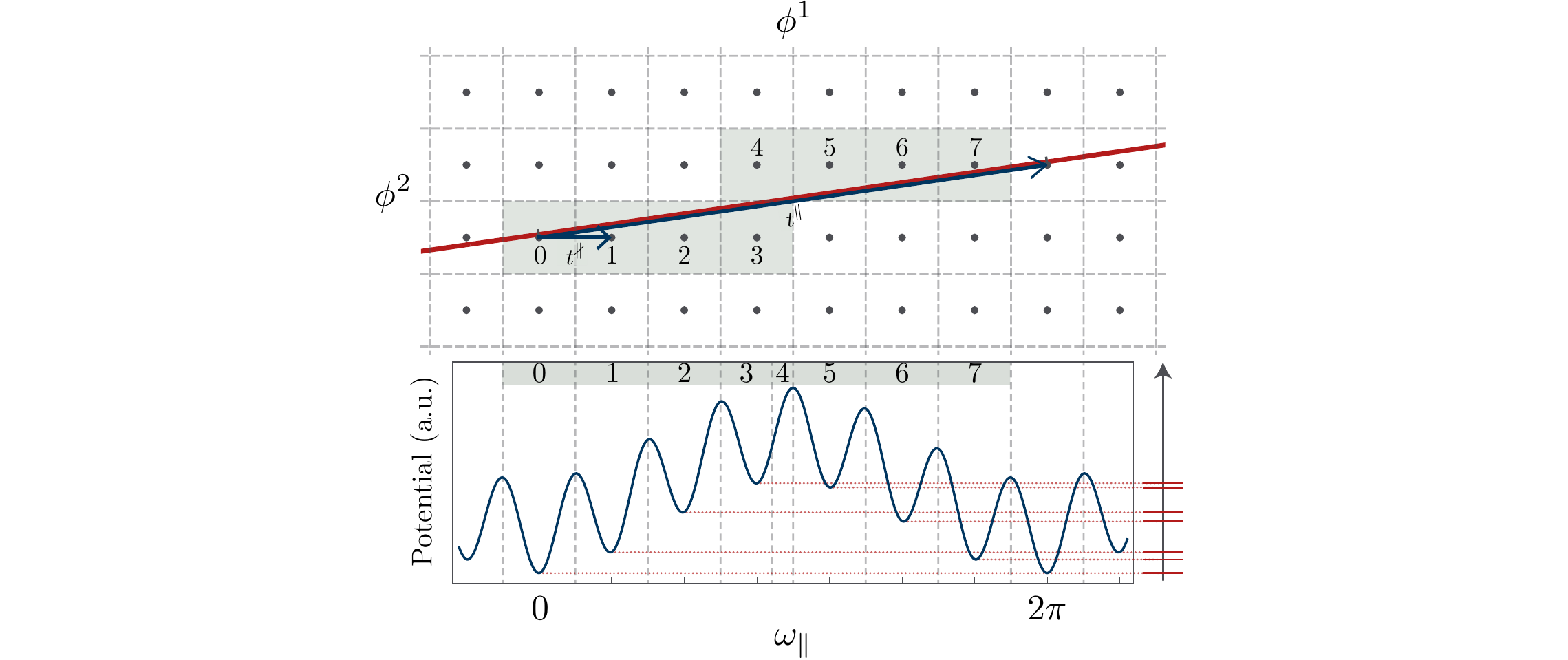}
\caption{\small \underline{Top}: Constraint surface (red line) along with the lattice $2\pi \mathbb Z^P$ (gray dots). Arrows show the aligned basis vectors $\bs t^\parallel$ and $\bs t^\nparallel$. \underline{Bottom}: Axion potential as a function of the aligned coordinate $\omega_\parallel$ (defined later in \eqref{omegadecomp}). Distinct tiles ${\mathcal T_{\bs n}}$ are numbered and shaded.}\label{fig:potplot}
\end{figure}

\begin{figure}
  \centering
  \includegraphics[width=1\textwidth]{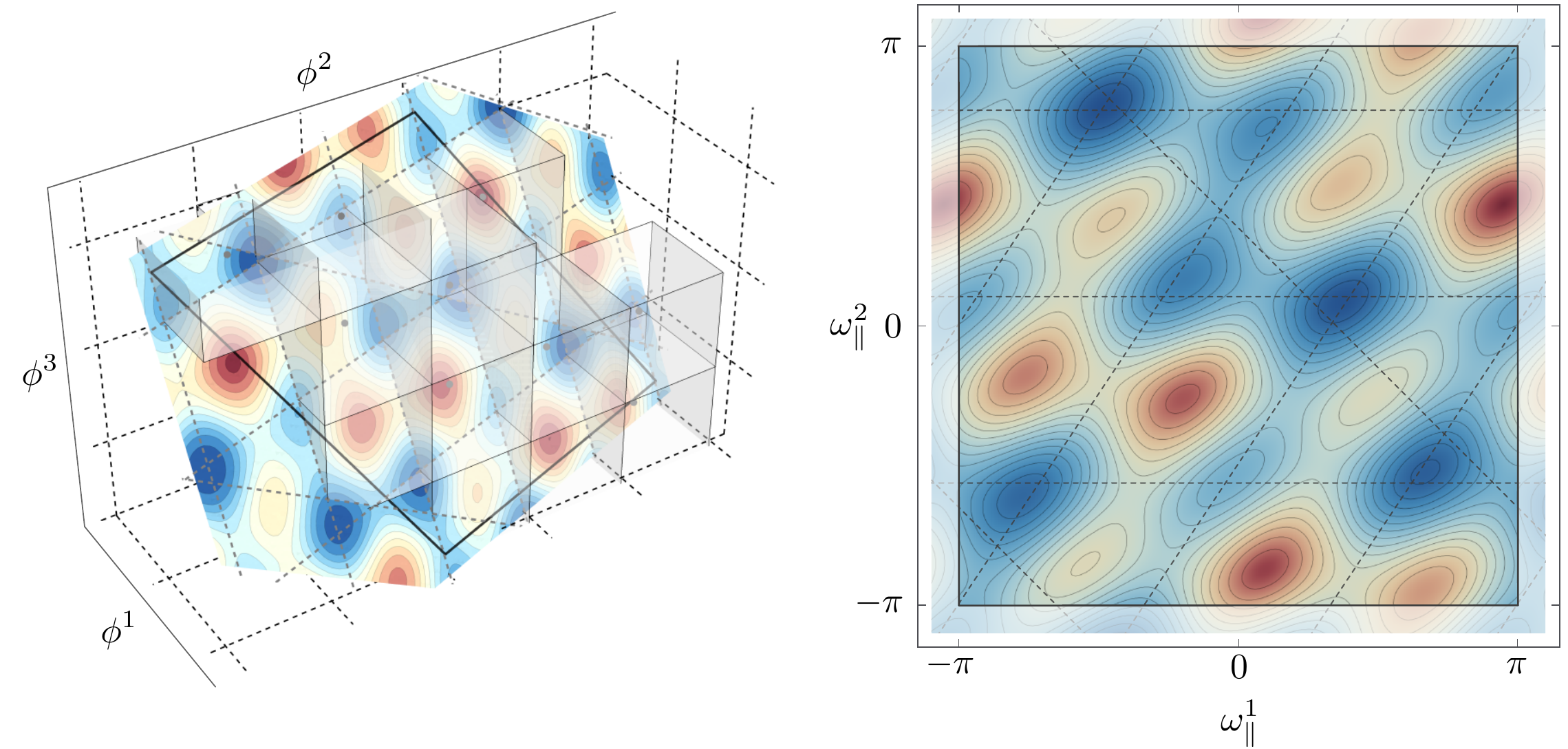}
  \caption{\small \underline{Left}: Contour plot of the auxiliary potential on the constraint surface $\Sigma$ in $\bs\phi$-coordinates, for an example with $N=2, P=3$. The solid black lines denote one periodic domain of the potential, while the dashed gray lines denote the boundaries of the tiles defined in \eqref{tiles} (the intersections of the cubical tiling \eqref{P-cube-doms} of the auxiliary potential with the constraint surface). The depicted cubes constitute a full set of those with distinct intersections with $\Sigma$. \underline{Right}: Contour plot of the physical potential and its tiles in aligned coordinates $\bs \omega_\parallel$ (defined later in \eqref{omegadecomp}). Opposing boundaries of the periodic domain are identified. \label{coordinateexamplephi}}
\end{figure}

To analyze \eqref{potentialnophases} it turns out to be useful to consider a set of $P$ real scalar fields $\bs \phi$, subject to an auxiliary potential 
\be
V_{\text{aux}}(\bs \phi) \equiv \sum_{I=1}^P \Lambda^4_I  \left[1-\cos(\phi^I) \right] \,.\label{V-aux}
\ee
Comparing to \eqref{potentialnophases} we observe that the argument of the $I$th cosine $({\mathbfcal Q} \bs\theta)^I$ has been replaced by an independent field $\phi^I$.
Hence the physical potential \eqref{potentialnophases}  is identical to \eqref{V-aux} if $\phi^I = ({\mathbfcal Q} \bs\theta)^I$, or more compactly 
\be\label{phionsigmadef}
\bs \phi|_\Sigma = {\mathbfcal Q} \bs\theta \,,
\ee 
where this equation defines a hyperplane $\Sigma$ in the auxiliary field space $\mathbb{R}^P$ (which we call $\bs \phi$-space). Specifically,
 note that
 \be \label{colsp}
{\mathbfcal Q} \bs\theta = \theta^1 \bs{\mathcal{Q}}_1+\theta^2 \bs{\mathcal{Q}}_2+\dots+\theta^N \bs{\mathcal{Q}}_N 
 \ee
is a  linear combination of the columns $\mathbfcal Q_j$ of $ \mathbfcal{Q}$.   The surface $\Sigma$ is the hyperplane spanned by these columns (the column space of $\mathbfcal Q \equiv \text{colsp}(\mathbfcal Q)$), and \eqref{potentialnophases} and \eqref{V-aux} coincide when $\bs \phi$ is constrained to $\Sigma$ :
\be 
V(\bs \theta) =V_{\text{aux}}\left( \bs \phi |_\Sigma \right) \,.  \label{Vpotentialnophasesaux}
\ee 
For this reason we refer to $\Sigma$ as the constraint surface (cf.~Figures \ref{fig:potplot} and \ref{coordinateexamplephi}).  An efficient way to impose this constraint on $\bs \phi$ is to introduce $P-N$ Lagrange multiplier fields into the action; we will do so explicitly in \S\ref{phases}. For $P \geq N$ the dimension of $\Sigma$ is $N$ if the columns of $\mathbfcal Q$ are linearly independent, and so the map is one-to-one. In this case $\mathbfcal Q$ is called full rank. We can assume this to be true without loss of generality and will do so from now on (cf. appendix \ref{app:massless}).

The utility of framing the problem in the extended $P$-dimensional space stems from the fact that the symmetries of $V_{\text{aux}}$ are manifest: $\phi^I \rightarrow \phi^I + 2 \pi n^I$ with $n^I \in \mathbb{Z}$, so that $V_{\text{aux}}$ is identical in  $P$-cubes of side-length $2\pi$ that we take to be centered on the sites of the scaled integer lattice $2 \pi \mathbb Z^P$. Each cube can be labeled by an integer $P$-vector $\bs n$:
\be 
 \{\bs \phi~:~\lVert \bs \phi-2\pi \bs n\rVert_\infty\le\pi\} \,, ~~\bs n\in \mathbb{Z}^P \,, \label{P-cube-doms}
\ee
where the $\ell_\infty$-norm of a vector returns its largest absolute component.\footnote{Clearly the $\ell_\infty$-norm is basis dependent. We denote the basis of a vector by its symbol, i.e. $\bs\phi$ is a vector in the standard basis for $\bs\phi$-space, while $\bs\theta$ is a vector in the standard basis for $\bs \theta$-space ($\mathbb{R}^N$).} Within a single $P$-cube  the potential $V_{\text{aux}}$ is relatively featureless and every $P$-cube contains a single minimum at its center $\bs \phi = 2 \pi \bs n$ where $V_{\text{aux}} = 0$. Points where $\Sigma$ passes through the center of a $P$-cube are therefore global minima of the physical potential. This set of points forms a sublattice $\mathscr{L}_\Sigma \equiv \Sigma\cap2\pi \mathbb{Z}^P$.  It is simple to see that this sublattice is rank $N$ if ${\mathbfcal Q}$ has integer entries and  full rank.\footnote{\label{Qintegerfootnote}One might worry that  $GL(N,\mathbb R)$ transformations of the axion fields do not preserve the fact that ${\mathbfcal Q}$ has integer entries. In fact, the necessary and sufficient condition on $\mathbfcal Q$ such that $\mathscr{L}_\Sigma = \text{colsp}(\mathbfcal Q) \, \cap \, 2 \pi \mathbb{Z}^P$ is rank $N$ is that $\Po = {\mathbfcal Q} ({\mathbfcal Q}^\top{\mathbfcal Q})^{-1}{\mathbfcal Q}^\top$, which is the orthogonal projector onto $\text{colsp}(\mathbfcal Q)$, contains only rational entries.  This property is preserved under $GL(N,\mathbb R)$ transformations.} The auxiliary potential is manifestly invariant under shifts between any pair of such points, and therefore so is the physical potential $V$. In other words, this sublattice defines the $N$ exact shift symmetries of \eqref{lagrtheta}.

The tiling of $\bs \phi$-space into $P$-cubes \eqref{P-cube-doms} provides a useful way to divide the physical field space into distinct regions.  The constraint surface $\Sigma$ slices across the cubes, and the regions of intersection of $\Sigma$ with various $P$-cubes are an $N$-dimensional tiling of $\Sigma$ (see Figure \ref{coordinateexamplephi}). Within each tile the potential is relatively smooth because none of the individual cosine terms in \eqref{potentialnophases} traverses its respective period. We can label each tile by the integer $P$-vector $\bs n$ of the corresponding $P$-cube \eqref{P-cube-doms}:
\be 
{\mathcal T_{\bs n}} =\{\bs \theta~:~\lVert  \mathbfcal{Q}\bs \theta-2\pi \bs n\rVert_\infty\le\pi\} \,, ~~\bs n\in  \mathbb{Z}^P \,. \label{tiles}
\ee
Not every integer $P$-vector $\bs n$ corresponds to a tile because some $P$-cubes do not intersect $\Sigma$. 

When some or all of the angles of $\Sigma$ with respect to the grid defined by \eqref{P-cube-doms} are small, one expects that at least some shifts from an initial tile to an adjacent one (adding 1 to one of the components of $\bs n$) will leave the physical potential close to invariant.   This is  the case, but we will see  in \S\ref{aligned-subsec} that we can define a different set of shifts that are in general closer to exact symmetries than these, and have the desirable property that they form a complete (but not overcomplete) basis for the set of all distinct tiles \eqref{tiles}.

\subsection{Exact periodicities and periodic domains}\label{periodic-doms-sec}
As discussed above, displacements between points in the sublattice $\mathscr{L}_\Sigma \equiv \Sigma\cap2\pi \mathbb{Z}^P$ leave $V_{\text{aux}}$ unchanged and lie within the constraint surface $\Sigma$. These  are the exact shift symmetries of the physical potential.

In general, a basis for a rank $M$ lattice is a set of $M$ linearly independent vectors with the property that \emph{integer} linear combinations generate all lattice points. Consider an $M$-parallelepiped, with edges defined by the basis vectors of a lattice. This parallelepiped is a periodic domain for the lattice and contains exactly one lattice point. A simple example is a $P$-hypercube in \eqref{P-cube-doms} with e.g.~$\bs n = \bs 0$, which is a periodic domain for the $P$-dimensional lattice $2 \pi \mathbb{Z}^P$.

We will use the notation $\{\bs t^\parallel_i\}$ to denote the $N$ integer vectors that generate the lattice $\Sigma\cap\mathbb{Z}^P$ (so that the vectors $\{2 \pi \bs t^\parallel_i\}$ generate $\mathscr{L}_\Sigma$).\footnote{Given a hyperplane and a lattice, it is non-trivial to find a basis for the sublattice resulting from their intersection. For a rank $M$ sublattice a set of $M$ linearly independent lattice vectors that lie within in the hyperplane do not in general generate all points in the sublattice. For instance, the columns of $\mathbfcal{Q}$ do not generally serve as a basis for $\mathscr{L}_\Sigma$. It is possible, however, to find the sublattice basis algorithmically, for instance with the extended LLL algorithm \cite{lllpackage}.} A single cell of this lattice sublattice is a region in which all distinct features of the potential are captured -- in other words, it is a periodic domain of the physical potential, and we can restrict our attention to one such cell.

Any basis  $\{ 2 \pi \bs t^\parallel_i\}$ for $\mathscr{L}_\Sigma$ forms a \emph{primitive set} for the full auxiliary lattice $2\pi \mathbb{Z}^P$ (see appendix \ref{app:lattice} for a proof). This means there exists a set of $P-N$ lattice vectors $\{\bs t^\nparallel_1,\dots,\bs t^\nparallel_{P-N}\}$ that are not parallel to $\Sigma$ and  that when combined with the $N$ vectors $\{\bs t^\parallel_i\}$  form a basis for $ \mathbb{Z}^P$. It will be important in a moment that the $P-N$ supplemental vectors are not unique. The only condition is that the matrix containing all $P$ basis vectors
\be
\big[ \bs t^\parallel_1 \dots\bs t^\parallel_{N}~ \bs t^\nparallel_1\dots\bs t^\nparallel_{P-N}\big]\label{t-basis}
\ee
is unimodular (determinant one with integer entries).  Whenever the vectors $\bs t^\nparallel_a$ are (at least somewhat) aligned with $\Sigma$, in a sense we shall make more precise below,  we refer to this basis for $ \mathbb{Z}^P$ as the \emph{aligned lattice basis}.

\subsection{Approximate symmetries and well-aligned theories}\label{aligned-subsec}
In general the transverse lattice vectors $\bs t^\nparallel_a$ will not be orthogonal to $\Sigma$. Their decomposition into $\Sigma$ and its orthogonal complement $\Sigma^\perp$ will be important. We label the orthogonal projectors onto these subspaces by $\Po$ and $\Pob$, respectively. Since $\Sigma = \text{colsp}(\mathbfcal Q)$, it follows that $\Sigma^\perp = \text{ker}(\mathbfcal Q^\top)$ and we have the following explicit form of the projectors in terms of the charge matrix,
\be\label{podef}
 \Po = \mathbbold{1}_P - \Pob = {\mathbfcal Q} ({\mathbfcal Q}^\top{\mathbfcal Q})^{-1}{\mathbfcal Q}^\top \,.
 \ee

Now consider a shift of the fields generated by the projection of a non-parallel lattice vector onto $\Sigma$ :
\be
\mathbfcal{Q}\bs\theta\rightarrow\mathbfcal{Q}\bs\theta+2\pi\Po\thinspace \bs t^\nparallel_a \,.
\ee
This shift is projected onto $\Sigma$ and hence corresponds to a physical shift of the potential $V = V_\text{aux}|_\Sigma$.  However it is not an exact symmetry because $\Po\thinspace \bs t^\nparallel_a $ is not  an integer vector. The amount by which this shift breaks the symmetry is proportional the projection of $\bs t^\nparallel_a$ onto $\Sigma^\perp$ :
\bea
V_{\text{aux}}(\bs \phi) & \rightarrow & V_{\text{aux}}(\bs \phi +2\pi \Po \bs t^\nparallel_a) \\
&=& V_{\text{aux}}(\bs \phi +2\pi  \bs t^\nparallel_a-2\pi\Pob  \bs t^\nparallel_a)  \\
&= & V_{\text{aux}}(\bs \phi -2\pi\Pob  \bs t^\nparallel_a) \,,
\eea
where in the second step we used that the potential is invariant under $\phi^I\rightarrow \phi^I+2\pi$. Therefore, if the integer vectors $\bs t^{\nparallel}_a$ can be chosen so that each component of $\Pob  \bs t^\nparallel_a$ is much less than one --  if $  \lVert\Pob  \bs t^\nparallel_a\rVert_\infty \ll 1$ -- the correction to each cosine term in \eqref{V-aux} is small and the shift $\bs \phi \rightarrow \bs \phi +2\pi \Po \bs t^\nparallel_a$ is an approximate symmetry.

To identify both the exact and approximate symmetries, we choose a basis \eqref{t-basis} for $\mathbb Z^P$ which is as aligned as possible with $\Sigma$.  The first $N$ vectors $\bs t^\parallel_i$ lie in $\Sigma$ (and are a basis for the lattice $\Sigma\cap\mathbb{Z}^P$), thus
\be 
\Pob \bs t^\parallel_i = \bs 0 \,, ~~\forall i \in \{1, \dots, N\}  \,.
\ee
These vectors generate the $N$ exact shift symmetries of the physical potential and any parallelepiped with the $\bs t_{i}^\parallel$ as edges is a periodic domain of the potential. 
The remaining $P-N$ vectors $\bs t^\nparallel_a$ should satisfy
\be \label{linf}
\left\Vert \Pob \bs t^\nparallel_a \right\Vert_\infty \,\text{are smallest possible}, ~~\forall a \in \{1, \dots, P-N\} \,.
\ee
The vectors $ 2\pi \bs P \bs t^\nparallel_a$ generate $P-N$ approximate symmetries of the potential.\footnote{Equation \eqref{linf} defines what is known as a reduced basis for the lattice generated by $\Pob$.  We are purposefully vague in the precise definition of ``smallest possible" and ``reduced": there are multiple definitions, such as Minkowski, LLL, or Rankin reduction. For example, depending on the precise application, one may be interested in a basis that aligns only some of the $P-N$ transverse directions. These details are irrelevant for the present discussion and we refer to the literature \cite{minkowski1911,Lenstra1982,Nguyen2004,Gama2006,Chen2011,Dadush:2013:ADS:2627817.2627896,Li:2014:ADS}. A particularly simple approximation is given by the Mathematica package for the extended LLL algorithm \cite{lllpackage}. We thank Liam McAllister and John Stout for discussion on this point.} The aligned lattice basis for a simple axion potential, one with $P=2$ and $N=1$, is shown in Figure \ref{fig:potplot}. We refer to theories where the orthogonal projections of all elements of the aligned basis are small as \textit{well-aligned}. That is, a well-aligned theory satisfies
\be \label{well-aligned-def}
\left\Vert \Pob \bs t^\nparallel_a \right\Vert_\infty\ll {1 \over P-N} \,,~~\forall a \in \{1, \dots, P-N\} \,
\ee
(cf.~\eqref{maxremainingphase} for the origin of the $1/(P-N)$ on the right-hand side).

To illustrate the utility of these approximate symmetries, suppose  $\bs \phi$ is chosen to be a global minimum of the potential that lies on $\Sigma$ (for instance $\bs \phi = \bs 0$). Repeated  shifts by $2\pi \Po \bs t^\nparallel_a$ then identify the approximate location (and determine the energy, as we  discuss in \S\ref{minimaquadsec}) of many physically distinct minima with slightly different properties (see Figure \ref{fig:potplot}). We will see later that in the random ensembles we study $\lVert\Pob  \bs t^\nparallel_a\rVert_\infty$ can generically be chosen so that it is exponentially small in $N$, for all $1 \leq a \leq P-N$.  More generally it can be small when $\det(\mathbfcal Q^\top \mathbfcal Q)$ is  large, since this determinant appears in the denominator of the projector. This allows us to locate and characterize  many minima in an otherwise intractably complex landscape very easily and to good accuracy. Furthermore, in well-aligned theories all $P$ phases $\delta^I$ can be eliminated to good accuracy (cf. \S\ref{phases}).

\subsection{Aligned coordinates and similar tiles} \label{similartilessec}
The tiling \eqref{tiles} of $\Sigma$ was introduced in \S\ref{sec:geompic} as a means of delineating relatively featureless sections of the potential by using the basic infrastructure provided by the auxiliary lattice. Here we show how this tiling enables the identification of many similar regions within one periodic domain of the physical potential.

As a preliminary illustration of a more general method, consider the periodic domain surrounding a global minimum of the potential (say the origin $\bs \theta = \bs 0$), and a $P$-cube centered at this position on $\Sigma$. Now choose a specific non-parallel lattice vector $\bs t^\nparallel_a$ and consider the set of $P$-cubes obtained by sequentially shifting the center of each cube by $2\pi \bs t^\nparallel_a$, together with the tiles defined as their intersections with $\Sigma$. In well-aligned theories the auxiliary potential evaluated on successive intersections in the list (and hence the physical potential in the corresponding tiles) will be very similar. Now, regardless of whether or not a model is well-aligned, after a certain number of shifts the $P$-cube will have receded far enough from $\Sigma$ that it no longer intersects it. If the number of shifts before this point is $m$, we  have identified $m$ distinct tiles that are labeled by consecutive integer multiples of $\bs t^\nparallel_a$.\footnote{Note that $m \sim 1/\Vert \Pob \bs t^\nparallel_a \Vert_\infty$, which is large in a well-aligned theory.}  These tiles may be scattered across multiple periodic domains because accumulating shifts eventually push part or all of the intersection out of the original periodic domain. Shifts by integer linear combinations of the $2\pi \bs t^\parallel_i$ can then be used to uniquely return all portions of tiles into the periodic domain containing the origin. All such tiles are distinct because they originated from distinct intersections with $\Sigma$. 
\begin{figure}
  \centering
  \includegraphics[width=1\textwidth]{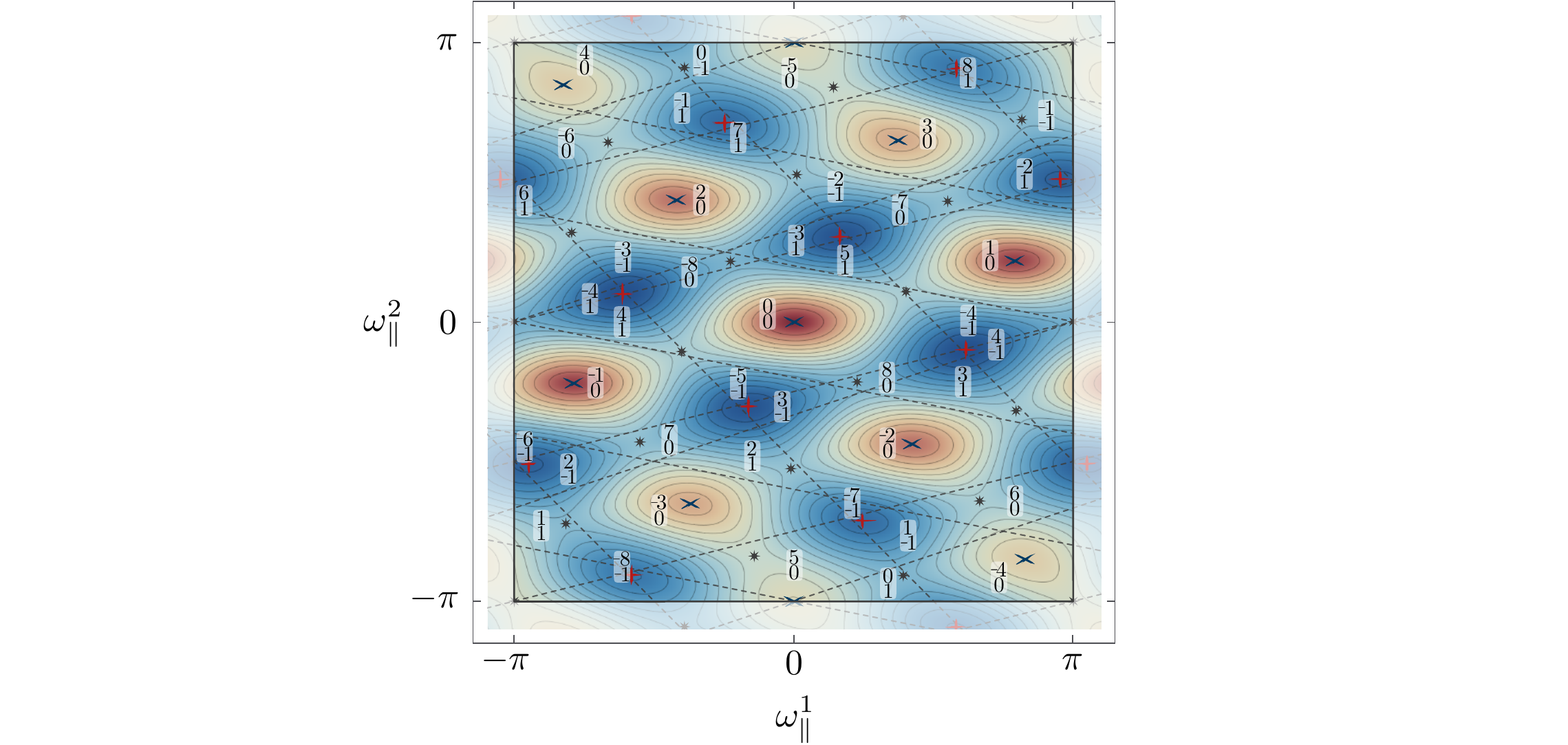}
  \caption{\small Illustration  of the tiling of the periodic domain of the physical potential induced by the intersections of the constraint surface with $P$-cubes surrounding lattice sites in the auxiliary lattice. We used $P=N+2=4$ and labeled each distinct intersection by a $P-N = 2$-vector. This example is well-aligned only in the direction $\bs t^{\nparallel}_1$ corresponding to the upper label (if both directions are well-aligned, the plot becomes too dense to be legible).\label{tiling}}
\end{figure}

A complete tiling of the periodic domain can be found by following a generalization of the procedure outlined above -- shifting the $P$-cube containing the global minimum by linear combinations of all non-parallel directions $2\pi \bs t^\nparallel_a$ , and scanning this space until all intersecting $P$-cubes are identified.  A more convenient labeling of the regions that tile the periodic domain is achieved by defining \emph{aligned coordinates} $\bs \omega$ for the auxiliary space:\footnote{It may be useful to reduce the basis for the lattice spanned by the $\bs t^\parallel_i$.}
\be \label{alignc}
\bs\phi\equiv \big( ~{\bs t^\parallel} ~\rvert ~{\bs t^\nparallel}~  \big)\thickspace \bs \omega \,.
\ee
Recall that the matrix appearing in \eqref{alignc}, which is identical to \eqref{t-basis}, is unimodular (determinant one) and thus has a unimodular inverse.  Integer $P$-vectors in $\bs \phi$-coordinates are in one-to-one correspondence with integer vectors in $\bs \omega$-coordinates. The components of $\bs \omega$ separate into components parallel and not parallel to $\Sigma$ :
\be \label{omegadecomp}
\bs \omega= \left(\begin{matrix}\bs{\omega}_\parallel\vspace{3pt}\\
\hline
\bs{\omega}_\nparallel
\end{matrix}\right) \,.
\ee
A shift by $2\pi$ of any of the first $N$ entries leaves the physical potential invariant, since this is a shift of $\bs \phi$ by a $2\pi \bs t^\parallel_i$. The periodic domain in $\bs \omega$-coordinates is simply an $N$-cube of side-length $2\pi$ in the $\bs \omega_\parallel$-plane. Since opposing sides of the periodic domain are identified, any point on $\Sigma$ is readily identified with a corresponding point in the central periodic domain $\bs \omega_\parallel ~(\text{mod}~2\pi \mathbb{Z}^N)$. It is likewise easy to recognize similar tiles in well-aligned theories: they correspond to  $P$-cubes with similar $\bs \omega_\nparallel$-coordinates. The above is illustrated in Figure \ref{tiling}. Note finally that an advantage of using aligned coordinates is that distinct tiles within one periodic domain are labeled by distinct integer $(P-N)$-vectors $\bs m$,\footnote{And vice-versa, modulo those related by $\bs m \leftrightarrow - \bs m$, as the $P$-cube grid and $N$-parallelepiped domains in $\Sigma$ are symmetric under the reflection about any of the $P$ Cartesian coordinate axes.} as $P-N$ is the amount of $\bs t^\nparallel_a$s and integral shifts along the $\bs \omega_\parallel$-directions do not change the tile. \\
 
\textit{Notational intermezzo -- }To simplify equations like \eqref{alignc}, from now on we adopt the notation $\trafo{\alpha}{\beta}$ for a transformation between the components of a vector in two bases. The subscript pair is to be read left to right: ``transform from $\bs \alpha$-coordinates to $\bs \beta$-coordinates". In this notation \eqref{alignc} becomes
$$
\bs \phi = \trafo{\omega}{\phi} \bs \omega \,.
$$
The inverse transformation is simply
$$
\bs \omega = \trafo{\phi}{\omega} \bs \phi \,.
$$
When the transformation is between spaces of equal dimension as here (so that $\bs T$ is square), $\trafo{\phi}{\omega} = (\trafo{\omega}{\phi})^{-1}$. When a transformation matrix's columns or rows separate in a useful way, as is the case here,
\be
\trafo{\omega}{\phi} = \big( ~{\bs t^\parallel} ~\rvert ~{\bs t^\nparallel}~  \big) \,,
\ee
the (rectangular) submatrices are labeled in the natural way :
 \be \label{abasis}
\big( \hspace{-3pt}\underbrace{\bs t^\parallel}_{\vspace{1pt}\equiv \trafo{\omega_\parallel}{\phi}} \vert \underbrace{\bs t^\nparallel}_{\vspace{1pt}\equiv \trafo{\omega_\nparallel}{\phi}} \hspace{-3pt}\big)\,.
\ee
We can apply this notation to the transformation (\ref{phionsigmadef}) between the $N$-vector $\bs \theta$ and the $P$-vector  $\bs \phi$ constrained to $\Sigma$, 
\be\label{thetaphitrafo}
\trafo{\theta}{\phi} =\mathbfcal Q\,,~~~\trafo{\phi}{\theta}= (\mathbfcal Q^\top \mathbfcal Q)^{-1} \mathbfcal Q^\top \,.
\ee
Combining the transformations (\ref{abasis}) and (\ref{thetaphitrafo}) we can also relate the coordinates $\bs \omega_\parallel$ and $\bs \theta$,
\be \label{moretrafos}
\trafo{\omega_\parallel}{\theta}=\trafo{\phi}{\theta}\trafo{\omega_\parallel}{\phi}\,,~~\trafo{\theta}{\omega_\parallel} = (\bs T_{\omega_\parallel\theta})^{-1}.
\ee
Finally, we can express the exact and approximate shift symmetries of the theory in terms of the $\bs \theta$-coordinates we started off with in \eqref{lagrtheta} (in \S\ref{periodic-doms-sec} and \S\ref{aligned-subsec} these were only expressed in $\bs \phi$-coordinates). The exact symmetries are given by
\begin{align}
	\bs \theta \rightarrow ~ &\bs \theta + 2 \pi \trafo{\phi}{\theta} \bs t^\parallel_i \\ = ~ &\bs \theta + 2 \pi (\mathbfcal Q^\top \mathbfcal Q)^{-1} \mathbfcal Q^\top \bs t^\parallel_i \,.
\end{align}
while the approximate symmetries are given by
\begin{align}
	\bs \theta \rightarrow ~ &\bs \theta + 2 \pi \trafo{\phi}{\theta} \Po \bs t^\nparallel_a \\ = ~ &\bs \theta + 2 \pi (\mathbfcal Q^\top \mathbfcal Q)^{-1} \mathbfcal Q^\top \bs t^\nparallel_a \,.
\end{align}

\begin{figure}
  \centering
  \includegraphics[width=1\textwidth]{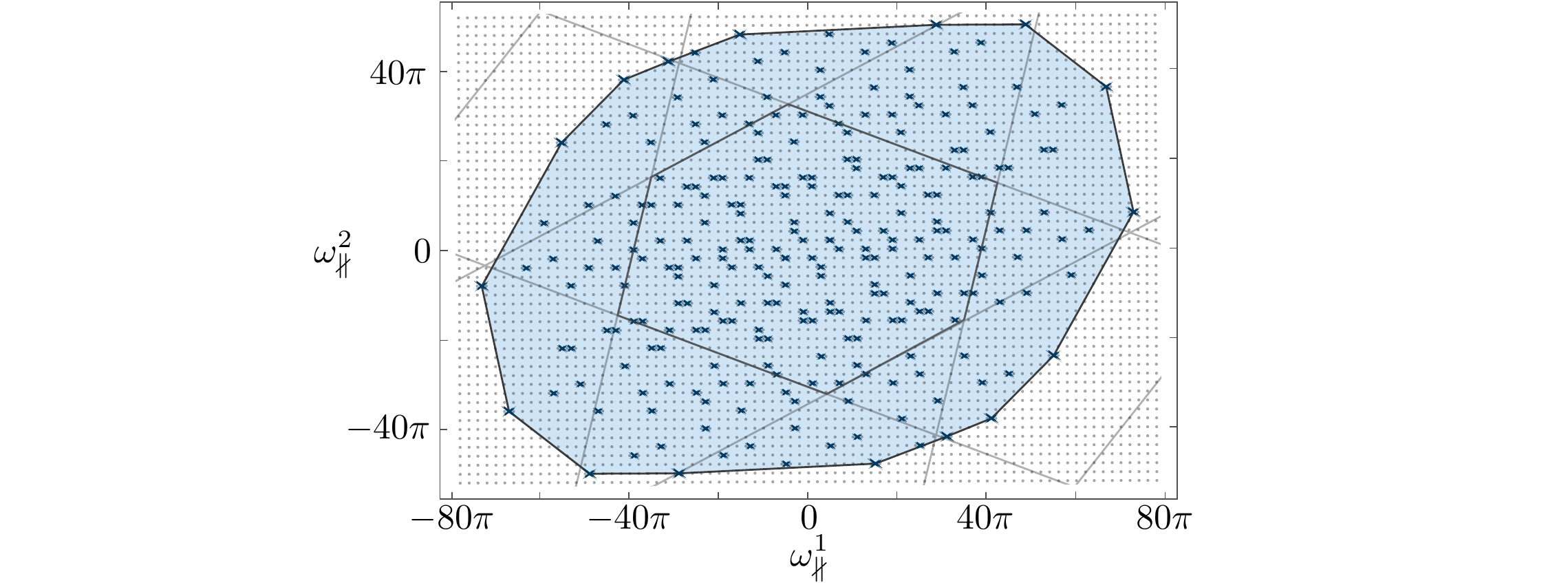}
  \caption{\small Illustration  of the $\bs \omega_\nparallel$-coordinates for every distinct tile on $\Sigma$, in an example where $P=N+2=8$. The central six-sided region denotes the area bounded by the simple sufficient condition \eqref{roughintersection}, while the full shaded region $\mathfrak C$ contains all lattice sites whose $P$-cubes intersect the constraint surface. Blue crosses (large and small) denote the coordinates of centers of cubes which have a vertex that lies on $\Sigma$. \label{convexhullandminima}}
\end{figure}

Returning to the tiling of the periodic domain, it is not hard to see that there are only finitely many tiles: consider sliding the center of a $P$-cube along any real linear combination of the $\bs t^\nparallel_a$ (which corresponds to a line emanating from the origin in the $(P-N)$-dimensional $\bs \omega_\nparallel$-space) -- a generalization of the illustration at the beginning of this section. As the perpendicular distance from the center of the cube to $\Sigma$ is ever-increasing along the line, eventually the cube will no longer intersect $\Sigma$. Thus there are only finitely many distinct tiles, which are labeled by certain integer $(P-N)$-vectors $\bs m$. We now argue that the allowed values for $\bs m$ lie in a particular (compact) convex region $\mathfrak C$ in $\bs \omega_\nparallel$-space. We define $\mathfrak C$ by the set of \textit{all} points in $\bs \omega_\nparallel$-space which correspond to centers of $P$-cubes in $\bs \phi$-space that have \textit{some} intersection with $\Sigma$ (note that this includes also vectors with irrational entries). This is the same as the collection of the $\bs \omega_\nparallel$-coordinates of all the points lying inside the $P$-cube of side-length $2 \pi$ centered at the origin in $\bs \phi$-space, or yet in other terms, the orthogonal projection of that $P$-cube onto $\Sigma^\perp$ (in $\bs \omega_\nparallel$-coordinates). It follows that distinct tiles are labeled by distinct integer vectors $\bs m \in \mathfrak{C}$ with\footnote{This is indeed a convex set: if $\trafo{\phi}{\omega_\nparallel} \bs \phi_1, \trafo{\phi}{\omega_\nparallel} \bs \phi_2 \in \mathfrak C$ then also $\lambda \trafo{\phi}{\omega_\nparallel} \bs \phi_1 + (1 - \lambda) \trafo{\phi}{\omega_\nparallel} \bs \phi_2 \in \mathfrak C$ since it is of the form $\trafo{\phi}{\omega_\nparallel} \left[ \lambda \bs \phi_1 + (1-\lambda) \bs \phi_2 \right]$ with indeed $\Vert \lambda \bs \phi_1 + (1-\lambda) \bs \phi_2 \Vert_\infty \leq \lambda \Vert \bs \phi_1 \Vert_\infty + (1-\lambda) \Vert \bs \phi_2 \Vert_\infty \leq \pi$.}
\bea\label{defboundaryC}
\mathfrak C = \left\{ \trafo{\phi}{\omega_\nparallel} \bs \phi \, \Big | \, \lVert \bs \phi\rVert_\infty\le \pi \right\} \,.
\eea
Quite clearly the extreme points\footnote{Extreme points of a convex set are points which are not interior points of any line segment belonging to the set.} of $\mathfrak C$ correspond to the projections of certain vertices of the $P$-cube. Since a compact convex set equals the (closed) convex hull\footnote{The convex hull of a set is the intersection of all convex sets containing that set. For the vertices of a polytope, the convex hull is the polytope.} of its extreme points by the Krein-Milman theorem \cite{minkowski1911,KreinMilman}, we have the following alternate definition of $\mathfrak C$ :
\begin{equation} \label{defboundaryC}
	\mathfrak C = \text{Conv} \left( \left\{ \trafo{\phi}{\omega_\nparallel}\bs e \, | \, e^I = \pm \pi, I \in  \left\{ 1 \dots P \right\} \right\} \right) \,,
\end{equation}
where $\text{Conv}( \, \cdot \, )$ denotes the convex hull of a set. $\mathfrak C$ is a polytope in the $(P-N)$-dimensional $\bs m$-space, illustrated in Figure \ref{convexhullandminima}.\footnote{In principle one could eliminate all the redundant points in the set of which the convex hull is being taken in \eqref{defboundaryC}, which correspond to vertices of cubes that lie on the constraint surface but which are not the only intersection of the $P$-cube with $\Sigma$ (cf. Figure \ref{convexhullandminima}), and maintain the same polytope $\mathfrak C$. The amount of remaining (extreme) points of $\mathfrak C$ is much less than the $2^P$ used in \eqref{defboundaryC}: we believe an upper bound scales only polynomially as $P^{P-N-1}$. However, we are not aware of a polynomial time algorithm that can find $\mathfrak C$.}

To recap, the aligned coordinates $\bs \omega$ are very convenient to identify a complete set of tiles covering exactly one periodic domain of the potential of \eqref{lagrtheta}.  First, it suffices to fix $\bs \omega_{\parallel} = \bs 0$.  This guarantees that only one periodic domain's worth of tiles will be counted, and  the value of $\bs \omega_{\parallel}$ at a point is irrelevant to how or if a $P$-cube centered there intersects $\Sigma$. Therefore, all distinct tiles are labeled by the $\bs \omega_\nparallel$-coordinates of the centers of their $P$-cubes, and there is some compact and convex region $\mathfrak C$ of $\bs \omega_{\nparallel}$-space that contains them all.

In general it is computationally very hard to identify the vertices that define the polytope ${\mathfrak C}$ in (\ref{defboundaryC}). A simple sufficient condition for a lattice site $2\pi {\bs m} $ to lie within the polytope is that its projections onto the $\bs \phi$-coordinate axes do not exceed $\pi$,
\be\label{roughintersection}
\lVert 2\pi \Pob\trafo{\omega_\nparallel}{\phi}{\bs m}\rVert_\infty\le \pi\,~~\Rightarrow ~~{\bs m}\in {\mathfrak C}\,,
\ee
while the inverse is not true. This subregion of the polytope is illustrated in Figure \ref{convexhullandminima}.

As mentioned above, another advantage of the $\bs \omega$-coordinates is that in well-aligned theories, regions of $\Sigma$ corresponding to intersections with $P$-cubes that are close together in $\bs \omega_\nparallel$-space will be nearly identical, because they differ by only a small number of shifts by approximate discrete symmetries. The corresponding regions may be very far apart even after modding to one periodic domain in $\bs \omega_\parallel$-space, because the shifts $2 \pi \Po \bs t^\nparallel_a$ can be very long.  Neighboring regions on $\Sigma$ are not in general similar, while specific distant regions are. This is illustrated in Figure \ref{coordinateexamplephi1}: there is no clear structure in the potential along an arbitrary ray in field space, but when considering lines that intersect widely separated tiles related by exact or approximate shift symmetries the potential becomes structured.

\subsection{Phases}\label{phases}
We now return to the phases $\bs \delta$ appearing in the original Lagrangian (\ref{lagrtheta}), with potential
\be \label{phasepot}
V = \sum_{I=1}^P \Lambda_I^4  \left[ 1-\cos \left({\mathbfcal Q}\bs\theta +\bs \delta \right)^I  \right].
\ee
Just as before, we promote the cosine arguments to $P$ independent fields $\phi^I$ that must be constrained to a hyperplane in order to reproduce the physical potential. That is, we require 
\be\label{phideltas}
\bs \phi = \mathbfcal Q \bs\theta + \bs \delta \,,
\ee
which defines a hyperplane parallel to $\Sigma = \text{colsp}(\mathbfcal Q)$, such that the constraint surface on which the auxiliary potential reproduces the physical potential (\ref{phasepot}) is $\Sigma+\bs\delta$.

To impose the constraint (\ref{phideltas}) in the action we introduce $P-N$ Lagrange multipliers $\nu_a$ :
\be \label{phipot22}
V=\sum_{I=1}^P \Lambda_I^4  \left[1-\cos(\phi^I ) \right]+\sum_{a=1}^{P-N} \nu_a \mathbfcal R^{a} ( \bs \phi -\Pob\bs\delta) \,.
\ee
Here $\mathbfcal R$ is any $(P-N) \times P$ matrix with the property that its row space  is $\Sigma^{\perp}$, the orthogonal complement to $\Sigma$. For instance, for $\mathbfcal R$ one could use any $P-N$ linearly independent rows of the matrix $\Pob = \mathbbold{1}_{P} -\Po$. The equations of motion for the $\nu_{a}$  constrain $\bs \phi - \Pob\bs\delta$ to be perpendicular to $\Sigma^{\perp}$; that is, they constrain $\bs \phi$ to lie in $\Sigma + \Pob\bs\delta = \Sigma + \bs \delta$. In checking this the identity $\mathbfcal R \Pob = \mathbfcal R $ is useful.

Since $\Pob \bs \delta$ is a vector in the $(P-N)$-dimensional subspace $\Sigma^\perp$, the projection in (\ref{phipot22}) has already removed all but $P-N$ of the original $P$ phases (this reduction is simply the obvious freedom to continuously redefine the $N$ fields $\bs \theta$ in  \eqref{phasepot}). We will now demonstrate that in  well-aligned  theories, the remaining  phases $\Pob \bs \delta$ can be reduced to small values using the approximate shift symmetries of the theory. Consider the shift $\bs\phi\rightarrow \bs\phi+2\pi \trafo{\omega_{\nparallel}}{\phi}\bs n_{\bs \delta}$, where $\bs n_{\bs \delta}$ is an arbitrary integer  $(P-N)$-vector.  This shift is an exact symmetry of the cosines, but affects the constraint terms in \eqref{phipot22} :
\be \label{phipot23}
V=\sum_{I} \Lambda_I^4  \left[1-\cos(\phi^I ) \right]+\sum_{a} \nu_a  \mathbfcal R^{a} \left( \bs \phi + 2\pi \trafo{\omega_{\nparallel}}{\phi}\bs n_{\bs \delta} - \Pob \bs \delta\right) \,.
\ee
To identify the integers $\bs n_{\bs \delta}$ that minimize the remaining phases in (\ref{phipot23}), recall the relation between the aligned coordinates $\bs \omega$ and $\bs \phi$-coordinates
\be
\bs \omega= \trafo{\phi}{\omega}\bs\phi\equiv\left(\begin{matrix}\trafo{\phi}{\omega_\parallel}\\\trafo{\phi}{\omega_\nparallel}\end{matrix}\right)\bs \phi\,.
\ee
Using these definitions, the vector $\bs n_{\bs \delta}$ that minimizes the remaining phases is
\be \label{ndelta}
\bs n_{\bs\delta} = \left[ {1\over 2\pi} \trafo{\phi}{\omega_{\nparallel}}  \Pob \bs \delta\right]_{\text{n.i.}}\,,
\ee
where $[\dots]_{\text{n.i.}}$ denotes the nearest integer vector. Using $\Pob = \Pob \trafo{\omega_{\nparallel}}{\phi} \trafo{\phi}{\omega_{\nparallel}}  \Pob$, one can see that this choice of $\bs n_{\bs\delta}$ reduces the phases to zero with an error bounded above by $\pi \Vert \Pob \trafo{\omega_\nparallel}{\phi} \Vert_\infty$, which is small when the theory is well-aligned.\footnote{\label{matrixinfnorm} The $\ell_\infty$-norm of a matrix $\bs A$ can be defined as the maximum absolute row sum, $\lVert \bs A \rVert_\infty = \max_i \left\{ \sum_j |A^i_{~j}| \right\}$. After the shift specified by \eqref{ndelta}, the remaining phase is $\bs \delta_{\text{r}} = 2 \pi \Pob \trafo{\omega_\nparallel}{\phi} \bs \alpha$ for some $(P-N)$-vector $\bs \alpha$ with $\lVert \bs \alpha \rVert_\infty \leq 1/2$. We may bound the magnitude of the largest component of this remaining phase by using the general inequality $\lVert \bs A \bs v \rVert_\infty \leq \lVert \bs A \rVert_\infty \, \lVert \bs v \rVert_\infty$ for any matrix $\bs A$ and vector $\bs v$: $\lVert \bs \delta_{\text{r}} \rVert_\infty \leq \pi \lVert \Pob \trafo{\omega_\nparallel}{\phi} \rVert_\infty$. To make the connection with our definition of well-aligned theories \eqref{well-aligned-def}, note $\lVert \Pob \trafo{\omega_\nparallel}{\phi} \rVert_\infty = \displaystyle\max_I \left\{ \displaystyle\sum_{a=1}^{P-N} \left| \left( \Pob \trafo{\omega_\nparallel}{\phi} \right)^I_{~a} \right| \right\} \leq \displaystyle\sum_{a=1}^{P-N} \displaystyle\max_I \left\{ \left| \left( \Pob \trafo{\omega_\nparallel}{\phi} \right)^I_{~a} \right| \right\} = $ $\displaystyle\sum_{a=1}^{P-N} \lVert \Pob \bs t^{\nparallel}_{a} \rVert_\infty$.}

Explicitly, the field redefinition $\bs\theta\rightarrow\bs\theta+\bs\theta_{\text{shift}}$ that reduces the phases, and the remaining phases are given with (\ref{thetaphitrafo}) and (\ref{ndelta}) by
\be\label{phasesrem}
\bs\theta_{\text{shift}}=\underset{[-\pi,\pi]}{\text{mod}}\,\left[\trafo{\phi}{\theta} (2\pi  \trafo{\omega_\nparallel}{\phi}\bs n_{\bs\delta}-\bs\delta) \right]\,,~~~\bs\delta_{\text{r}}=\underset{[-\pi,\pi]}{\text{mod}}\,\left[\mathbfcal Q\bs\theta_{\text{shift}}+\bs\delta\right]\,.
\ee
From now on, to simplify the discussion we will focus on well-aligned theories where we can neglect the phases. As we shall see in \S\ref{randomsec} this  holds to good accuracy for large classes of axion theories with $P\lesssim 2N$.  An explicit example illustrating the use of the aligned lattice basis and phase reduction can be found in appendix \ref{coordinateexample}.

\section{Minima and saddle points}\label{seccminima}
In the previous section we identified the most suitable basis to identify the symmetries, treating all $P$ terms in the potential on equal footing. These  symmetries can  be employed to systematically explore all potential minimum locations, which in principle yields all minima to arbitrary accuracy. In typical applications, however, it may be more efficient to include other data as well, such as the scale of each of the non-perturbative terms. In order to find the stable minima, for example, non-perturbative terms that are entirely subleading will have essentially no effect on the  location of the minimum, but may split the degeneracy between minima as discussed in \cite{Bachlechner:2015gwa} and reviewed in \S\ref{bandstructuresec}. We will now discuss how the minima of a given axion theory can be determined to various degrees of accuracy.

\subsection{Systematics of all minima}\label{vacuumsystematics}
In \S\ref{symm} we described how to decompose the field space into tiles, each of which corresponds to an intersection of  $\Sigma$ with a distinct $P$-cube in $\bs \phi$-space centered on a lattice point of $2 \pi \mathbb Z^{P}$. The auxiliary potential $V_{\text{aux}}(\bs \phi)$ is identical inside all $P$-cubes, but they can have  a distinct (or empty) intersection with the constraint surface, and therefore contain a distinct region of the physical potential $V(\bs\theta)$.

Just as already done in \S\ref{phases} we introduce $P-N$ Lagrange multipliers $\nu_a$ that enforce the constraint of the auxiliary potential to $\Sigma$. To find extrema on $\Sigma$ within a tile  labeled by $\bs m$ we then minimize the potential, 
\be \label{}
V=V_{\text{aux}}(\bs \phi)+\sum_{a=1}^{P-N} \nu_a \mathbfcal R^{a}   \bs \phi  \,,
\ee
within the corresponding $P$-cube $\bs\phi=2\pi \trafo{\omega_\nparallel}{\phi}\bs m+\delta{\bs\phi}$, where $\lVert\delta{\bs\phi}\rVert\le \pi$. Requiring a vanishing gradient gives
\bea\label{optimization}
 \Lambda_I^4  \sin\left(\delta{\phi}^I \right) + \left( \bs \nu \mathbfcal R \right)^I &=& 0 \,, \,  \,  \, \forall I \in \{1, \dots, P \} \,, \nonumber\\
\mathbfcal R \thinspace(2\pi \trafo{\omega_\nparallel}{\phi}\bs m+\delta{\bs\phi})  &=& \bs 0 \,.
\eea
Since $ \mathbfcal R $ is a set of row vectors that span $\Sigma^{\perp}$, the first condition is the requirement that the gradient of $V_{\text{aux}}$ is perpendicular to $\Sigma$ -- in other words, that the gradient projected onto $\Sigma$ vanishes. The second condition ensures that the point is in $\Sigma$. Solving the optimization problem (\ref{optimization}) within all $P$-cubes labeled by $\bs m \in \mathfrak{C}$ yields all distinct extreme points, including all minima. 

\subsection{Neighboring minima}
The aligned basis is ideally suited to identify similar regions of the axion potential. These tiles need not be close to one another in the physical field space, as we saw in \S\ref{tiles}. Recall that this is because the $\bs t^\nparallel_a$ may contain large integers and therefore generate a large separation between the tiles' associated $P$-cubes (in $\bs \phi$-coordinates). This means that similar tiles are not generally immediate neighbors. For the purpose of this paper, we define immediately neighboring minima to be those whose $P$-cubes share a face or corner.

A minimum located at $\bs \phi$, which is within in the $P$-cube labeled by $\bs n =[\bs \phi/(2\pi)]_{\text{n.i.}}$, has $3^P$ neighboring $P$-cubes labeled by
\be\label{neighboringdomains}
\bs  n_{\text{neighbor}}= \bs n+\bs e\,,~~~e^I\in\{0,\pm1\} \,,
\ee
only some of which have a non-vanishing intersection with $\Sigma$. In order to identify all immediately neighboring minima we have to consider all neighboring $P$-cubes (\ref{neighboringdomains}) that do intersect $\Sigma$. That means we need to consider all vectors $\bs e$ that correspond to lattice sites which satisfy $2\pi \trafo{\phi}{\omega_\nparallel}\Pob (\bs n+\bs e)\in \mathfrak C$ (cf.~\eqref{defboundaryC}). Again, a simple sufficient condition is given by
\be
\Vert 2\pi \Pob (\bs n+\bs e)\Vert_\infty<\pi\,.
\ee
The precise  locations of neighboring  minima can be found by solving the optimization problem \eqref{optimization} with $\bs \phi= 2\pi (\bs n+\bs e)+\delta \bs \phi$ for each candidate $\bs e$.

\subsection{Minima to quadratic order} \label{minimaquadsec}
\begin{figure}
  \centering
  \includegraphics[width=1\textwidth]{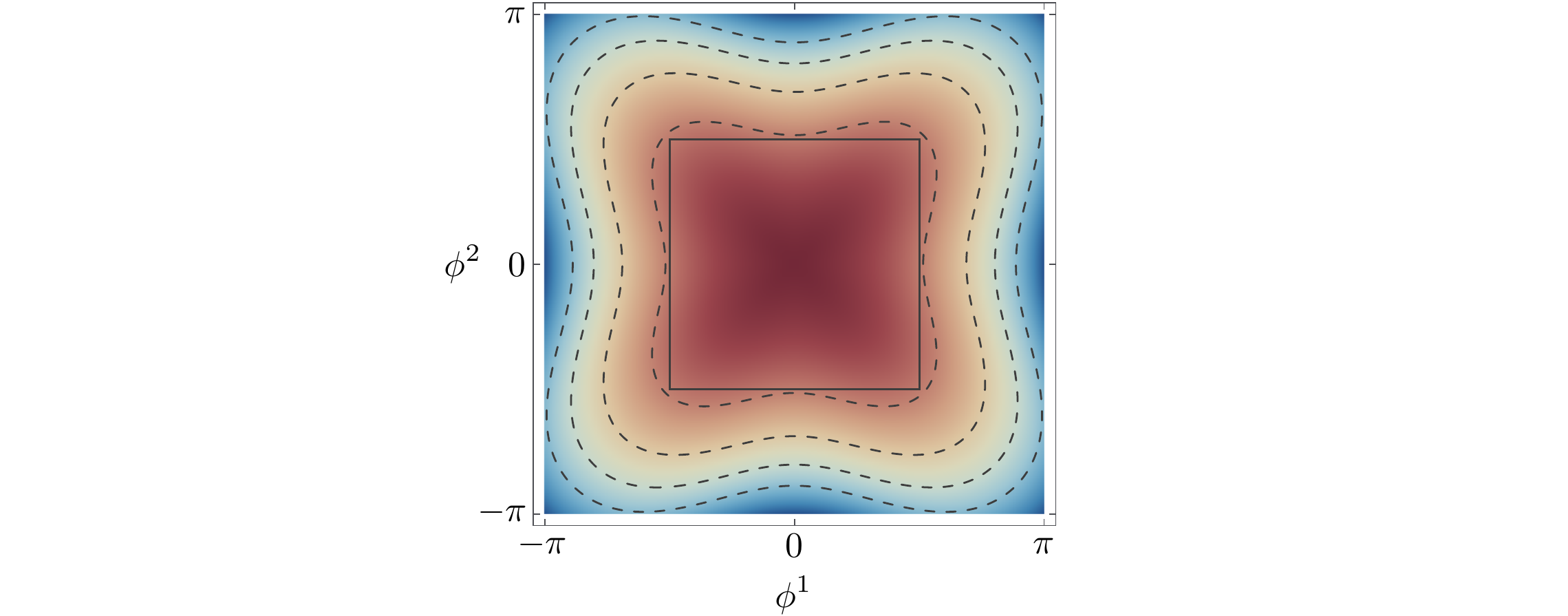}
  \caption{\small Contour plot of the relative error made by expanding  $\cos (\phi^1) + \cos (\phi^2)$ to quadratic order. The dashed lines indicate, from inner to outer, the levels of $0.25$, $.5$, $0.75$ and $1$ relative error, respectively. The maximum relative error is $1.5$. The quadratic domain is indicated by the gray square of side length $\pi$ while the periodic domain is the whole box.  \label{quadraticdomain}}
\end{figure}

In principle we can solve (\ref{optimization}) and determine the location of all minima in the  theory. However, although we can certainly solve  (\ref{optimization}) in any one $P$-cube to arbitrary accuracy, the very large number of domains make this impossible at large $P$.  Instead, we will employ a quadratic expansion of the potential and the approximate symmetries to find an analytic expression for the approximate location of many minima. 

The auxiliary potential has one single minimum located at the center of each $P$-cube, around which we can use a quadratic expansion. We will refer to the (somewhat arbitrary) region within which the quadratic expansion is a good approximation as the {\it quadratic domain}. Since the non-perturbative potential consists of simple cosines, we define the quadratic domain as
\be\label{quadraticdomainphi}
-{\pi\over 2}\le \phi^I\le{\pi\over 2}\,,~~~\forall I \in \{ 1,\dots,P \} \,,
\ee 
such that the relative error made never exceeds $25\%$ within that region. This choice might change if the underlying periodic function deviates from a cosine, but we chose it with some foresight in a way that this region will, in well-aligned theories, capture many minima of the full non-linear potential. The periodic and quadratic domains along with the relative error made by approximating cosines by a quadratic function are illustrated in Figure \ref{quadraticdomain}.

The quadratic expansion dramatically simplifies the problem of finding minima. Consider a small displacement $\delta \bs \phi$ from the auxiliary lattice point $2\pi\trafo{\omega_\nparallel}{\phi} \bs m$. The potential in the corresponding quadratic domain evaluates to
\be\label{quadraticconstrained}
V_{\delta \bs \phi}={1\over 2}\delta\bs\phi^\top \diag(\Lambda_I^4)\delta\bs \phi+\bs\nu^\top \mathbfcal R (2\pi\trafo{\omega_\nparallel}{\phi} \bs m+\delta\bs \phi) +\mathcal{O}(\delta\bs\phi^4)\,,
\ee
so a vanishing gradient is implied by the conditions
\bea
\diag(\Lambda_I^4)\delta\bs \phi+  \mathbfcal R^\top \bs \nu\nonumber &=& \bs 0\,,\\
\mathbfcal R(2\pi\trafo{\omega_\nparallel}{\phi} \bs m+\delta\bs \phi)  &=& \bs 0 \,.
\eea
Solving this system of equations for the location of a minimum $\bs\phi_{\bs m}$ on the constraint surface gives\footnote{For non-vanishing phases $\bs\delta_\text{r}$ in (\ref{phasesrem}) the minima are located at $\bs\phi_{\bs m}+\bs\delta_\text{r}$, and correspondingly at potentials $(\delta\bs\phi_{\bs m}-\bs\delta_\text{r})^\top \diag{(\Lambda_I^{4})}(\delta\bs\phi_{\bs m}-\bs\delta_\text{r})/2$, where we defined $\delta\bs\phi_{\bs m}=2\pi\bs\Delta^\perp\trafo{\omega_\nparallel}{\phi} \bs m$. Note the relation $\bs\phi=\trafo{\omega_\parallel}{\phi}\bs\omega_\parallel+\bs\delta_\text{r}$, which implies that the  locations of minima in $\bs\omega_\parallel$-coordinates remain unchanged in the presence of small phases.}
\be\label{deltaphieq}
\bs\phi_{\bs m}=2\pi(\mathbbold 1-\bs\Delta^\perp)\trafo{\omega_\nparallel}{\phi} \bs m\,,
\ee
where $\mathbbold 1-\bs\Delta^\perp$ is a non-orthogonal projector onto the constraint surface: 
\be
(\bs\Delta^\perp)^2 = \bs\Delta^\perp\equiv \diag{(\Lambda_I^{-4})}\mathbfcal R ^\top\left[ \mathbfcal R \,  \diag{(\Lambda_I^{-4})}\mathbfcal R^\top\right]^{-1}\mathbfcal R\,.
\ee
Scanning over all $\bs m$, we can check which $\bs\phi_{\bs m}$ lie within the quadratic domain of their respective lattice site $2\pi\trafo{\omega_\nparallel}{\phi} \bs m$; namely, within the intersection of $2P$ half-planes,
\begin{figure}
\includegraphics[width=1\textwidth]{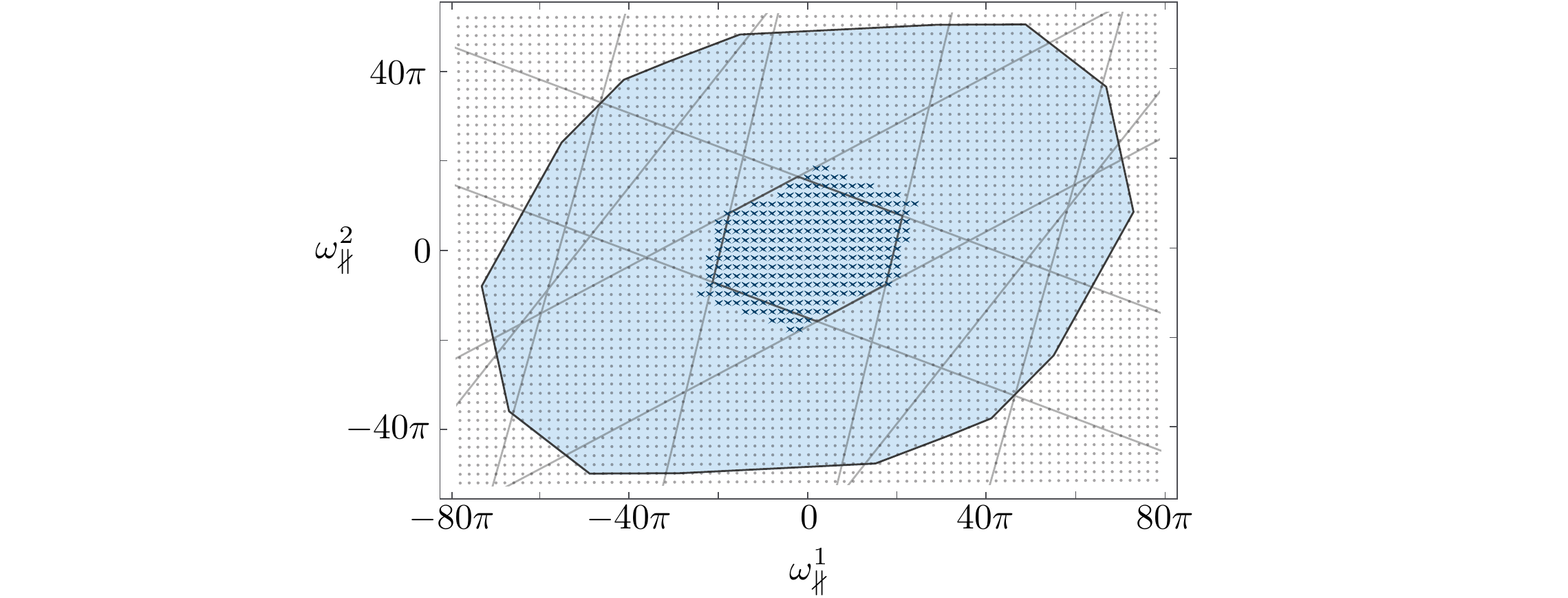}
\caption{\small The blue shaded region illustrates the polytope $\mathfrak C$ in $\bs \omega_\nparallel$-coordinates containing all lattice points corresponding to distinct tiles for an example with $P=N+2=8$ (cf.~Figure \ref{convexhullandminima} where the same region is shown). The central six-sided region is bounded by  the $2P$ half-planes that determine the validity of the quadratic approximation \eqref{boundariesofquadraticdomain}. Tiles that contain a minimum as determined by numerical minimization are denoted by blue crosses, and nearly perfectly overlap with the quadratic region.\label{aligneddomainfig}}
\end{figure}
\be\label{boundariesofquadraticdomain}
\Vert 2\pi\bs\Delta^\perp \trafo{\omega_\nparallel}{\phi} \bs m\Vert_\infty \leq {\pi\over 2} \,.
\ee
For these minima we will have succeeded at finding an approximate  location of the constrained system. The  energy density is given by 
\be \label{analyticVenergies}
V(\bs\phi_{\bs m})\approx {2\pi^2}\bs m^\top\left[ \trafo{\phi}{\omega_\nparallel}\diag{(\Lambda_I^{-4})} \trafo{\phi}{\omega_\nparallel}^\top\right]^{-1}\bs m\,,
\ee
where we used the specific choice $\mathbfcal R=\trafo{\phi}{\omega_\nparallel}$ to simplify the expression.

In Figure \ref{aligneddomainfig} we illustrate  the tiles in which there is a minimum located by numerically minimizing the potential, along with those for which the quadratic approximation predicts a minimum (i.e.~predicts a minimum located within the quadratic domain where the approximation is self-consistent). In general these sets of points are not immediately related, but for explicit examples we typically found a substantial overlap.

Finally, let us comment on the special case of equal scales $\Lambda_I=\Lambda$ and $P=N+1$.
In this case the energies of the minima are
\be\label{minimapnplus1}
V(\bs\phi_{m})\approx{2\pi^2}\Lambda^4 \left({c \, m\over \sqrt{\det\mathbfcal Q^\top \mathbfcal Q}}\right)^2\,,~~~\forall |m|<{\cal N}_{\text{vac}}\,.
\ee
where $c$ is an integer.  The derivation of this result can be found in appendix \ref{311deriv}.
The approximate signs in \eqref{minimapnplus1} denote the quadratic approximation which is valid for ${\cal N}_{\text{vac}} \sim  \sqrt{\det\mathbfcal Q^\top \mathbfcal Q} $ minima. No other approximations are made in (\ref{minimapnplus1}).

\subsection{A uniform sample over all minima} \label{unisample}
\begin{figure}
  \centering
  \includegraphics[width=1\textwidth]{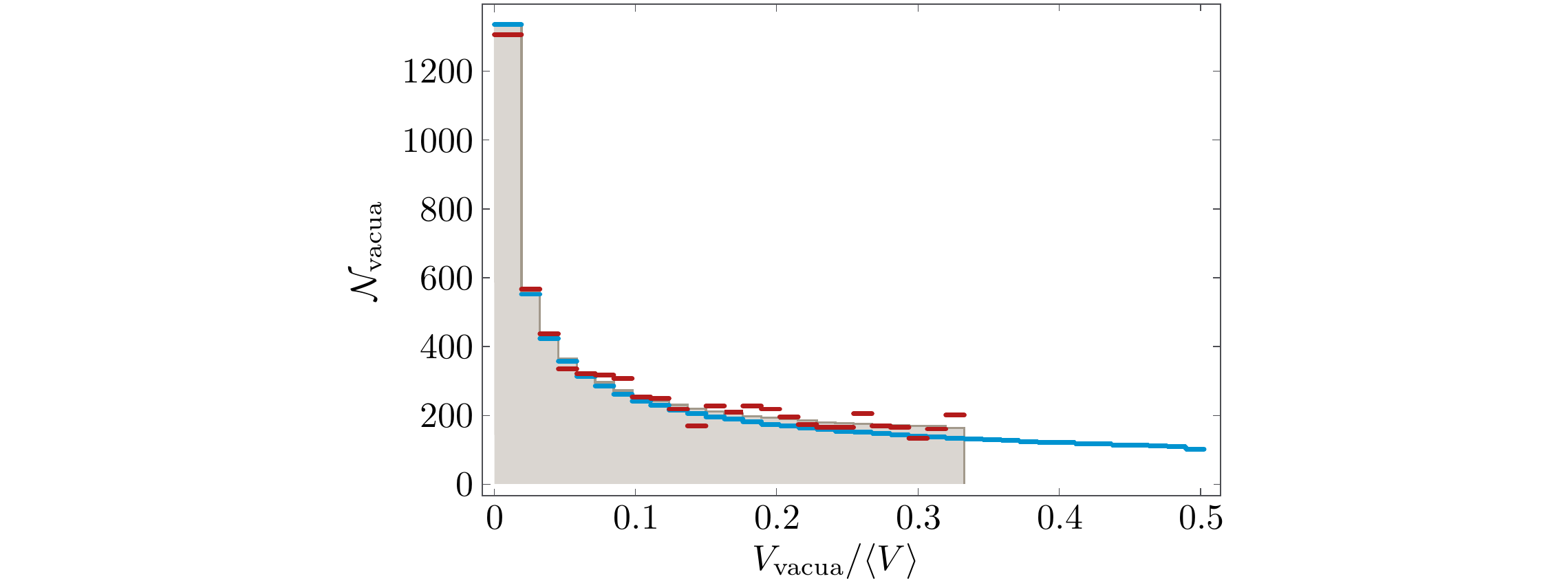}
  \caption{\small  Histogram showing the number of minima over the energy density, for an example with $N=5$, $P=6$. Shaded region: the exact distribution of all minima; red/dark: the distribution obtained via random sampling; blue/light: the distribution obtained via quadratic approximation. \label{vacuumdist}}
\end{figure}

The most direct approach to find all minima  is to consider the $(P-N)$-dimensional polytope $\mathfrak C$ defined in (\ref{defboundaryC}).  This region contains all vectors $\bs m\in \mathbb Z^{P-N}$ that label the  tiles of one periodic domain of the potential constrained to $\Sigma$ (see Figure~\ref{tiling}). Minimizing the potential in all the tiles in $\mathfrak C$ covers one entire periodic domain of the potential, yielding the set of all minima. However, the number of tiles is typically exponential in $N$ or $P$, so even at moderately large values of these parameters this comprehensive approach becomes intractable.

Instead, we can take advantage of the approximate symmetries of the potential that we identified. The auxiliary potential has  minima only at the centers of each $P$-cube, which suggests that many tiles containing minima of the physical potential will occur in those $P$-cubes for which the constraint surface $\Sigma$ passes through the quadratic domain near the center of the cube.  Such tiles lie within a connected region in $\bs m$-space; that is, a subset ${\mathfrak R}\subset \mathfrak C$.   (This compactness property only applies for the specific tiling and labeling of the tiles defined by the aligned coordinates.)  The region  ${\mathfrak R}$ contains exponentially fewer lattice sites than $\mathfrak C$, so the problem of comprehensively sampling ${\mathfrak R}$ is much less computationally intensive, but still requires a number of computations that is exponential in $P$.

We can further simplify the problem if we are interested only in statistical properties of the potential for which a relatively small, but representative sample of distinct tilings suffices. Such a representative sample can be obtained by uniformly sampling over lattice sites contained within the polytope $\mathfrak C$, for example by performing a random walk that samples the polytope in a time polynomial in $P-N$ \cite{CIS-115135,Cousins:2015:BKG:2746539.2746563,matlabsampling}. A simpler but much more computationally intensive mechanism to uniformly sample a polytope would be to define a simple region that fully contains the polytope and sample that, rejecting any sample that is not contained in the polytope. Again, in order to determine a statistical sample of most of the \emph{minima} it suffices to sample only the region ${\mathfrak R}$.
The sampling techniques above apply for both the non-linear optimization problem of \S\ref{vacuumsystematics} and the analytic result for the energies of the minima in the quadratic approximation, (\ref{analyticVenergies}). 

We illustrate the distribution of energies at the minima for a specific theory obtained via three different approaches in Figure \ref{vacuumdist}. The probability distribution of the energy density obtained by sampling a small number of all minima agrees well with the exact distribution. Furthermore, as expected, the quadratic approximation works best for relatively low minima, and becomes increasingly inaccurate for higher minima.

It is possible for the physical potential to have minima in tiles for which $\Sigma$ does not intersect the quadratic domain (that is, tiles that are outside ${\mathfrak R}$).  However such minima are rare, at least in the well-aligned regime $N \gg 1, P-N \ll N$.  This can be understood qualitatively as follows.  Each $P$-cube of sidelenght $2\pi$ can be decomposed various regions -- the quadratic domain at the center $\lVert\bs\phi \lVert_\infty < \pi/2$ (which contains a fraction $2^{-P}$ of the volume of the cube), surrounded by rectilinear regions defined by allowing some of the components of $\bs \phi$ to exceed $\pi/2$ in magnitude.  The Hessian of the auxiliary potential  \eqref{V-aux} is 
\be \label{Haux}
\left(H_\text{aux}\right)^I_{~J}  = \delta^I_{~J} \, \Lambda_I^4 \cos (\phi^I) \,.
\ee
This is positive definite precisely in the quadratic domain around the center.  Because the Hessian of the physical potential is a projection of $\bs H_\text{aux}$ onto $\Sigma$, any critical point of the physical potential that occurs in the quadratic domain must be a minimum.  For each component of $\bs \phi$ that lies outside the quadratic domain, the auxiliary Hessian matrix \eqref{Haux} has an additional negative eigenvalue.  Critical points of the physical potential in such regions can  be minima (rather than saddle points) only if the negative eigenvalue(s) of the Hessian are projected out when the potential is constrained to $\Sigma$.  For $P-N \ll N$, only a small fraction $P-N/P$ of the $P$ directions are  projected out and therefore it is unlikely (or impossible if $P-N$ is less than the number of negative modes)  that an critical point outside the quadratic domain will be a minimum.  We have verified this expectation numerically.   Therefore, in this regime the great majority of the exact minima lie within  ${\mathfrak R}$.

\subsection{Saddle points and maxima}
The  cosine function changes sign under a half-period shift, $\cos (\phi) = - \cos (\phi + \pi)$, so the physical characteristics of maxima  mirror those of minima.  Due to the ``1''s in \eqref{lagrtheta}, the global maximum has energy $V_\text{max} \approx \sum_{I} 2 \Lambda^{4}_{I}$.  (As explained below, this would be an equality if the phases were exactly $\delta^I = \pi$, and is a very good approximation in well-aligned theories.)
 All the techniques we apply to minima carry over to maxima and other critical points almost unchanged.  In particular, the locations and energies of maxima and saddles can be found efficiently by using the approximate symmetries.
In fact, the property of ``well-alignedness'' that allowed us to reduce the phases to values very close to zero allows us to set the phases to anything we like, subject to errors of order those in our original procedure.  In other words, we can set the phases $\bs \delta = \bs \delta_{\text{arb}} + {\mathcal O}\big( \Vert \Pob \trafo{\omega_\nparallel}{\phi} \Vert_\infty \big)$, where the $P$-vector $\bs \delta_{\text{arb}}$ denotes any arbitrary phases.  Geometrically, this is possible because in well-aligned theories  the angles between $\Sigma$ and the grid in $\bs \phi$-space are small, so  that $\Sigma$ approaches very close to every distinct point in $\bs \phi$-space.  

For studying maxima it is convenient to set all phases as close as possible to $\pi$.    Relating $\bs \phi$ and $\bs \theta$ in this way corresponds to shifting the centers of the $P$-cube tiling of $\bs \phi$-space so that they fall on global maxima rather than global minima -- that is, the new cubes are centered on the corners of the original ones.  With this change nearly every equation in this paper carries over unchanged  or with the obvious changes from minima to maxima.  In particular maxima have a  quadratic domain defined in the same way as for minima in  \eqref{quadraticdomainphi}, as the cube of side-length $\pi$ surrounding a now maximum of $V_{\text{aux}}$, and they have identical statistics for their number, energies (except subtracted from $V_\text{max}$ rather than added to $V_\text{min}$, etc.

This trick of setting the phases to a desired value is also useful for studying saddle points of any given degree.  For instance, to study saddles of degree one (critical points where the Hessian has 1 negative eigenvalue and $N-1$ positive eigenvalues) we should set one phase equal to $\pi$ and the rest as close as possible to zero.  These points are the centers of the faces of the original cubes, and are points where the auxiliary Hessian \eqref{Haux} has precisely one negative mode (and the auxiliary potential has a degree one saddle).  Tiles where $\Sigma$ passes through the quadratic domain of these points  often contain degree one saddles of the physical potential, and tiles that do not may not, for the same reason described in the previous subsection for the case of minima.  Such degree one saddles occur between tiles that correspond to $P$-cubes that are neighbors along a face, and play a crucial role in the analysis of tunneling transitions (cf. \cite{Bachlechner:2017zpb}).

\subsection{Estimates in random ensembles} \label{randomsec}
In the previous section we discussed how to systematically enumerate and locate all minima of a given axion theory. We found an analytic expression for their energy densities in the quadratic approximation. We now turn to a discussion of the expected number and distribution of minima in ensembles of random axion theories.

We define these theories by ensembles of random integer charge matrices $\mathbfcal Q$ and energy scales $\Lambda_I^4 $, as discussed in \S\ref{intro}. To repeat our assumptions, $\mathbfcal Q$ is a  $P \times N$ matrix of independent, identically distributed random integer entries with vanishing mean and standard deviation $\sigma_{\mathcal Q}$. We assume the universal limit of random matrix theory such that the precise distribution (including the fact that the entries of $\mathbfcal Q$ are integer) becomes irrelevant and all expectation values only depend on $\sigma_{\mathcal Q}$. This assumption roughly holds whenever $\gtrsim 3/N$ of the entries of the charge matrix are non-vanishing, and the distribution of the entries is not heavy-tailed. The field space metric is irrelevant for the discussion in this section.

In general  it is a very difficult task to analytically obtain the distribution of minima, or even the number of stable minima. Even in the quadratic approximation this problem amounts to determining the number of lattice sites within a non-trivial high-dimensional polytope defined by (\ref{boundariesofquadraticdomain}). In this section we therefore mostly restrict our attention to the simplest case of one single auxiliary field, $P=N+1$ and equal scales $\Lambda_I=\Lambda$, unless otherwise noted. We will find that the energy density (\ref{minimapnplus1}) is valid for super-exponentially many minima.

\subsubsection{The quantity and energies of minima for $P=N+1$} \label{nondegenerateminima}
We now determine the number of distinct minima ${\cal N}_\text{vac}$ that are well-approximated by the quadratic expansion in (\ref{minimapnplus1}). This count is simply given by the number of sites of the $P$-dimensional, rank $P-N=1$ sublattice $\delta\bs\phi_{m}=2\pi\Pob \trafo{\omega_\nparallel}{\phi}m$ that are contained within the quadratic domain $\Vert\delta\bs\phi_{m}\Vert_\infty \leq \pi/2$.  When all phases in the original Lagrangian exactly vanish there exists a two-fold degeneracy of all minima. If the phases do not precisely vanish, they can be absorbed  up to a finite remainder that is typically of order the change of $\delta\bs\phi$ between similar minima, see \S\ref{phases}. To further simplify the problem we assume identical scales $\Lambda_I=\Lambda$. The number of distinct minima in the quadratic domain is then simply
\be
{\cal N}_{\text{vac}}= {1\over 2 \, \Vert \Pob \trafo{\omega_\nparallel}{\phi} \Vert}_{\infty} \, .
\ee
Note that since $\mathbfcal Q$ has independent, identically distributed random entries,   $\Pob$ projects onto a random direction that is isotropically distributed, hence the vector $\Pob\trafo{\omega_\nparallel}{\phi}$ is isotropically distributed, with (\ref{cdef}) its two-norm is given by
\be\label{normpt}
\Vert\Pob\trafo{\omega_\nparallel}{\phi}\Vert_2^2={c^2\over \det\mathbfcal Q^\top \mathbfcal Q} \,.
\ee
and we defined the positive integer $c$ as in (\ref{cdef}). A vector that is distributed isotropically on the sphere consists of independent, normally distributed entries. Matching the expected norm of that vector to (\ref{normpt}) then determines the distribution of the entries,
\be\label{rqm1dist}
(\Pob\trafo{\omega_\nparallel}{\phi})^I\in{\cal N}\left(0, {c\over\sqrt{P \det\mathbfcal Q^\top \mathbfcal Q}}\right)\,,~~~\forall I \in \left\{ 1,\dots,P \right\} \,,
\ee
where ${\cal N}(0,\sigma)$ denotes a normal distribution of mean zero and standard deviation $\sigma$. It is now straightforward to evaluate the median of the largest absolute entry of $\Pob\trafo{\omega_\nparallel}{\phi}$, which yields the number of minima as
\be\label{nminimamedian}
{\cal N}_{\text{vac}}\approx {\sqrt{P}\over 2 \ell(P)} \sqrt{\det\mathbfcal Q^\top \mathbfcal Q}\,,
\ee
where $ \ell(N) \equiv \sqrt{2} \, \text{erf}^{-1}(2^{-1/N})$  is the median largest absolute entry of an $N$-vector with entries that are unit normal distributed. In (\ref{nminimamedian}) we used that $c$ is an order one integer, which we confirmed  in extensive simulations for the ensembles under consideration.

The matrix $\mathbfcal Q^\top \mathbfcal Q$ is a real Wishart matrix, the determinant of which is distributed as the product of $P-N$ chi-squared random variables with $P$, $P-1$, $\dots$, $P-N+1$ degrees of freedom, respectively \cite{goodman1963}, which gives for our case
\be
\langle \det{\mathbfcal Q^\top \mathbfcal Q}\rangle=\sigma_{\mathcal Q}^{2(P-1)} {P!\over 1!}\,. 
\ee
Finally, we find a simple expression for the expected number of minima,
\be\label{nminimamedian2}
{\cal N}_{\text{vac}}\approx {\sigma_{\mathcal Q}^{P-1}\over 2 \ell(P)} \sqrt{PP!} \,,
\ee
which is exponentially large in $N$ in the universal regime, where at least a fraction $3/N$ of the entries in $\mathbfcal Q$ are non-vanishing.  To give a sense of  these numbers, with $P=N+1=150$, and $\sigma_{\mathcal Q} = 1$, one obtains ${\cal N}_{\text{vac}} \approx 10^{131}$. Note that this result was only derived for $P-N=1$, but as we  discuss below we expect similar results to hold more generally (see \eqref{Nvacmult}). The scaling with $N$ is identical to that observed in \cite{Bachlechner:2015gwa} for a specific case where $P\gg N$. 

Finally, let us estimate the energy levels at which the minima arise. To that end, we will assume that the distribution of minima can be well-approximated by all lattice sites that lie within the quadratic domain, i.e. $|\delta\phi^I|\le\pi/2$. Since the displacements $\delta\bs\phi$ are proportional to $\Pob\trafo{\omega_\nparallel}{\phi}$, which by (\ref{rqm1dist}) is roughly normal distributed, we can easily estimate the typical magnitude of the entries of the displacement vector, when the largest component is $\pi/2$, 
\be
|\delta\phi^I|\approx {\pi\over 2 \ell(P)}\,.
\ee
The maximum energy density in a minimum is therefore well-approximated by\footnote{Note that this expression applies for general $P$.}
\be \label{vacuumenergyest}
 \text{max}~(V_{\text{vac}})\approx{1\over 2}\left({\pi\over 2\ell(P)}\right)^2\langle V\rangle\approx 0.14~\langle V\rangle\,,
\ee
where we used $\ell(N) \approx 3$ for $N \sim 100$ in the last approximation, and used the mean of the potential $\langle V\rangle =\sum_{i=1}^P\Lambda_I^4 $. Since the potential is simply quadratic the median energy density is given by
\be
\text{median}~(V_{\text{vac}})\approx {1\over 4} \text{max}~(V_{\text{vac}})\approx 0.034~\langle V\rangle\,.
\ee

\subsubsection{Hessian eigenvalues}
Beyond their energies, another interesting characteristic of critical points  is the spectrum of  eigenvalues of the Hessian.  If the kinetic matrix is $K_{ij} = f^{2} \delta_{ij}$, the canonically normalized fields are $\bs \Theta \equiv f \bs \theta$. Defining $\bold Q \equiv \bs K^{-1/2} \mathbfcal Q = \mathbfcal Q/f$, the potential in canonically normalized coordinates is
\be \label{canpot}
V(\bs \Theta) = \sum_{I=1}^P \Lambda_I^4 \left[ 1 - \cos\left( \bold Q \bs\Theta \right)^I \right]\,.
\ee
The eigenvalues of the Hessian of this potential at a critical point are then the masses of the canonical fields at that point.    In \S\ref{seccdiameters} we will perform a more detailed analysis of the size of the tiles surrounding minima and the masses of the canonically normalized fields for various less trivial choices of kinetic matrix.

Expanded around a minimum labeled by $\bs m$, the Hessian of \eqref{canpot} is 
\be \label{Hphys}
{\bs H}= \bold Q^\top \text{diag} \left( \Lambda_I^4 \right) \bold Q + {\mathcal O}[(\Po^\perp \bs m)^2]\,.
\ee
For simplicity let us take all $\Lambda_{I} = \Lambda$.  In that case  $\bold Q^\top \bold Q$ is a Wishart matrix, and at large $N$ the empirical density (i.e. the amount of eigenvalues in a small interval) follows the Marchenko-Pastur distribution \cite{MP1967}. For a Wishart matrix with standard deviation 1, the mean smallest eigenvalue is $ \mathcal{O}(1 /N) $ while the largest is $ \mathcal{O}( N) $ (if $P=N+1$ the precise values are $1/4N$ and $4N$, respectively). The Marchenko-Pastur distribution has a sharp peak near the minimum and a long tail to larger values.  The mean and median are both of order $N$.  Putting the dimensions back in, this means the masses will range from 
$$
{ \sigma_{\mathcal Q}^2 \over N} \left({\Lambda^{2} \over f}\right)^2  \simleq m^{2} \simleq N \sigma_{\mathcal Q}^2  \left({\Lambda^{2} \over f}\right)^2\,.
$$

As mentioned previously, the Hessian on a degree one (one negative mode) saddle is of interest for questions involving tunneling from one minimum to another \cite{Bachlechner:2017zpb}. Such saddles are most easily analyzed by setting one phase to $\pi$ and the rest to zero.  This corresponds to considering points where $\Sigma$ passes close to the center of one face of the $P$-cube (note that such points are degree one saddles of $V_{\text{aux}}$).  Since $\bold Q$ is isotropic it does not matter which phase we set to $\pi$.  Choosing the first one, it is easy to see that \eqref{Hphys} becomes
\be \label{H2}
{\bs H} \approx \Lambda^{4} \, \bold Q^\top \text{diag} (-1, 1, 1, \dots) \bold Q  \, .
\ee
This is not a Wishart matrix and we are unaware of any analytic results for its eigenvalue spectrum.  It has at most one negative eigenvalue.  If  the smallest eigenvalue $\lambda_{\text{min}}$ turns out to be positive, this critical point is in fact a minimum rather than a saddle.  Numerically we established that the mean and standard deviation of the minimum eigenvalue are
$$
\langle \lambda_{\text{min}}(\bs H) \rangle \approx - {N \over 2} \sigma_{\mathcal Q}^2  \left({\Lambda^{2} \over f}\right)^2 \,, \,\,\,\, \, \, \, \, \, \, \sigma_{\lambda_{\text{min}}(\bs H)} \approx \sqrt{3 \over 2 N }\, \left| \langle  \lambda_{\text{min}}(\bs H)  \rangle \right|\,.
$$  
Hence, at large $N$ is is extremely unlikely that there is no negative eigenvalue and the would-be degree one saddle is actually a minimum.  This at least partially confirms the expectation explained in \S\ref{unisample}, that most local minima occur in tiles where the constraint surface intersects the quadratic domain of the  minimum of  $V_\text{aux}$, rather than in neighboring regions such as these.
Similarly most saddles of degree $k$ will occur in regions where $\Sigma$ intersects the quadratic domain of a degree $k$ saddle of $V_\text{aux}$.  The relation between the number of saddles ${ \cal N}(k)$ of degree $k$ and the number of minima can then be estimated from \eqref{H2}:
$$
{\cal N}(k) \approx {{P}\choose{k}} {\cal N}_{\text{vac}}\,.
$$

\subsubsection{Neighboring minima}
In the previous section we estimated the total number of distinct, non-degenerate minima in the entire potential. For some questions one might however only be interested in the immediate neighborhood of one particular minimum. We therefore turn to determining the expected number of immediate neighboring minima, including degenerate ones. To allow for a simple estimate, consider all  sites neighboring the origin, $\bs n=\bs 0+\bs e$, with unit or zero entries for $\bs e$ as in (\ref{neighboringdomains}). Let us count the immediate neighbors that lead to a minimum within the quadratic domain (\ref{quadraticdomainphi}),
\be
\lVert 2 \pi \Pob \bs e \rVert_\infty \leq {\pi \over 2} \,. 
\ee
We verified numerically that in the universal limit the entries of the orthogonal projector matrix $\Pob$ have variance $(P-N)/P^2$. Using the central limit theorem and denoting the number of non-vanishing entries of $\bs e$ by $n_{\bs e}$, we can approximate
\bea
\left( \Pob \bs e \right)^I \in {\cal N}(0,\sqrt{n_{\bs e}(P-N)}/P)\,.
\eea
The median largest entry of $\Pob \bs e$ evaluates to $\ell(P) \sqrt{n_{\bs e}(P-N)}/P$, such that for $P\lesssim 2N$ a large fraction of the $3^P$ neighbors are stable minima, i.e. $\lVert\Pob \bs e\lVert_\infty\ll \pi$.  Therefore, not only is the total number of minima extremely large, but each minimum has a vast number number of immediately neighboring minima.

\subsubsection{Phases}\label{phasesestimate}
\begin{figure}
  \centering
  \includegraphics[width=1\textwidth]{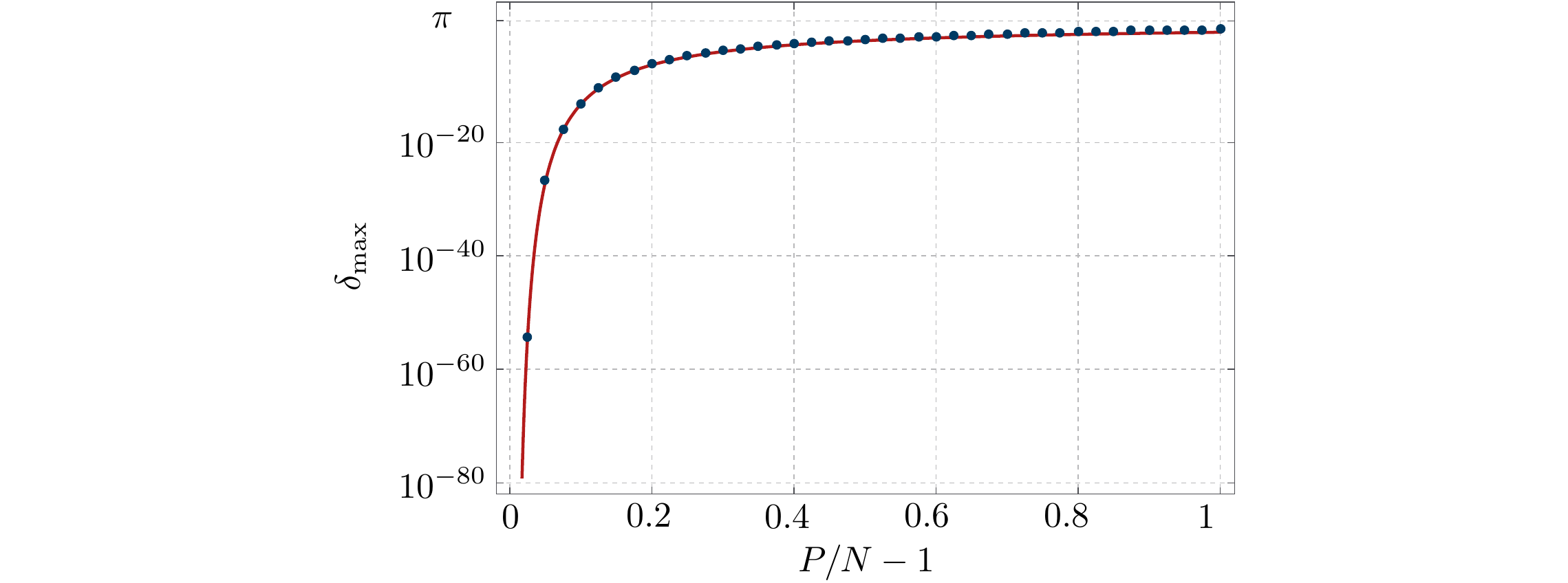}
  \caption{\small Ensemble average of largest  phase (\ref{maxremainingphase}) as a function of $P$, with $N=20$ and $\sigma^2_{\mathcal Q}=2/3$. The solid line is  the estimate (\ref{phaseestimate}). \label{phasesfig}}
\end{figure}

As discussed in \S\ref{phases}, in well-aligned axion theories the $N$ exact and $P-N$ approximate shift symmetries allows one to set  $N$ phases to precisely to zero and make the remaining $P-N$ phases very small. The accuracy to which all phases can be eliminated, as measured by the largest remaining phase $\delta_\text{max}$, depends on how aligned the basis is:
\be\label{maxremainingphase}
\delta_{\text{max}} \leq \pi \left\Vert \Pob \trafo{\omega_\nparallel}{\phi} \right\Vert_\infty \leq \pi (P-N) \, \max_a \left\{ \lVert \Pob \bs t^\nparallel_a \rVert_\infty \right\} \,.
\ee
 In order to get some analytical intuition for how well-aligned theories in our ensemble tend to be, let us assume that the vectors $\Pob \bs t^\nparallel_a$ are orthogonal and are the shortest they could possibly be, as in (\ref{normpt}), and that  the volume of the cubic, but arbitrarily oriented periodic domain of the lattice generated by $\Pob$ is given by $\left( \det\mathbfcal Q^\top \mathbfcal Q \right)^{-1/2}$. These assumptions yield the distribution for the components of all projections,
\be\label{rqm1dist2}
(\Pob \bs t^\nparallel_a)^I\in{\cal N}\left(0, { (\det\mathbfcal Q^\top \mathbfcal Q)^{- \frac{1}{2(P-N)}}\over\sqrt{P}}\right)\,,~~~\forall I \in \left\{ 1,\dots,P \right\} \,,~~a \in \left\{ 1,\dots,P-N \right\} \,.
\ee
This gives an upper bound on the number of minima in the quadratic domain,
\be \label{Nvacmult}
{\cal N}_\text{vac}\lesssim \left({\sqrt{P}\over 2\ell(P)}\right)^{P-N}\sqrt{\det\mathbfcal Q^\top \mathbfcal Q}\approx\left({\sqrt{P}\over 2\ell(P)}\right)^{P-N}\sigma_{\mathcal Q}^N \sqrt{P!\over (P-N)!}\,,
\ee
reproducing (\ref{nminimamedian2}) in the special case $P=N+1$. 

We can furthermore easily obtain the largest components of the orthogonal projections of the aligned lattice basis,
\be\label{phaseestimate}
\max_a \left\{ \lVert \Pob \bs t^\nparallel_a \rVert_\infty \right\} \sim { \ell(P[P-N])\over \sqrt{P}(\det\mathbfcal Q^\top \mathbfcal Q)^{{1\over 2(P-N)}}}\approx  { \ell(P[P-N])\over \sqrt{P}}\left({1\over \sigma_{\mathcal Q}^{2N} }{(P-N)!\over P!}\right)^{{1\over 2(P-N)}}\,.
\ee
Using Stirling's approximation we observe that when random matrix universality applies the basis is well-aligned for  $P\approx N$, and for order unity $\sigma_{\mathcal Q}$ the basis ceases to be well-aligned with growing $P$ at $P\lesssim 2N$.  We illustrate how the largest phase increases with the number of non-perturbative terms along with the analytic estimate (\ref{phaseestimate}) in Figure \ref{phasesfig}.

\subsection{Band structure of subleading terms} \label{bandstructuresec}
Finally, let us address the last feature of the axion Lagrangian that we ignored so far; the subleading terms in the non-perturbative potential, denoted only by ellipses in  (\ref{lagrtheta}). Explicitly, we have the full axion potential
\be
V=\sum_{I=1}^P \Lambda_{I}  \left[1-\cos\left({\mathbfcal Q} \bs\theta \right)^I \right]+V_{\text{sl}}(\bs\theta)\,,
\ee
where we introduced a subleading potential $-\Lambda_{\text{sl}}^4 \le V_{\text{sl}} \le \Lambda_{\text{sl}}^4$ of scale $ \Lambda_{\text{sl}}^4$ that is negligible compared to the leading $P$ terms in the non-perturbative potential. Remember that we chose coordinates such that $\theta^i\rightarrow \theta^i+2\pi$ are the discrete shift symmetries respected by the full theory, such that also $V_{\text{sl}}(\bs\theta)$ breaks any larger symmetries respected by the $P$ leading terms to those fundamental symmetries. If there are any shift symmetries respected by the $P$ leading terms that are broken by the subleading potential this will result in a multiplicative increase in the number of distinct minima, as discussed in \cite{Bachlechner:2015gwa}. This effect is related to, but distinct from the mechanism discussed thus far.

\begin{figure}
\centering
\includegraphics[width=1\textwidth]{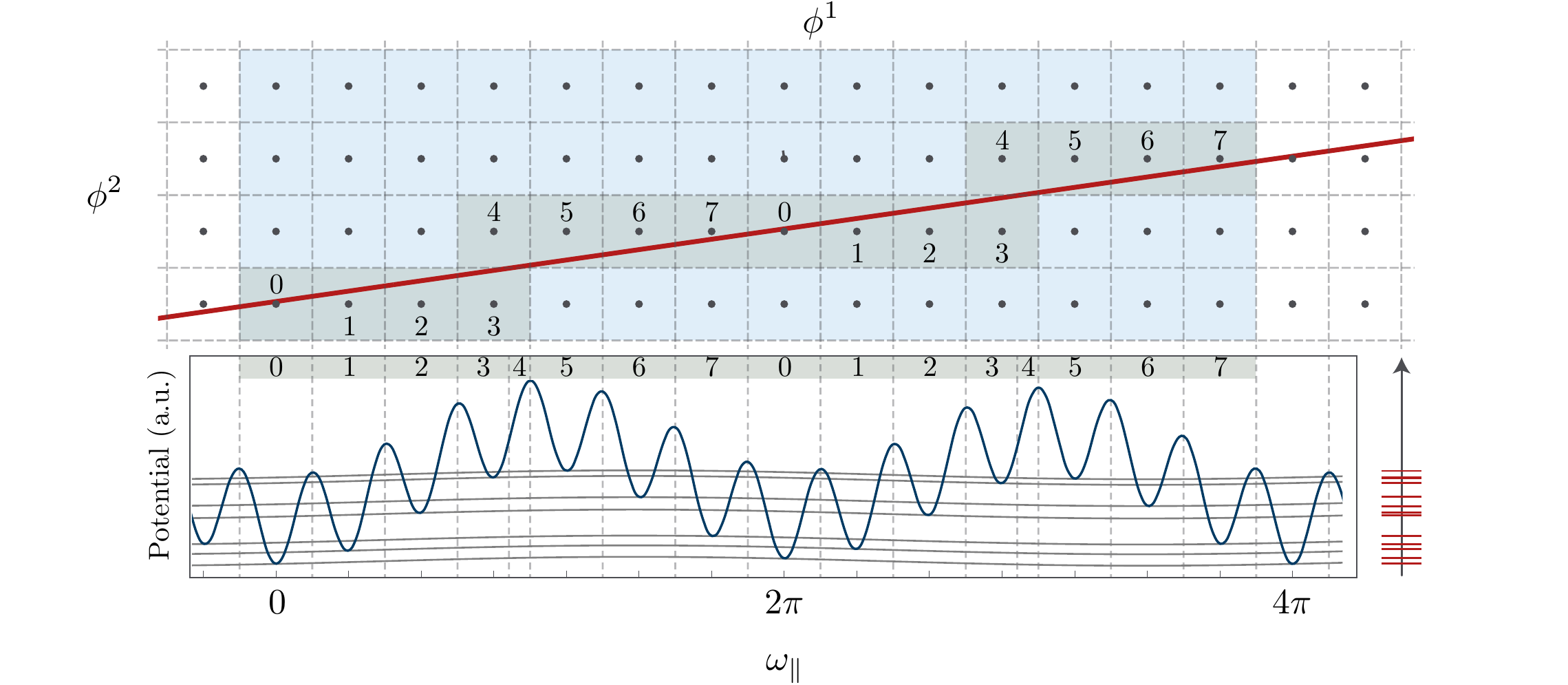}
\caption{\small \underline{Top}: Constraint surface (red line) along with the lattice $2\pi \mathbb Z^P$ (gray dots). Distinct tiles of the leading cosine terms are numbered and shaded dark, while the periodicity of the subleading terms is indicated by the light shading. \underline{Bottom}: Axion potential. The subleading terms further split the energies in the minima of the leading terms.}\label{bandstructurefig}
\end{figure}

The leading potential is invariant under the $P$ shifts $\left( \mathbfcal Q \bs \theta \right)^I \rightarrow \left( \mathbfcal Q \bs \theta \right)^I + 2\pi$, which generates an $N$-dimensional lattice denoting the shift symmetries in terms of the $\bs \theta$-coordinates. In the notation of \S\ref{symm} a basis for this lattice is given with (\ref{moretrafos}) by $\mathbfcal B = \trafo{\omega_\parallel}{\theta}$, i.e. the leading potential is (minimally) invariant under the $N$ shifts $\bs \theta \rightarrow \bs \theta + 2\pi \mathbfcal B_i$. The subleading potential, however, is only invariant under shifts on the integer lattice $2\pi \mathbb Z^N$. This means that the periodic domain of the full potential contains ${\mathcal N}_{\text{sl}}= 1 / \sqrt{\det \mathbfcal B^\top \mathbfcal B}$ periodic domains of the leading potential (as this is the inverse volume of that domain). If the leading $P$ terms in the non-perturbative potential contain ${\cal N}_{\mathcal Q}$ minima, each of these minima degenerates into ${\mathcal N}_{\text{sl}}$ distinct minima due to the further symmetry breaking in the subleading potential. The total number of minima therefore becomes
\be
{\cal N}_\text{vac}={\mathcal N}_{\text{sl}}\times{\cal N}_{\mathcal Q}\,.
\ee
We illustrate this energy level splitting in Figure \ref{bandstructurefig}.

Of course we could have included the charges of the subleading terms in the $P$ rows of the leading potential and found the corresponding aligned basis that includes all possible charges in the theory. However, the subleading potential is irrelevant for all practical purposes when identifying approximate shift symmetries of the potential and therefore would only introduce a spurious complication of the computational problem by increasing the dimensionality $P$ of the auxiliary lattice. When identifying the shift symmetries according to \S\ref{symm} it is therefore important to identify which terms in (\ref{lagrtheta}) can safely be ignored for a problem at hand.

\section{Aligned axion diameters}\label{seccdiameters}
In this section we provide a systematic discussion of the theory in the vicinity of local minima; that is, within the tiles ${\mathcal T_{\bs n}}$.  Recall that the tiles are defined as regions within which none of the individual terms in the potential exceeds its maximum (see \S\ref{intro}), and therefore define the characteristic scale on which the potential changes.  Within each of these tiles the potential is relatively flat and hence provides for a natural environment to study large field inflation. Several specific  cases were previously studied in the literature \cite{Kim:2004rp,Dimopoulos:2005ac,Choi:2014rja,Long:2014dta,Bachlechner:2014hsa, Czerny:2014xja,Tye:2014tja,HT1,HT2,Ali:2014mra,Bachlechner:2014gfa, Burgess:2014oma,Madison,Madrid,Shiu:2015xda,Palti:2015xra,Kappl:2015esy}. In this section we describe a systematic approach to determine the size of an arbitrary tile. We restrict our discussion to well-aligned theories (cf.~\S\ref{symm}) where we can set all $P$ phases in \eqref{lagrtheta} to zero to good accuracy by a shift of $\bs \theta$.

\begin{figure}
\centering
\includegraphics[width=1\textwidth]{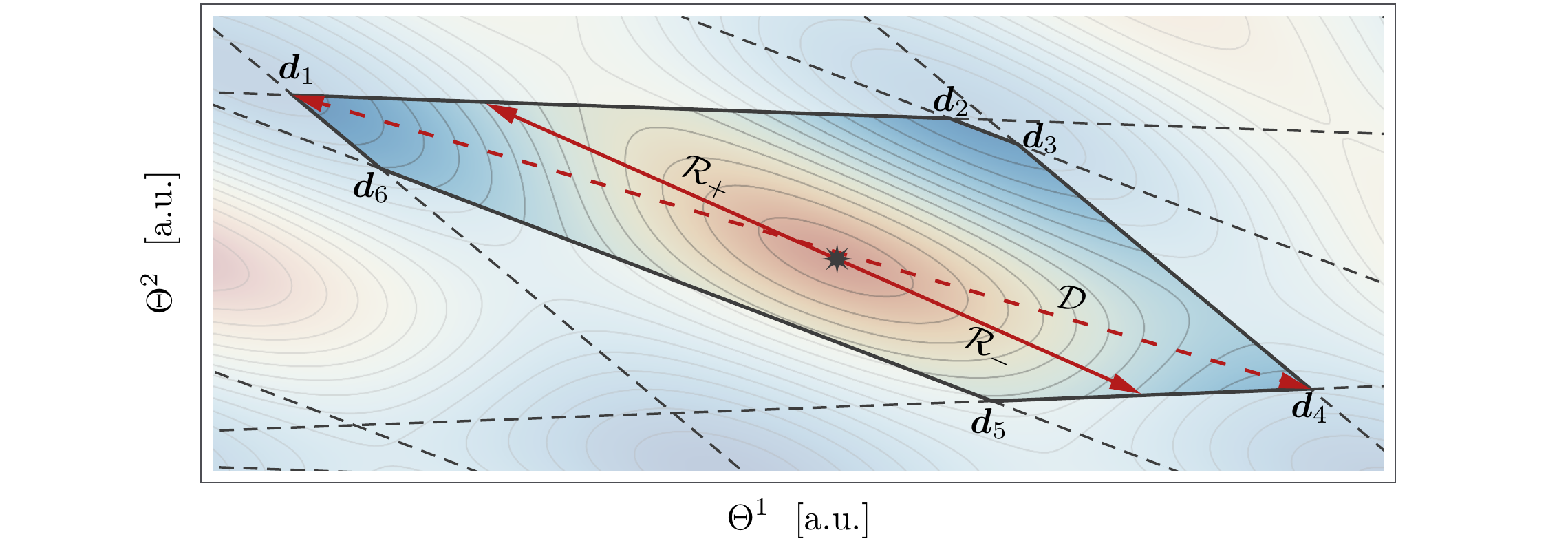}
\caption{\small Illustration of the tile ${\mathcal T_{(3\,0)^\top}}$ of the potential shown in Figure \ref{tiling} in canonical fields $\bs \Theta$. The solid arrows denote the field ranges ${\cal R}_\pm$ along the two lightest directions in the vicinity of the minimum, while the dashed arrow denotes the diameter ${\cal D}$ of the tile.\label{canonicalpolytopefigure}}
\end{figure}

Recall the Lagrangian (\ref{lagrtheta}) of a well-aligned axion theory,
\be\label{lagrthetasecdiameters}
{\mathcal L}={1\over 2} \partial\bs\theta^\top \bs K \partial\bs\theta-\sum_{I=1}^P \Lambda_I^4  \left[1-\cos\left({\mathbfcal Q}\bs\theta  \right)^I \right]\,,
\ee
where we retain only $P \geq N$ leading terms in the potential. In this section we are interested in invariant field space distances, so it is convenient to introduce canonically normalized fields $\bs \Theta$,
\begin{equation} \label{canonicalcoordinates}
	\bs \Theta \equiv \sqrt{\bs K} \bs \theta\,,
\end{equation}
where $\sqrt{\bs K}$ is the positive matrix square root.\footnote{The matrix square root satisfies  $\sqrt{\bs K}\sqrt{\bs K} =\bs K$, and is related to the matrix $\bold S_{\bs K}$ containing the (column) eigenvectors of $\bs K$ and its eigenvalues $f_{i}^2$ by $\sqrt{\bs K}=\bold S_{\bs K} \,\text{diag}(f_i)\,\bold S_{\bs K}^\top \,$.} The Lagrangian in canonically normalized coordinates reads
\begin{equation} \label{ThetaTheory}
	\mathcal L = {1\over 2}\partial \bs\Theta^\top \partial \bs\Theta - \sum_{I=1}^P \Lambda_I^4 \left[ 1 - \cos\left( \bold Q \bs\Theta  \right)^I \right] \,,
\end{equation}
where the canonical charges are related to the integer charges by
\begin{equation} \label{canonicalcharges}
	\bold Q \equiv \mathbfcal Q \, \bs K^{-1/2} \,.
\end{equation}
In canonically normalized coordinates the tiles \eqref{tiles} are given by
\be \label{nfundamentaldomain}
{\mathcal T_{\bs n}}=\{\bs \Theta~:~\lVert \bs {\bold Q}\bs\Theta-2\pi  \bs n\rVert_\infty\le\pi\}\,.
\ee
The tiles are polytopes in $N$ dimensions defined by the intersection of $2P$ half-planes\footnote{Note that ${\mathcal T_{\bs n}}$ is indeed a polytope, i.e. a finite volume subset of $\mathbb{R}^N$ bounded by hyperplanes of codimension one, since $\bold Q$ is full rank.},  and spherical shells determine the surfaces of constant invariant distance to the center of the sphere. We illustrate this polytope in Figure \ref{canonicalpolytopefigure}.

\subsection{Diameters and field ranges in well-aligned theories} \label{diameterFRsec}
To characterize the scale of these domains we will consider two distinct measures of size. One is the \textit{diameter} ${\cal D}_{\bs n}$, that is, the length of the longest straight line contained in the tile ${\mathcal T_{\bs n}}$. It is clear that this line will run between two vertices of the polytope, and that the tile with the largest diameter is the one at the origin $\bs n = \bs 0$. If we denote the vertices of the polytope by $\bs d_{\bs n, l}$, we have therefore the corresponding diameter
\be\label{diametersformal}
{\cal D}_{\bs n}=\underset{l,k}{\text{max}} ~\lVert\bs d_{\bs n , l}-\bs d_{\bs n , k} \rVert_2 \le {\cal D}_{\bs 0} =  2\,\underset{l}{\text{max}} ~\lVert\bs d_{\bs 0 , l}\rVert_2\,.
\ee
Note that the diameters of the tiles depend only on the charge and kinetic matrices of the theory (not on the couplings $\Lambda_I^4$).

The expression (\ref{diametersformal}) defines a unique size for every tile, but in practice there are exponentially many vertices, making it hard to evaluate. Furthermore, the low energy physics in the vicinity of a minimum is generally dominated by the lightest degrees of freedom, which do not necessarily coincide with the axis of largest diameter. This motivates our second characterization of the scale (which does depend on the couplings), namely the {\it field range} ${\cal R}_{\bs n\pm}$ within $\mathcal{T}_{\bs n}$ along the lightest direction -- the line defined by the eigenvector $\hat{\bs \Psi}_{\bs H}$ of smallest eigenvalue of the Hessian ${\bs H}$ at the minimum in the tile. The Hessian is given by
\be \label{Hessiann}
{\bs H}= \bold Q^\top \text{diag} \left( \Lambda_I^4 \right) \bold Q + {\mathcal O}[(\Po^\perp \bs n)^2]\,.
\ee
More precisely we define two field ranges ${\cal R}_{\bs n}(\pm \hat{\bs \Psi}_{\bs H}) \equiv {\cal R}_{\bs n\pm}$ as the canonically normalized field space distance between a minimum at $\bs \Theta_{\bs n}$ and the boundary of the corresponding tile in the least massive directions $\pm \hat{\bs \Psi}_{\bs H}$. Solving the equation defining the boundary of the tile,
\be
\left\lVert \pm {\cal R}_{\bs n\pm} \bold Q\hat{\bs \Psi}_{\bs H} + \underset{[-\pi,\pi]}{\text{mod}}\,{\bold Q} \bs\Theta_{\bs n} \right\rVert_\infty=\pi\,,
\ee
yields the field ranges
\be \label{generalFRlightest}
{\cal R}_{\bs n}(\pm \hat{\bs \Psi}_{\bs H}) = \min_I \left\{ {\pi\over |(\bold Q \hat{\bs \Psi}_{\bs H})^I|} \mp {\underset{[-\pi,\pi]}{\text{mod}}\,(\bold Q \bs\Theta_{\bs n})^I \over (\bold Q \hat{\bs \Psi}_{\bs H})^I}\right\} \,.
\ee
Note that \eqref{generalFRlightest} holds for the field range ${\cal R}_{\bs n}(\hat{\bs \Theta})$ along an arbitrary  direction $\hat{\bs \Theta}$. 

In well-aligned theories the sizes of exponentially many tiles ${\mathcal T_{\bs n}}$ are well-approximated by the size of the tile ${\mathcal T_{\bs 0}}$ containing the origin $\bs \Theta = \bs 0$ (or, when $P = N$, this is the only tile), so we will focus our attention on this last tile in the remainder of this section. In ${\mathcal T_{\bs 0}}$ the expressions for the diameter and field ranges simplify. For any unit $N$-vector $\hat{\bs \Theta}$ (in particular the lightest directions $\pm \hat{\bs \Psi}_{\bs H}$) we have
\be\label{diammin}
{\cal R}_{\bs 0} (\hat{\bs \Theta}) = {\cal R}_{\bs 0} (-\hat{\bs \Theta}) = {\pi\over \lVert \bold Q \hat{\bs \Theta} \rVert_\infty} \,,
\ee
which indeed follows from the general expression \eqref{generalFRlightest}. The diameter can alternatively be expressed as
\be\label{D0alt}
\mathcal{D}_{\bs 0} = \max \left\{  2 {\cal R}_{\bs 0} (\hat{\bs \Theta}) ~ \Big| ~ \hat{\bs \Theta} \in S^{N-1} \right\} = \max \left\{ {2 \pi \over \lVert \bold Q \hat{\bs \Theta} \rVert_\infty} ~ \Big| ~ \hat{\bs \Theta} \in S^{N-1} \right\} \,.
\ee
This last expression for the diameter can be used to derive bounds on it in an arbitrary theory, namely\footnote{See appendix \ref{D0boundsappendix} for a short derivation.}
\begin{equation} \label{D0bounds}
	\frac{2 \pi}{\lambda_\text{min}  (|\bold Q|)}< \mathcal{D}_{\bs 0} \leq \frac{2 \pi \sqrt{P}}{\lambda_\text{min}  (|\bold Q|)} \,,
\end{equation}
where $\lambda_\text{min}  (|\bold Q|)$ denotes the smallest eigenvalue of the matrix $| \bold Q | \equiv \sqrt{\bold Q^\top \bold Q}$, i.e. it is the smallest singular value of $\bold Q$. 

\subsection{N-flation, lattice and kinetic alignment} \label{explicitexamplessec}
\begin{figure}
\centering
\includegraphics[width=1\textwidth]{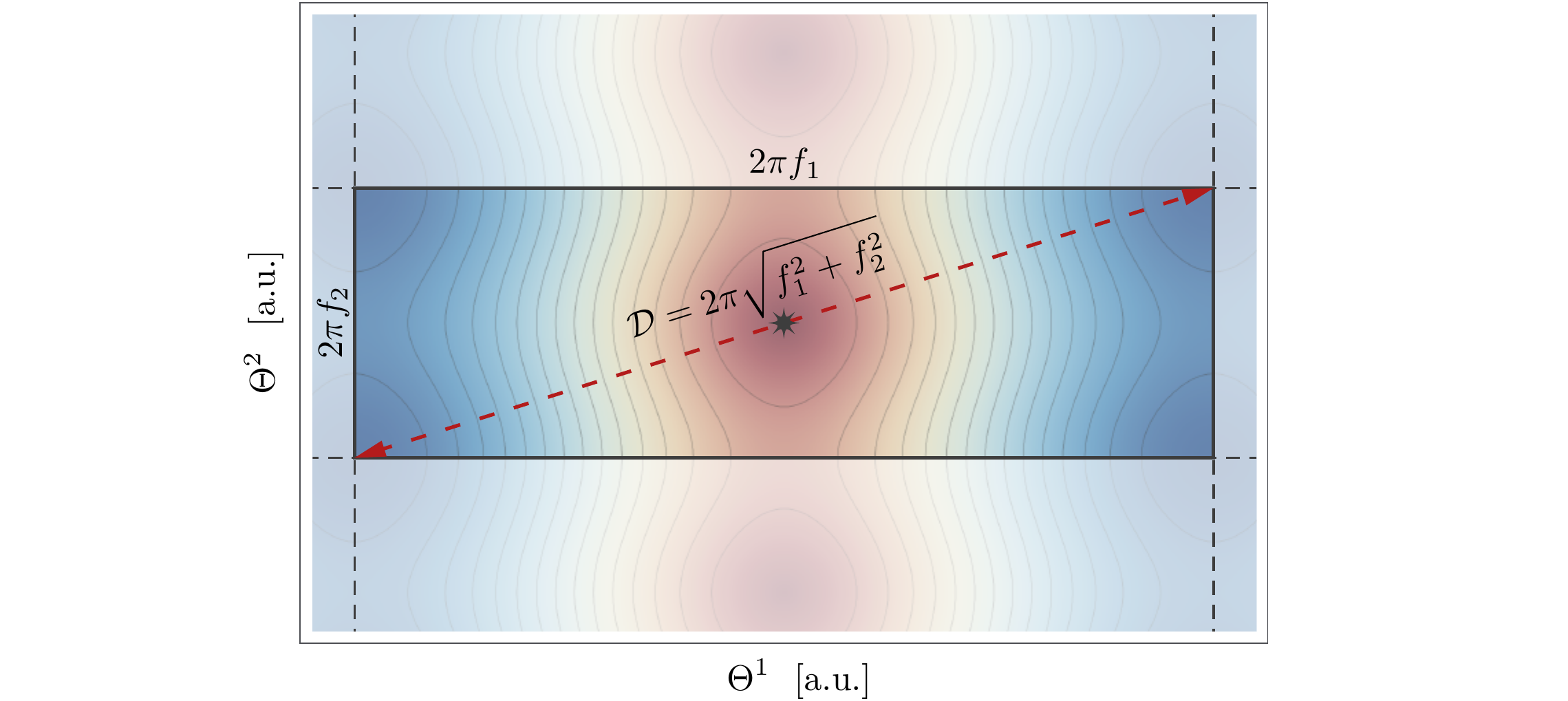}
\caption{\small Illustration of the tile ${\mathcal{T}}$ for N-flation. The diameter is the Pythagorean sum of the $f_I$.}\label{Nflationfig}
\end{figure}

In the previous sections we defined two notions of size of the axion field space in the vicinity of minima. To illustrate these definitions we now apply them to three special cases, focussing in particular on how the diameter in multi-axion theories may be enhanced compared to the single-axion diameter $2 \pi f$. We will consider N-flation \cite{Dimopoulos:2005ac}, lattice\footnote{The term ``lattice alignment'' is used synonymous with ``KNP alignment'' and should not be confused with the ``aligned lattice basis'' introduced in \S\ref{periodic-doms-sec}. The two terms refer to unrelated mechanisms.} (or KNP) alignment \cite{Kim:2004rp}, and finally kinetic alignment \cite{Bachlechner:2014hsa}. Each of these models was originally restricted to $P=N$ non-perturbative terms, but we will generalize the main ideas behind lattice and kinetic alignment to well-aligned theories with $P \geq N$. As mentioned in \S\ref{diameterFRsec} in well-aligned theories it suffices to consider the representative tile $\mathcal{T}_{\bs 0}$ around the origin, which we will do in the following.

\subsubsection{N-flation}
As our first example we consider the $N$-axion theory with diagonal kinetic matrix $\bs K =\text{diag}(f_I^2)$ and $P=N$ trivial charges ${\mathbfcal Q}= \mathbbold{1}_N$ \cite{Dimopoulos:2005ac}. In terms of canonical coordinates the Lagrangian is given by
\be
\mathcal L = {1\over 2}\partial \bs\Theta^\top \partial \bs\Theta - \sum_{I=1}^N \Lambda_I^4 \left[ 1 - \cos\left( { \Theta^I \over f_I} \right) \right]\,.
\ee
This theory has only one  distinct tile and has one minimum at the origin $\bs \Theta = \bs 0$, as illustrated in Figure \ref{Nflationfig}. The mass matrix is diagonal, $\bs H=\text{diag}(\Lambda_I^4/f_I^2)\propto \mathbbold{1}_N$, where for simplicity we selected the scales $\Lambda_I$ such that all masses are equal. 

The tile consists of an $N$-dimensional hyperrectangle  with side-lengths $2 \pi f_I$, which yields the diameter as the Pythagorean sum
\be\label{diameternflation}
{\cal D}=2\pi \sqrt{\sum_I f_I^2}\,.
\ee
For fixed $f_\text{max} \equiv \max_I \{ f_I \}$, the largest possible diameter is obtained when all metric eigenvalues are equal, $ f_I=f_\text{max}\,$: ${\cal D} = 2 \pi \sqrt{N} f_\text{max}$.  By contrast, if there are large hierarchies in the $f_{I}$ the diameter is ${\cal D} \simgeq 2\pi f_\text{max}$. Since all directions are equally massive, the lightest direction is degenerate and the field ranges accessible from the minimum at the origin are just half of the diameter, ${\cal R}_{\pm} = {\cal D}/2$. In the cosmological context this scenario is known as {\it N-flation}, a particular realization of assisted inflation \cite{Liddle:1998jc}: while none of the individual fields $\Theta^I$ traverse a displacement larger than $f_\text{max}$, the simultaneous displacement of $N$ fields realizes an invariant field range parametrically as large as $\sqrt{N} f_\text{max}$.

\subsubsection{Lattice alignment}\label{latticealignmentsec}
\begin{figure}
\centering
\includegraphics[width=1\textwidth]{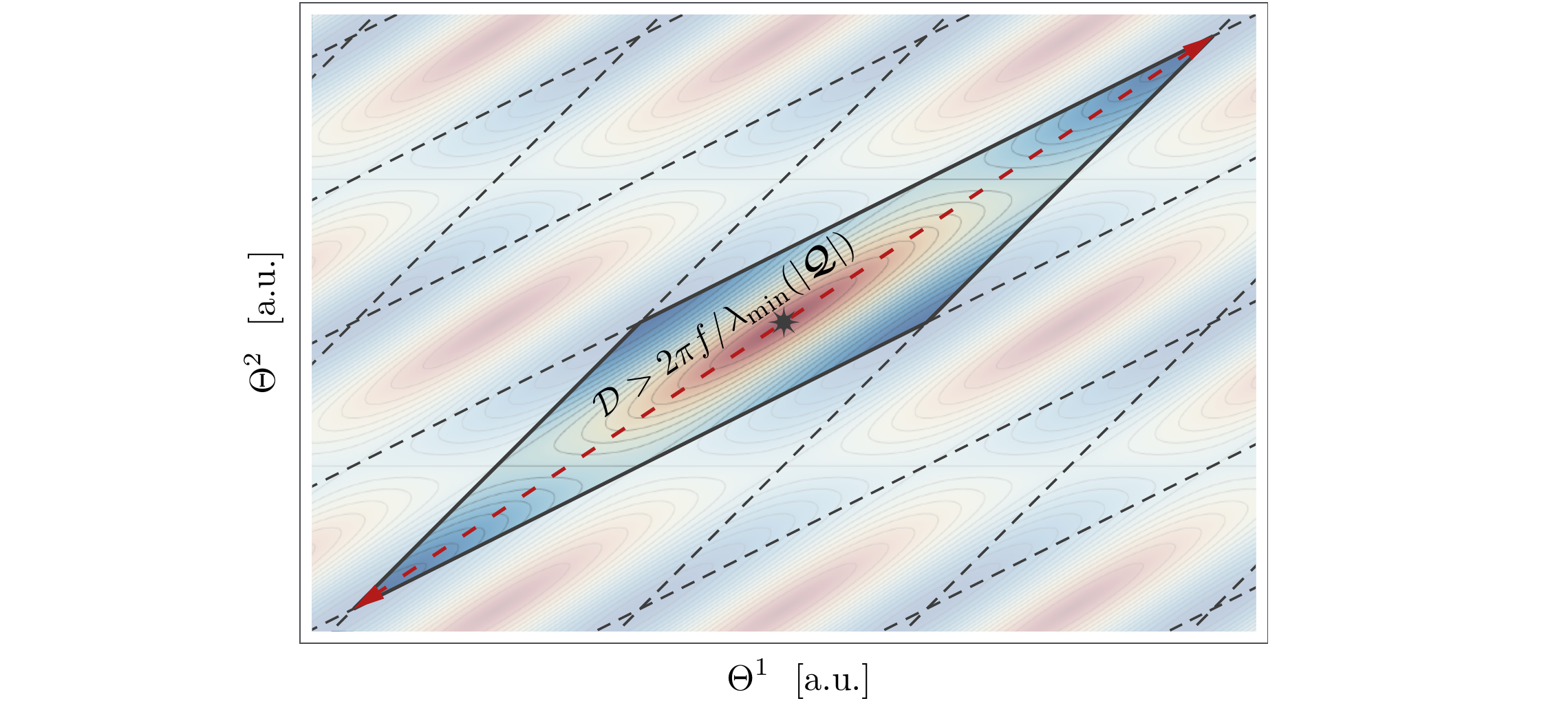}
\caption{\small Illustration of the tile ${\mathcal{T}}$ for lattice alignment. The diameter is enhanced by the inverse of the smallest singular value of the canonical charge matrix.  }\label{latticealignmentfig}
\end{figure}

We now consider the \textit{lattice alignment} (or KNP) mechanism, first discussed by Kim, Nilles and Peloso \cite{Kim:2004rp} for the special case $N=P$. Lattice alignment relies on a small singular value of the charge matrix. For simplicity, we assume a kinetic matrix proportional to the identity, $\bs K=f^2  \mathbbold{1}_N$, while the $P \geq N$ integer charges are left general,
\be\label{lagrlatticealingment}
\mathcal L = {1\over 2}\partial \bs\Theta^\top \partial \bs\Theta - \sum_{I=1}^P \Lambda_I^4 \left[ 1 - \cos \frac{\left(\mathbfcal Q \bs\Theta \right)^I}{f} \right]\,.
\ee
Recall the general bound \eqref{D0bounds} on the diameter of the tile $\mathcal{T}_{\bs 0}$,
\begin{equation}
\mathcal{D}_{\bs 0}>\frac{2 \pi f}{ \lambda_\text{min}(|\mathbfcal Q|)} \,.
\end{equation}
Lattice alignment is the observation that one can arbitrarily enhance the field range in these theories compared to the single-field $2 \pi f$ by decreasing the smallest eigenvalue $\lambda_\text{min}$ of $|{\mathbfcal Q}|=\sqrt{\mathbfcal Q^\top \mathbfcal Q}$.\footnote{Practically this can be achieved by having some columns of $\mathbfcal Q$ be nearly degenerate, meaning that their normalized variants have an overlap nearly equal to one (recall that $\mathbfcal Q$ is integer-valued).} As $\lambda_\text{min}$ decreases, the lower bound on the diameter increases. We illustrate this phenomenon in Figure \ref{latticealignmentfig}.

\subsubsection{Kinetic alignment} \label{kineticalignmentsec}
Finally let us discuss models with \textit{kinetic alignment} \cite{Bachlechner:2014hsa}. In the original discussion one assumed $P = N$, a trivial charge matrix $\mathbfcal Q = \mathbbold{1}_N$ and a general kinetic matrix $\bs K$, but the definition of kinetic alignment can just as easily be given in the more general context of well-aligned theories with $P \geq N$ and $\bs K, \mathbfcal Q$ unspecified. Note that the upper bound in \eqref{D0bounds} is saturated if
\begin{equation} \label{sufficientsaturationcondition}
	\lVert \bold Q \hat{\bs \Psi}_{|\bold Q|} \rVert_\infty = \frac{1}{\sqrt{P}} \lVert \bold Q \hat{\bs \Psi}_{|\bold Q|} \rVert_2 \,,
\end{equation}
in other words, if the direction defined by $\bold Q \hat{\bs \Psi}_{|\bold Q|}$ aligns with a diagonal of the $P$-cube, and $\hat{\bs \Psi}_{|\bold Q|}$ denotes the eigenvector of $|\bold Q|$ with smallest eigenvalue. For this to be possible it is in particular necessary that the constraint surface $\Sigma$ contains a diagonal of the $P$-cube. The sufficient condition \eqref{sufficientsaturationcondition} to saturate the upper bound in \eqref{D0bounds} is the extension of the original kinetic alignment proposal to arbitrary well-aligned theories with $P \geq N$. In models with (perfect) kinetic alignment we therefore have a diameter
\begin{equation} \label{generalkineticalignment}
	\mathcal{D} = \frac{2 \pi}{\lVert \bold Q  \hat{\bs \Psi}_{|\bold Q|} \rVert_\infty} = \frac{2 \pi \sqrt{P}}{\lambda_\text{min}(|\bold Q|)} \,.
\end{equation}
One might naively expect that alignment with a diagonal requires a great amount of fine-tuning in the canonical charge matrix $\bold Q$. In large dimensions $N,P \gg 1$, however, the converse is true: a $P$-cube has many more vertices ($2^P$) than faces ($2P$). Therefore, an isotropically oriented  vector within an isotropically oriented constraint surface $\Sigma$ is much more likely be pointing towards a vertex of a $P$-hypercube than towards a face. In \S\ref{randommechanismssec} we will make this expectation more precise and demonstrate that in broad classes of random axion theories the relation \eqref{sufficientsaturationcondition} is indeed approximately satisfied, and kinetic alignment is generic.
\begin{figure}
\centering
\includegraphics[width=1\textwidth]{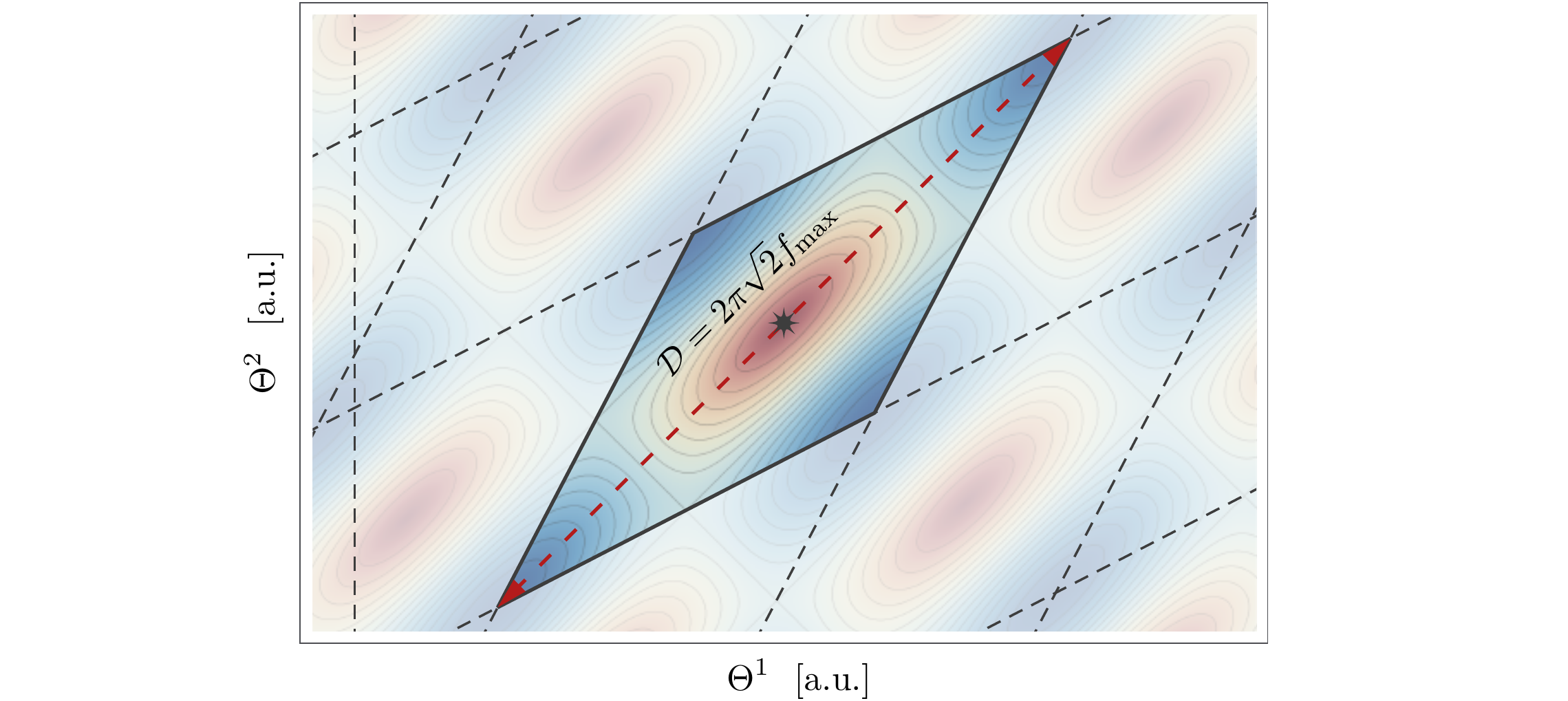}
\caption{\small Illustration of the tile ${\mathcal{T}}$ for kinetic alignment in the original model where $P = N, \mathbfcal Q = \mathbbold{1}_N$. The diameter is enhanced by $\sqrt{N}$ relative to $f_\text{max}$.}\label{kineticalignmentfig}
\end{figure}

For completeness let us consider the original model where $P = N$, $\mathbfcal Q = \mathbbold{1}_N$. Here $\bold Q = \bs K^{-1/2}$, and the eigenvector $\hat{\bs \Psi}_{\bs K^{-1/2}}$ is equal to the eigenvector corresponding to the largest eigenvalue $f_\text{max}^2$ of $\bs K$. The vector $\bold Q \hat{\bs \Psi}_{\bs K^{-1/2}} = \hat{\bs \Psi}_{\bs K^{-1/2}} / f_\text{max}$ points towards a diagonal when $\hat{\bs \Psi}_{\bs K^{-1/2}}$ does, which is the condition for (perfect) kinetic alignment as given in \cite{Bachlechner:2014hsa}. In this case
\begin{equation} \label{originalkineticalignment}
	\mathcal{D} = 2 \pi \sqrt{N} f_\text{max} \,.
\end{equation}
Note that here the diameter only depends on the largest metric eigenvalue $f_\text{max}^2$, and is independent of all other $f_I \le f_\text{max}$. When there are large hierarchies in the metric eigenvalues, the diameter (\ref{originalkineticalignment}) in kinetically aligned theories is larger by a factor of $\sqrt{N}$ relative to the diameter (\ref{diameternflation}) of the N-flation scenario. The enhancement of the diameter by $\sqrt{N}$ relative to $f_\text{max}$ was originally referred to as kinetic alignment, which we illustrate in Figure \ref{kineticalignmentfig}. More generally we have \eqref{generalkineticalignment}: the enhancement of the diameter by $\sqrt{P}$ relative to the inverse of the smallest singular value of $\bold Q$.

\subsection{Alignment in random ensembles}\label{randommechanismssec}
We now discuss diameters and field ranges in ensembles of random axion theories.\footnote{This was previously considered in \cite{Bachlechner:2014gfa}, but important aspects were missed that we discuss here.} The Lagrangian is given by
\begin{equation} \label{wellalignedtheory}
	\mathcal L = {1\over 2}\partial \bs\theta^\top\bs K \partial \bs\theta - \sum_{I=1}^P \Lambda_I^4 \left[ 1 - \cos\left( \mathbfcal Q \bs\theta  \right)^I \right] \,,
\end{equation}
and we study random ensembles of kinetic matrices $\bs K$, integer charges $\mathbfcal Q$, and dynamical scales $\Lambda_I^4$ introduced in \S\ref{intro} and used already in \S\ref{randomsec}.  We work at $2N\gtrsim P \geq N \gg 1$, where the theory \eqref{lagrtheta} is generically very well-aligned so that it is consistent set the phases to zero in \eqref{wellalignedtheory}. Furthermore as in the previous section we restrict our attention to the tile around the global minimum at $\bs \Theta = \bs 0$, because at least for minima in the quadratic regime (see \S\ref{unisample}) the diameters and field ranges along particular directions are similar up to ${\cal O}(1)$ factors.

Before discussing the details we first briefly review the main results for diameters in random axion theories.  With (\ref{generalkineticalignment}) the diameter of a tile in a well-aligned theory is given by
\be\label{d0rough}
{\cal D} \approx  {2 \pi \sqrt{P} \over \lambda_\text{min}(|\bold Q|)}\,,
\ee
where, again, $|{\bold Q}|=\sqrt{\bold Q^\top \bold Q}$. The diameter is enhanced by $\sqrt{P}$  due to the fact that a random $P$-vector is very likely to be aligned with a vertex of the $P$-cube periodic domain of the auxiliary lattice rather than with one of its faces (kinetic alignment, see \S\ref{kineticalignmentsec}), as well as by $1/\lambda_\text{min}(|\bold Q|)$ (lattice alignment, cf. \S\ref{latticealignmentsec}).

In the universal limit the matrix $\bold Q^{\top}\bold Q$ resembles a Wishart matrix so we expect its eigenvalue distribution to  depend only on $P$, $N$ and the scale of the charges. In the simple case of $\bs K = f^2 \mathbbold{1}_N$ and random $\mathbfcal Q$ this scale is $(\sigma_{\mathcal Q} / f)^2$. We can substitute a naive random matrix theory expectation \cite{silverstein1985} for the smallest eigenvalue in (\ref{d0rough}) and obtain 
\be\label{d0estimate}
{\cal D} \approx 2\pi \sqrt{P} {f\over \sigma_{\mathcal Q}}{1\over \sqrt{P} (1 - \sqrt{N/P})} \approx  2\pi \sqrt{P} {f\over \sigma_{\mathcal Q}} \frac{2 \sqrt{N}}{P - N} \,,
\ee
where the last approximate equality is valid when $P-N \ll N$. 

For aligned theories where $P\approx N$, there are three parametric enhancements that each can scale as $\sim \sqrt{N}$. The first factor of $\sqrt{P} \approx \sqrt{N}$ in (\ref{d0estimate}) is due to kinetic alignment and  depends on the fact that the canonical charge matrix is isotropic. The second factor can arise from the sparsity of the charge matrix, encoded in $\sigma_{\mathcal Q}$, which may be as small as $\approx 3/\sqrt{N}$, while retaining universality. The last factor is due to the eigenvalue distribution of a Wishart matrix and leads to two different parametric scalings: when $P - N = \text{constant}$ and $N$ is large, we have a third parametric enhancement of $\sqrt{N}$, while for $P-N\propto N$ and $N$ large the third term decreases the diameter parametrically as ${1/\sqrt{N}}$. Using the least possible entries in the integer charge matrix we therefore have the following scaling with $N$:
\be \label{diametersummary}
{\cal D} \lesssim \begin{cases} N^{3/2} f \,,~~~\text{for}~~P-N=\text{constant}\,,\\ N^{1/2} f\,,~~~\text{for}~~P-N\propto N\,,\end{cases}
\ee
both valid when $N$ is sufficiently large.

Even though these naive expectations are very crude, they turn out to accurately represent the mean diameter in a broad class of random models as we show below. Furthermore, we will find that the field range $\mathcal R_{\bs 0\pm}$ along the lightest direction scales with $N$ in a manner very similar to the diameter, and that this scaling is robust even when there are large hierarchies present in the dynamical scales $\Lambda_I^4$. The simple expectation (\ref{diametersummary}) from random matrix theory can then be compared to fundamental theories with axions, such as  compactifications of string theory \cite{Long:2016jvd}.

Finally, via \eqref{d0rough} the results \eqref{d0estimate} and \eqref{diametersummary} can also be applied to determine the scaling of the smallest eigenvalue $m^2$ of the Hessian matrix \eqref{Hessiann} -- that is, the mass-squared of the lightest field around a minimum. Up to $\mathcal{O}(1)$ factors and with all $\Lambda_I = \Lambda$ equal,
\begin{equation}
	m \approx 2 \pi \sqrt{P} \, \frac{\Lambda^2}{\mathcal{D}} \,.
\end{equation}

\subsubsection{Diameter estimates} \label{diametersec}
To estimate the diameter in random axion theories, recall from \S\ref{kineticalignmentsec} that perfect kinetic alignment implies that the diameter $\mathcal{D}_{\bs 0}$ of the tile is given by twice the field range along the eigenvector $\hat{\bs\Psi}_{|\bold Q|}$. In the random theories we introduced (modulo a caveat on the kinetic matrices $\bs K$ that we will discuss) we claim that that kinetic alignment is well-satisfied at large $P$. More precisely
\be\label{Dpsi2}
{2\pi\over \lVert{\bold Q} \hat{\bs\Psi}_{|\bold Q|} \rVert_\infty} \approx {2\pi\sqrt{P}\over \ell(P)}{1\over \lVert{\bold Q \hat{\bs\Psi}_{|\bold Q|}} \rVert_2}
\ee
is satisfied with ever-increasing probability as $P \rightarrow \infty$.\footnote{Recall the definition of $\ell(P)$ in (\ref{nminimamedian}),  $\ell(P) = \sqrt{2} \, \text{erf}^{-1}(2^{-1/P}) \approx \sqrt{2 \log P}$ at large $P$\,.} This hinges on the following fact: the images of eigenvectors of $\bold Q^\top \bold Q$ under $\bold Q$ are uniformly distributed on the unit $P$-sphere and therefore their entries are approximately normally distributed. In other words they are \textit{delocalized}.\footnote{In fact, the eigenvectors of $\bold Q^\top \bold Q$ themselves are delocalized (in particular $\hat{\bs\Psi}_{|\bold Q|}$), but this is of subordinate relevance.} The asymptotic exactness of the relation \eqref{Dpsi2} as $P \rightarrow \infty$ provides us with a reliable lower bound on the diameter $\mathcal{D}_{\bs 0}$ at large $P$,
\be\label{Dpsi3}
\mathcal{D}_{\bs 0} \geq 2 \mathcal{R}_{\bs 0} (\hat{\bs\Psi}_{|\bold Q|}) \rightarrow {2\pi\sqrt{P}\over \ell(P) \sigma_{\mathcal Q} \, \lambda_\text{min}(|\hat{\bold Q}|)} \,.
\ee
Here we have extracted a scale $\sigma_{\mathcal Q}$ from the entries in $\bold Q = \mathbfcal Q \, \bs K^{-1/2}$ via the definition $\hat{\bold Q} = \hat{\mathbfcal Q} \, \bs K^{-1/2}$, where the entries of $\hat{\mathbfcal Q}$ are distributed according to $\mathcal{N}(0,1)$ in the universal regime. This separates a trivial scaling factor $\sigma_{\mathcal Q}$ appearing in $\lambda_\text{min}(|\mathbfcal Q|)$ from its more intrinsic scaling properties with $N,P$. The lower bound \eqref{Dpsi3} is significant because it differs from an upper bound on $\mathcal D_{\bs 0}$ (cf. \S\ref{diameterFRsec}) only by the logarithmic factor $\ell(P)$ :
\be\label{Dpsi4}
\mathcal{D}_{\bs 0} \leq {2\pi\sqrt{P}\over \sigma_{\mathcal Q} \, \lambda_\text{min}(|\hat{\bold Q}|)} \,.
\ee

An intuitive understanding of \eqref{Dpsi2} was given in \S\ref{kineticalignmentsec}: in a large-dimensional $P$-cube the number of vertices vastly outnumbers the number of faces, thus it is much more likely for a vector to (approximately) point towards a vertex than towards a face. More quantitatively, the matrix $\bold Q^\top \bold Q$ is rotationally invariant (i.e. its form is preserved under $\bold Q \rightarrow \bs O \bold Q$ with $\bs O$ a $P \times P$ orthogonal matrix) so from general considerations in random matrix ensembles \cite{Rudelson1} we expect the eigenvectors of $\bold Q^\top \bold Q$ (including $\hat{\bs \Psi}_{|\bold Q|}$) to be delocalized (see also \cite{Bachlechner:2014rqa}). Furthermore, provided $\bs K$ does not introduce significant anisotropy, the vector $\bold Q \hat{\bs \Psi}_{|\bold Q|}$ will be delocalized as well.

In the following three sections we will verify the delocalization of $\bold Q \hat{\bs \Psi}_{|\bold Q|}$ in various ensembles of kinetic matrices. Having established this, we will use \eqref{Dpsi3} to analytically obtain a reliable lower bound on the diameter of $\mathcal{T}_{\bs 0}$ in the different ensembles. We will examine two distinct regimes :
\begin{itemize}
\item ``hard edge" : as $N \rightarrow \infty$, $P - N$ is held fixed,
\item ``soft edge" : as $N \rightarrow \infty$, $N/P$ is held fixed.
\end{itemize}

\subsubsection{Unit metric} \label{unitK}
\begin{figure}
  \centering
  \includegraphics[width=1\textwidth]{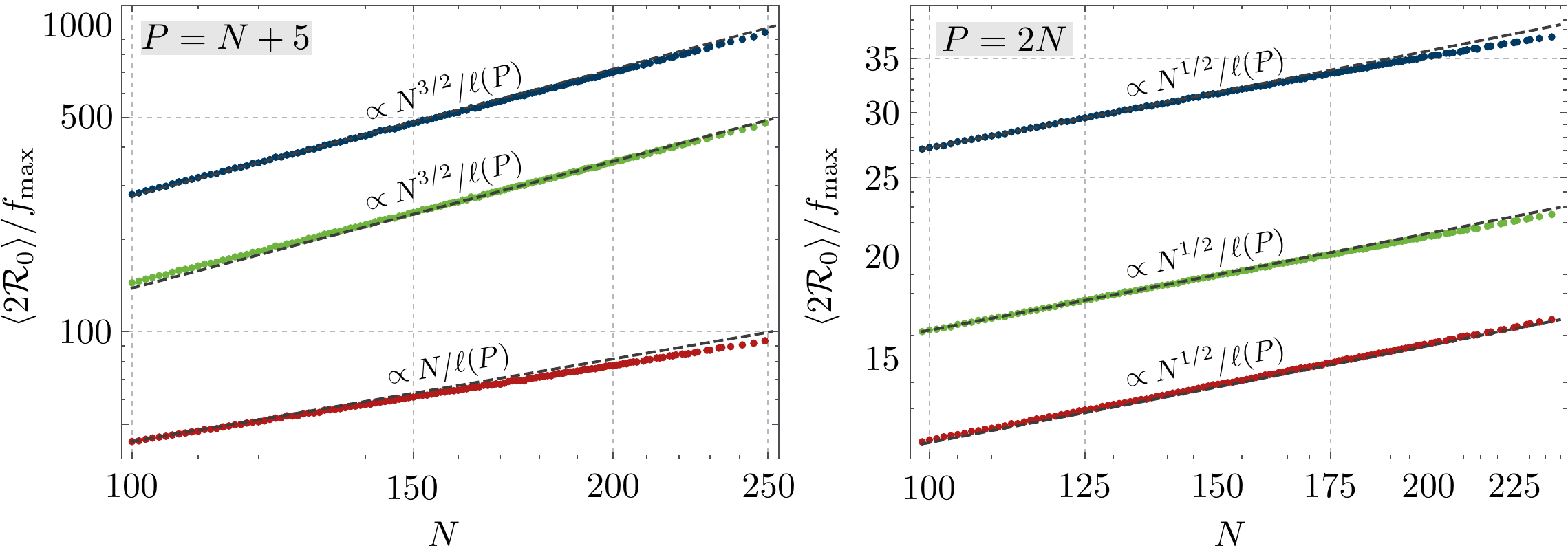}
  \caption{\small Mean field range along $\hat{\bs \Psi}_{|\bold Q|}$ for three ensembles of kinetic matrices $\bs K$. The field range along $\hat{\bs \Psi}_{|\bold Q|}$ is a significant lower bound on the diameter $\mathcal{D}_{\bs 0}$ as explained in \S\ref{diametersec}.  \underline{Left}: the hard edge, $P=N+5$. \underline{Right}: soft edge, $P=2N$. Top, middle and bottom (blue, green and red) lines denote the unit kinetic matrix, Wishart and inverse Wishart ensembles with largest eigenvalue $f_\text{max}^2$ (set to 1 here). We chose $\sigma_{\mathcal Q}=7/N$. Dashed lines show the analytic scaling. \label{FRdonotvarylambdaN+nu}}
\end{figure}

In random axion theories where $\bs K = f^2 \, \mathbbold{1}_N$ with $f$ a fixed scale, we can formulate the most precise results. In this case $\bold Q^\top \bold Q = (\sigma_{\mathcal Q} / f)^2 \hat{\mathbfcal Q}^\top \hat{\mathbfcal Q}$, is well-described by a Wishart matrix. At large $N$, the eigenvector $\hat{\bs \Psi}_{|\bold Q|}$ is delocalized and we further verified numerically that the $P$-vectors $\bold Q \hat{\bs \Psi}_{|\bold Q|}$ are delocalized as well. The field range along $\hat{\bs\Psi}_{|\bold Q|}$ is therefore given via (\ref{Dpsi3}) by
\begin{equation} \label{fieldrangeK=1}
	\mathcal{R}_{\bs 0}(\hat{\bs\Psi}_{|\bold Q|}) \approx {\pi\sqrt{P} f \over \ell(P) \sigma_{\mathcal Q} \, \lambda_\text{min}(|\hat{\mathbfcal Q}|)} \,.
\end{equation}

For any specific $N,P$ the probability distribution of the smallest eigenvalue in the Wishart ensemble can be calculated (either recursively \cite{edelman1988, edelman1991} or directly \cite{Wirtz2015}). When $P-N$ is held fixed as $N$ tends to infinity, the asymptotics of its mean satisfy $\langle \lambda_\text{min} (\hat{\mathbfcal Q}^\top \hat{\mathbfcal Q}) \rangle \sim 1 / N$ as $N \rightarrow \infty$, hence the name ``hard edge" statistics, as the smallest eigenvalue approaches the constraint that the matrix is positive definite. The knowledge of the distribution of the smallest eigenvalue can be translated to calculate the probability distribution of a lower bound on the diameter $\mathcal{D}_{\bs 0}$ via \eqref{fieldrangeK=1}. In general only the first $P - N$ moments of the probability distribution of the diameter along $\hat{\bs\Psi}_{|\bold Q|}$ are finite, while higher moments diverge.\footnote{In particular for $P = N$ the distribution of the diameter along $\hat{\bs\Psi}_{|\bold Q|}$ is heavy-tailed. In that case one can show that the median diameter behaves as $c(1,0) (2 \pi f N / \ell(P) \sigma_\mathcal{Q})$ with $c(1,0) = ( \sqrt{\log 4 + 1} - 1 )^{-1} \approx 1.84$.} More specifically, at large $N$, we find for the $z$th moment ($z \leq P-N$):
\begin{equation} \label{unitKFR}
	\langle \mathcal{R}_{\bs 0}(\hat{\bs\Psi}_{|\bold Q|})^z \rangle \sim c(z,P-N) \left( \frac{\pi N f}{\ell(P) \sigma_\mathcal{Q}} \right)^z ~~~ \text{ as } N \rightarrow \infty \text{ with } P-N =\text{ constant} \,,
\end{equation}
for some constants $c(z,P-N)$.\footnote{We found $c(1,1) = \sqrt{\pi / 2} \approx 1.25, ~ c(1,3) = \sqrt{\pi} e \left[ 3 I_1(1) - I_0(1) \right] / 3 \sqrt{2} \approx 0.49, ~ c(2,3) = \left( e^2 - 5 \right) / 8 \approx 0.30$ and $c(3,3) = \sqrt{\pi} e \left[ 5 I_0(1) - 9 I_1(1) \right] / 15 \sqrt{2} \approx 0.28$. To our knowledge there is no closed-form expression for $c(z,P-N)$ -- generic values must be determined numerically. This can be done algorithmically \cite{edelman1988}.} As $\sigma_\mathcal{Q} \sim 1 / \sqrt{N}$ in sparse models, we see the mean field range along $\hat{\bs \Psi}_{|\bold Q|}$ scales as $N^{3/2}$ for all $P-N > 0$, up to a logarithmic correction factor.\footnote{This parametrically improves the lower bound on the diameter of the tile obtained in \cite{Bachlechner:2014rqa} for this random model, where for $P-N > 0$ one found a lower bound that scales only linearly with $N$.} The standard deviation also exhibits this $N^{3/2}$ scaling with $N$.\footnote{However, the probability distribution on $\mathcal{R}_{\bs 0}(\hat{\bs\Psi}_{|\bold Q|})$ is super-exponentially suppressed at small values, and only power-like suppressed at large values. Thus the field range along $\hat{\bs \Psi}_{|\bold Q|}$ may easily become larger than the mean, but not smaller. This holds in the hard edge limit for all $P \geq N$. For the $P=N$ case see also the discussion in appendix A of \cite{Bachlechner:2014rqa}.} It may be instructive to recapitulate how the $N^{3/2}$ scaling arises, namely as the product of three factors $N^{1/2}$ with different origins: one comes from the alignment of $\bold Q \hat{\bs \Psi}_{|\bold Q|}$ along a diagonal of the $P$-hypercube (kinetic alignment), another from the square root of the smallest eigenvalue of a Wishart matrix scaling like $1/\sqrt{N}$ in the large $N$-limit where $ P - N$ is held fixed (lattice alignment), and finally a $\sqrt{N}$ arising from the assumed sparsity of the charge matrices.

For soft edge statistics where $N / P$ is held fixed, we may use the result of \cite{silverstein1985}, which implies $\langle \lambda_\text{min}(|\hat{\mathbfcal Q}|) \rangle \rightarrow ( \sqrt{P/N} - 1 ) \sqrt{N}$. We only discuss the mean field range along $\hat{\bs \Psi}_{|\bold Q|}$ in this case. We find
\begin{equation}
	\langle \mathcal{R}_{\bs 0}(\hat{\bs\Psi}_{|\bold Q|}) \rangle \sim \frac{\pi f}{\ell(P) \sigma_\mathcal{Q} (1 - \sqrt{N/P})} ~~~ \text{ as } N \rightarrow \infty \text{ with } N/P ~\text{constant} \,.
\end{equation}
So in sparse models where the amount of non-perturbative effects scales linearly with $N$, the mean field range along $\hat{\bs\Psi}_{|\bold Q|}$ (and hence, typically, the diameter $\mathcal{D}_{\bs 0}$) is enhanced only by the minimal amount $N^{1/2}$ compared to the single-axion $f$.

\subsubsection{Wishart metric} \label{wishartK}
Here we discuss the random ensemble where $\bs K$ is a Wishart matrix, constrained to have largest eigenvalue equal to a fixed scale $f_\text{max}^2$. Specifically, we draw the entries of a matrix $\bs A \in \mathbb{R}^{N \times N}$ from a normal distribution with zero mean and unit variance, and form the combination $\bs A^\top \bs A$. After, we rescale this matrix to have largest eigenvalue $f_\text{max}^2$.

As in the case of unit kinetic matrix, we find the eigenvector $\hat{\bs \Psi}_{|\bold Q|}$ to be delocalized to good accuracy for all $P \geq N$. For constant $P -N$ as $N \rightarrow \infty$, we numerically established that
\begin{equation}
	\lambda_\text{min} (| {\mathbfcal Q} \, \bs K^{-1/2} |) \approx \frac{2}{f_\text{max}} \, \lambda_\text{min} ( |{\mathbfcal Q}| )
\end{equation}
to good approximation (in the distributional sense). For fixed $N/P$, we found
\begin{equation}
\left \langle \frac{1}{\lambda_\text{min} ( |{\mathbfcal Q} \, \bs K^{-1/2} |)} \right\rangle = g(N/P) \left\langle \frac{1}{\lambda_\text{min} (| {\mathbfcal Q}| )} \right\rangle f_\text{max}
\end{equation}
for some profile $g(N/P)$ which decreases monotonically from 1 at $N/P = 0$ to 1/2 at $N/P = 1$. This allows us use the results of the $\bs K \propto \mathbbold{1}$ ensemble discussed in the previous section. Thus, in the hard edge limit, the field range along $\hat{\bs \Psi}_{|\bold Q|}$ satisfies
\begin{equation} \label{wishartKFR}
	\langle \mathcal{R}_{\bs 0}(\hat{\bs\Psi}_{|\bold Q|})^z \rangle \sim c(z,P-N) \left( \frac{\pi N f_\text{max}}{2 \ell(P) \sigma_\mathcal{Q}} \right)^z ~~~ \text{ as } N \rightarrow \infty \text{ with }  P-N \text{ constant} \,,
\end{equation}
while for soft edge asymptotics we find
\begin{equation}
	\langle \mathcal{R}_{\bs 0}(\hat{\bs\Psi}_{|\bold Q|}) \rangle \sim \frac{\pi \, g(N/P) f_\text{max}}{\ell(P) \sigma_\mathcal{Q} (1 - \sqrt{N/P})} ~~~ \text{ as } N \rightarrow \infty \text{ with } N/P  \text{ constant}  \,.
\end{equation}
As in the ensemble with $\bs K \propto \mathbbold{1}$, the mean field range along $\hat{\bs \Psi}_{|\bold Q|}$ scales as $N^{3/2}$ in the hard edge limit and as $N^{1/2}$ for soft edge statistics, up to a logarithmic correction factor.

\subsubsection{Heavy-tailed metric} \label{invwishartK}
Finally we consider an example of a heavy-tailed ensemble of kinetic matrices $\bs K$, i.e., large fluctuations of its eigenvalues are polynomially suppressed (as opposed to exponentially, as in the Wishart ensemble). Specifically, we consider inverse-Wishart matrices $\bs K$, with largest eigenvalue rescaled to $f_\text{max}^2$. (We rescale the combination $(\bs A^\top \bs A)^{-1}$ of the previous section.)\footnote{With this rescaling the eigenvalues of $\bs K$ no longer follow a heavy-tailed distribution. However, we will see that the field range distribution along $\hat{\bs \Psi}_{|\bold Q|}$ is qualitatively different in this ensemble compared to the others we have discussed, and we will argue that this is precisely due to the heavy-tailed character of the non-rescaled eigenvalues of $\bs K$.}

Once again $\hat{\bs \Psi}_{|\bold Q|}$ is delocalized to good approximation, and we established numerically that the mean field range along $\hat{\bs \Psi}_{|\bold Q|}$ qualitatively behaves as
\begin{equation}
	\langle \mathcal{R}_{\bs 0}(\hat{\bs\Psi}_{|\bold Q|}) \rangle \propto \frac{\sqrt{N} f_\text{max}}{\ell(P) \sigma_\mathcal{Q}} ~~~ \text{ as } N \rightarrow \infty \text{ with } P-N \text{ constant}  \,,
\end{equation}
and as
\begin{equation}
	\langle \mathcal{R}_{\bs 0}(\hat{\bs\Psi}_{|\bold Q|}) \rangle \propto \frac{f_\text{max}}{\ell(P) \sigma_\mathcal{Q}} ~~~ \text{ as } N \rightarrow \infty \text{ with } N/P \text{ constant}  \,.
\end{equation}
In the hard edge case, a factor of $\sqrt{N}$ is lost compared to the unit and Wishart kinetic matrix ensembles because the distribution of the smallest singular value of the canonical charge matrix is qualitatively different. In particular, we found
\begin{equation}
	\frac{f_\text{max}}{\sigma_\mathcal{Q}} \left\langle \lambda_\text{min} ( | \mathbfcal Q \, \bs K^{-1/2} | ) \right\rangle = \mathcal{O}(1) ~~~ \text{ as } N \rightarrow \infty \text{ with } P - N \text{ fixed}. 
\end{equation}
An intuitive explanation of this goes as follows: in the ensemble where the kinetic matrix is a rescaled Wishart matrix,  $\bs K = \bs A^\top \bs A / \lambda_\text{max} (\bs A^\top \bs A )$, the largest eigenvalue $\lambda_\text{max}$ is not too different from a typical eigenvalue; the ratio $\lambda_\text{max} / \text{median}_i(\lambda_i)$ is of order 1. This is because large eigenvalues occur with exponentially small probability. So a typical eigenvalue of $\bs K$ is expected to be broadly distributed on the interval $[0,f_\text{max}^2]$. In other words,
\begin{equation}
	f_\text{max}^2 \, \lambda_\text{min} ( \bs K^{-1/2} {\mathbfcal{Q}}^\top {\mathbfcal Q} \bs K^{-1/2} ) \approx \lambda_\text{min} ( {\mathbfcal{Q}}^\top {\mathbfcal Q} ) \approx \sigma_\mathcal{Q}^2 / N \,.
\end{equation}
In the ensemble $\bs K = \left( \bs A^\top \bs A \right)^{-1} / \lambda_\text{max} [ ( \bs A^\top \bs A )^{-1} ]$, on the other hand, the largest eigenvalue is on average much larger than a typical eigenvalue. So eigenvalues of $\bs K$ will be small ($\ll f_\text{max}^2$) with high probability. With this one can appreciate how $f_\text{max} \langle \lambda_\text{min} \rangle / \sigma_{\mathcal Q} = \mathcal{O}(1)$ in this ensemble.

For soft edge statistics the reason for the reduction of the expected field range down to $N^{1/2}$ is the same as in the other ensembles: $\langle {\lambda}_\text{min} \rangle \propto N$. We summarize these results together with the mean field range behaviour along $\hat{\bs \Psi}_{|\bold Q|}$ in the unit and Wishart kinetic matrix ensembles (sections \ref{unitK} and \ref{wishartK}) in Figure \ref{FRdonotvarylambdaN+nu}.

\subsubsection{Variable couplings: dynamic alignment} \label{dynamicalignmentsec}
\begin{figure}
  \centering
  \includegraphics[width=1\textwidth]{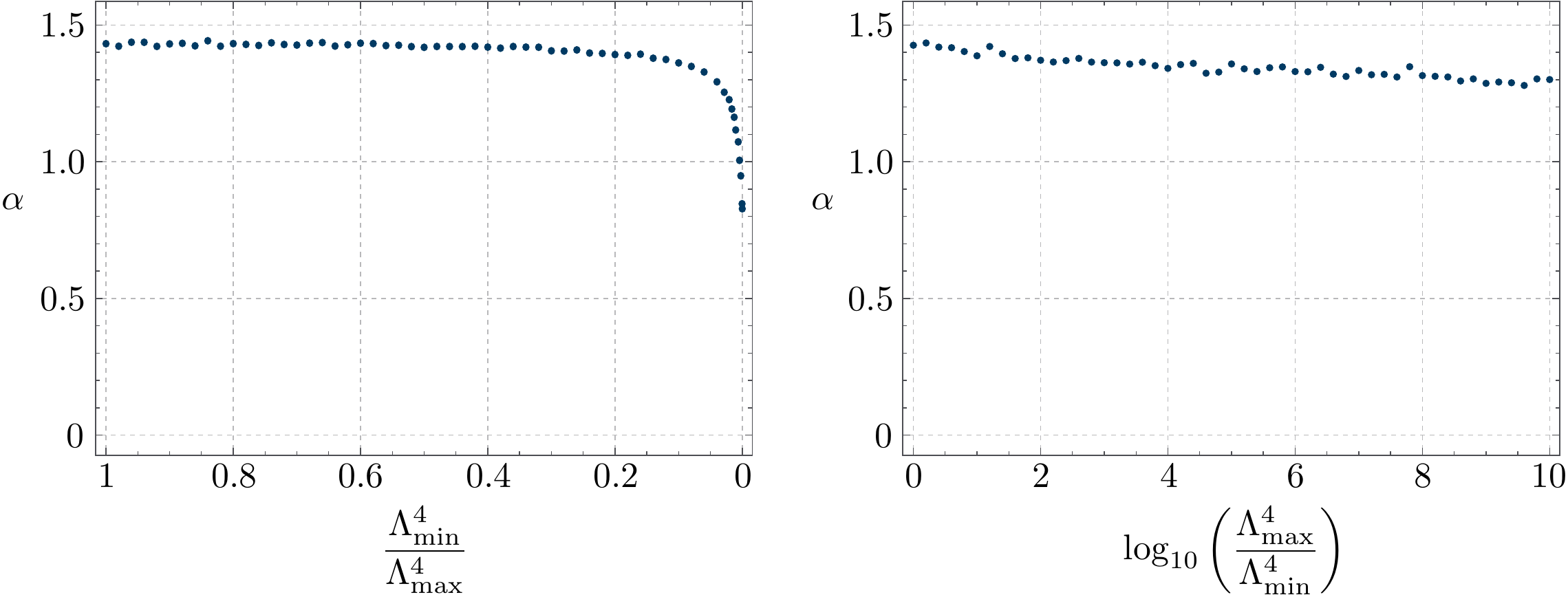}
  \caption{\small Scaling of mean field range along the lightest direction $\hat{\bs \Psi}_{\bs H}$ with $N$, ${\cal R}(\hat{\bs \Psi}_{\bs H}) / f \propto N^\alpha/\ell(P)$, for the hard edge case $P = N + 5$. \underline{Left}: $\Lambda_I^4$ are uniformly distributed over $[\Lambda_\text{min}^4,\Lambda_\text{max}^4]$. \underline{Right}: $\log_{10}(\Lambda_\text{max}^4/\Lambda_I^4)$ are uniformly distributed over $[0,10]$. The scaling is extracted from sampling the range $100\le N\le 200$. Observe that even when large hierarchies are present in the couplings, the scaling behavior with $N$ of the mean field range along the lightest direction may remain very similar to the scaling behavior of the mean diameter. \label{dynamicalignmentfig}}
\end{figure}

To investigate the dynamics we consider the field range along the lightest direction $\hat{\bs \Psi}_{\bs H}$ emanating from $\bs \Theta = \bs 0$, as discussed in \S\ref{diameterFRsec}. From $\bs H = \bold Q^\top \text{diag} \left( \Lambda_I^4 \right) \bold Q$ we observe that if the couplings $\Lambda_I^4$ are not all equal, this is not the same direction as the previously considered $\hat{\bs \Psi}_{|\bold Q|}$ (which we showed was well-aligned with the direction providing the actual diameter $\mathcal{D}_{\bs 0}$ of the tile $\mathcal{T}_{\bs 0}$ due to kinetic alignment). However if the two directions are sufficiently aligned the available field range within the tile along each will be similar. Below we illustrate in two specific examples how much the couplings may deviate from overall equality before this alignment fails and the expected field ranges along $\hat{\bs \Psi}_{|\bold Q|}$ and $\hat{\bs \Psi}_{\bs H}$ become parametrically different in $N$. In these examples we find that $\hat{\bs \Psi}_{|\bold Q|}$ and $\hat{\bs \Psi}_{\bs H}$ remain well-aligned although the couplings may differ from one another to a certain degree. While a general analysis of the alignment between the lightest and the kinematic direction lies beyond the scope of this work,\footnote{An interesting question is whether there exists a simple criterion on the distribution of the couplings $\Lambda_I^4$ that assures the vectors $\hat{\bs \Psi}_{|\bold Q|}$ and $\hat{\bs \Psi}_{\bs H}$ are aligned. A useful definition of ``aligned" would relate the field ranges $\mathcal{R}(\hat{\bs \Psi}_{|\bold Q|})$ and $\mathcal{R}(\hat{\bs \Psi}_{\bs H})$ as $N,P \rightarrow \infty$.} the insensitivity of this alignment to the distribution of couplings has been called \textit{dynamic alignment} in a previous discussion \cite{Bachlechner:2014gfa}. 

For simplicity we consider ensembles with trivial kinetic matrix, $\bs K = f^2 \mathbbold{1}$, and consider two qualitatively different hierarchies in the couplings. These are illustrated in Figure \ref{dynamicalignmentfig}. In a first example, assume the couplings $\Lambda_I^4$ are uniformly distributed on the interval $[\Lambda_\text{min}^4,1]$. As we dial down $\Lambda_\text{min}^4$ from one to zero, we expect the field range along $\hat{\bs \Psi}_{\bs H}$ to diminish with respect to the field range along $\hat{\bs \Psi}_{|\bold Q|}$. In a second example, consider the case where the couplings are log-uniformly distributed on the interval $[10^{-10},1]$. Although the couplings may wildly differ from one another, the expected field range along the lightest direction is very robust.

\section{Conclusions} \label{conc}

In this paper we  presented the details of a novel formalism that allows us to analyze a class of periodic functions of many variables, focusing  on those of the form \eqref{symm}.  Our technique  identifies the exact periods, breaks up a unit cell of the resulting lattice into conveniently labelled tiles, and makes it possible to identify \emph{approximate} shift symmetries.  This last feature is very powerful, because (at least in the large $N$ random ensembles we consider) these approximate symmetries are extremely close to exact.  As a result if we analyze one region of the function, the results can be translated to exponentially many other regions with exponential accuracy.  In particular the number of critical points is exponentially large, and the spacing of their energy levels  is exponentially fine.

We employ this technology to determine the number and characteristics of critical points of the potential \eqref{symm}, and to analyze the vicinity of a typical minimum.  For $N \sim 100$ there are generically (in our ensembles) an enormous number of distinct minima, each with a unique vacuum energy.  This number can be larger than $10^{120}$ and the distribution of  energies is smooth, so this theory provides values for the vacuum energy consistent with observation even if the energy scales in the potential are close to the Planck scale \cite{Bachlechner:2015gwa, Bachlechner:2017zpb}.  Furthermore we find that there is a  range of masses  for the canonically normalized fields that is enhanced by powers of $N$. The lightest of these provide long gentle slopes that may turn out to be suitable for large-field inflation if the energy scales are high (or small field inflation if they are lower) \cite{Bachlechner:2017zpb}.  Lastly, the characteristics of the critical points are such that the barriers between basins of attractions of adjacent minima tend to be quite thin.  We will explore the consequences of this for tunneling transitions in \cite{bejk2}, where we will also discuss the natural candidate for dark matter that arises in these theories.


\section*{Acknowledgements}
We thank Nick Cook, Frederik Denef, Mario Kieburg, Liam McAllister, Arman Mimar, Ruben Monten, Tom Rudelius, Geoff Ryan, Marjorie Schillo and John Stout for valuable discussions. OJ is supported by a James Arthur fellowship. The work of MK is supported in part by the NSF through grant PHY-1214302, and he acknowledges membership at the NYU-ECNU Joint Physics Research Institute in Shanghai. The work of TB and KE was supported by DOE under grant no. DE-SC0011941.

\appendix
\section{Separating flat directions in $V$} \label{app:massless}
In this appendix we discuss how to separate  flat directions in $V$ (that is, directions in field space with exactly zero potential) from the non-flat directions.   Such directions are present when the charge matrix $\boldsymbol{\mathcal{Q}}$ has rank $R < N$.  This can happen either when $P < N$ (in which case necessarily $R \leq P < N$) or because $\boldsymbol{\mathcal{Q}}$ is not full rank.  After the separation procedure described in this appendix, we are left with $N-R$ massless fields decoupled from a reduced theory of $R$ axions with  a full rank charge matrix, to which the  techniques in the bulk of our paper apply.

We start with the $N$-axion theory \eqref{lagrtheta},
\be \label{lagrthetamassless}
{\mathcal L}={1\over 2} \partial\bs\theta^\top \bs K \partial\bs\theta-\sum_{I=1}^P \Lambda_I^4  \left[1-\cos\left(\mathbfcal Q\bs\theta + \bs \delta \right)^I \right]+\dots\,,
\ee
and change coordinates to canonically normalized fields $\bs \Theta = \sqrt{\bs K} \bs \theta$, in which the charge matrix takes the form $\bold Q = \mathbfcal Q \, \bs K^{-1/2}$. Suppose the rank of $\boldsymbol{\mathcal{Q}}$ is $R < N$. This happens either when $P < N$, or when $P \geq N$ but not all columns of $\mathbfcal{Q}$ are linearly independent. Then there are $L = N - R$ flat directions; moving along these directions does not change $V$. In other words, the null space of $\boldsymbol{\mathcal{Q}}$, $\text{ker}(\mathbfcal{Q})$, is $L$-dimensional, which is the same as the dimension of $\text{ker}(\bold Q)$. Find an orthonormal basis $\boldsymbol{t}_1, \boldsymbol{t}_2, \dots, \boldsymbol{t}_L$ of $\text{ker}(\bold Q)$, and extend it to a basis of $\mathbb{R}^N$ by the adherence of $R$ vectors $\boldsymbol{t}_{L+1}, \boldsymbol{t}_{L+2}, \dots , \boldsymbol{t}_N$ (note these are generally not integer-valued vectors). Now define new coordinates $\bs{\Omega}$ via the rule
\begin{equation}
	\boldsymbol{\Theta} = \trafo{\Omega}{\Theta} \boldsymbol{\Omega} = \big( \trafo{\Omega_L}{\Theta} \, \vert \, \trafo{\Omega_R}{\Theta} \big) \left( \begin{matrix} \bs \Omega_L \\ \bs \Omega_R \end{matrix} \right) \,,
\end{equation}
where we have split the $N$-vector $\bs \Omega$ into a piece of length $L$ and a piece of length $R$, and the matrices $\trafo{\Omega_L}{\Theta}$ and $\trafo{\Omega_R}{\Theta}$ are composed by placing the $\bs t_{1 , \dots, L}$ respectively the $\bs t_{L+1 , \dots, N}$ on consecutive columns. Note that $\trafo{\Omega}{\Theta}$ is an orthogonal matrix. In these coordinates the flat directions are manifestly separated from the non-flat ones. Indeed, since
	\begin{equation} \label{relationAA}
	\bold Q \boldsymbol{\Theta} = \bold Q \trafo{\Omega_R}{\Theta} \boldsymbol{\Omega}_R \equiv \bold Q_R \boldsymbol{\Omega}_R \,,
	\end{equation}
only the $R$ fields $\boldsymbol{\Omega}_R$ appear in the potential. Furthermore the kinetic term reads
\begin{equation}
	\frac{1}{2} \partial \bs \Omega_L^\top \partial \bs \Omega_L + \frac{1}{2} \partial \bs \Omega_R^\top \partial \bs \Omega_R \,,
\end{equation}
such that the massless fields $\bs \Omega_L$ decouple. In \eqref{relationAA} $\bold Q_R$ is a full rank $P \times R$ matrix (but it is not integer-valued, in general). For the final step, note that there exists an invertible $R \times R$ matrix $\bs R$ such that $\bold Q_R \bs R$ is an integer-valued matrix.\footnote{This is because $\text{colsp}(\bold Q_R) = \text{colsp}(\mathbfcal Q)$, and the projector onto a linear subspace is basis-independent. So we know the projector onto $\text{colsp}(\bold Q_R)$ has rational entries, implying the existence of $\bs R$ (see also footnote \ref{Qintegerfootnote}).} Transforming to coordinates $\bs \Omega_R = \bs R \, \bs \Xi_R$, the Lagrangian in $\bs \Xi_R$-coordinates has a kinetic matrix $\bs R^\top \bs R$ and an integer-valued, full rank charge matrix $\bold Q_R \bs R$. As $P \geq R$, there are effectively more non-perturbative contributions to the potential than axions. So one can apply the techniques developed in the body of this work to the reduced system of $R$ axions $\bs \Xi_R$.

\section{Eliminating very massive axions}\label{app:classical}
In this appendix we discuss how to eliminate axions that receive large masses, e.g. due to their coupling to classical sources. At low energies these axions are effectively fixed to a certain value. Specifically let us assume the $N$-axion theory \eqref{lagrtheta},
\be \label{lagrthetamassive}
{\mathcal L}={1\over 2} \partial\bs\theta^\top \bs K \partial\bs\theta-\sum_{I=1}^P \Lambda_I^4  \left[1-\cos\left(\mathbfcal Q \bs\theta + \bs \delta \right)^I \right]+\dots\,,
\ee
is supplemented with $L < N$ such classical sources, where the rows of a full rank $L \times N$ matrix $\mathbfcal C$ specify which axion combinations couple to each source,\footnote{In general identical combinations may couple to more than one source, implying that $\mathbfcal C$ would not be full rank, or $L$ may be greater than $N$. These cases are easily dealt with.} and that these directions are fixed according to
\begin{equation} \label{massiveconstraints}
	\mathbfcal{C} \bs \theta = \bs \delta_{\mathcal{C}} \,,
\end{equation}
where $\bs \delta_{\mathcal{C}}$ is a certain $L$-vector. We would like to perform a (linear) coordinate transformation $\bs \theta \rightarrow \bs \xi$ that disentangles the massive directions from the others in \eqref{lagrthetamassive}. In order to retain the same discrete shift symmetries in the $\bs \xi$-basis as in the $\bs \theta$-basis, such a transformation must be unimodular. To construct it, note that the directions unaffected by the classical sources lie in $\text{ker}(\mathbfcal{C})$, which has dimension $N - L \equiv R$. The intersection $\text{ker}(\mathbfcal{C}) \cap \mathbb{Z}^N$ is thus a lattice of rank $R$.\footnote{We assume the projector onto $\text{ker}(\mathbfcal{C})$ contains only rational entries.} Extend a basis $\boldsymbol{t}_1, \boldsymbol{t}_2, \dots, \boldsymbol{t}_R$ of this lattice to a basis for $\mathbb{Z}^N$ by adding $L$ integer vectors $\boldsymbol{t}_{R+1}, \boldsymbol{t}_{R+2}, \dots, \boldsymbol{t}_N$ (see appendix \ref{app:lattice} for a proof that this can always be done). The $N \times N$ matrix $\trafo{\xi}{\theta} = \left( \trafo{\xi_R}{\theta} \, \, | \, \, \trafo{\xi_L}{\theta} \right)$ is unimodular, where $\trafo{\xi_R}{\theta}$ ($\trafo{\xi_L}{\theta}$) is formed by placing the $\boldsymbol{t}_1, \boldsymbol{t}_2, \dots, \boldsymbol{t}_R$ ($\boldsymbol{t}_{R+1}, \boldsymbol{t}_{R+2}, \dots , \boldsymbol{t}_N$) on consecutive columns. If we denote the first $R$ components of the $N$-vector $\bs{\xi}$ by $\bs{\xi}_R$ and the final $L = N - R$ by $\bs{\xi}_L$, and relate the coordinates $\bs{\xi}$ to $\bs \theta$ by $\bs \theta = \trafo{\xi}{\theta} \bs \xi$, we have $\mathbfcal C \bs \theta = \mathbfcal C \trafo{\xi_L}{\theta} \bs \xi_L$, and thus via \eqref{massiveconstraints}
\begin{equation}
	\bs \xi_L = ( \mathbfcal C \trafo{\xi_L}{\theta} )^{-1} \bs \delta_{\mathcal{C}} \,.
\end{equation}
The Lagrangian \eqref{lagrtheta} then effectively becomes
\be
{\mathcal L} = {1\over 2} \partial\bs\xi_R^\top \bs K_R \partial\bs\xi_R - \sum_{I=1}^P \Lambda_I^4  \left[ 1 - \cos\left(\mathbfcal Q_R \bs\xi_R + \bs \delta_R \right)^I \right] + \dots\,,
\ee
where
\begin{align}
	\bs K_R &= \trafo{\xi_R}{\theta}^\top \, \bs K \, \trafo{\xi_R}{\theta} \,, \\
	\mathbfcal Q_R &= \mathbfcal Q \, \trafo{\xi_R}{\theta} \,, \\
	\bs \delta_R &= \bs \delta + \mathbfcal Q \, \trafo{\xi_L}{\theta} ( \mathbfcal C \trafo{\xi_L}{\theta} )^{-1} \bs \delta_{\mathcal{C}}.
\end{align}
The axions that couple to classical sources have been eliminated while preserving the original form of the theory.

\section{Extending a sublattice basis}\label{app:lattice}
In this appendix we prove that any basis for the rank $N$ sublattice defined by the intersection of an $N$-dimensional linear subspace $\Sigma$ with the integer lattice $\mathbb Z^{P}$ can be extended to a basis for the full integer lattice. In particular, one can always supplement the $N$ $P$-vectors $\bs t^\parallel_i$ with $P-N$ additional vectors  $\bs t^\nparallel_a$ to form a basis for $\mathbb Z^{P}$ (cf.~\S\ref{symm}).

Before giving the proof, it is perhaps worth giving an example of a sublattice that \emph{cannot} be extended this way.  First, recall that since we are discussing lattices, one should consider only  linear combinations of the basis vectors with integer coefficients. Now consider the rank one sublattice of $\mathbb Z^{2}$ that is the even integers along the $x$-axis; that is, the sublattice generated by the vector $(2,0)$.  It is clear that this cannot be extended to a basis for $\mathbb Z^{2}$ by the addition of any vector.  However, note that this sublattice is \emph{not} the intersection of any linear subspace with $\mathbb Z^{2}$ -- the intersection of the $x$-axis with $\mathbb Z^{2}$ is  generated by the vector $(1,0)$.

The proof  is as follows:\footnote{MK would like to thank Arman Mimar for explaining this to him.} every rank $P$ lattice $\mathscr{L}$ can be thought of as a finitely generated free Abelian group (under addition of the lattice vectors).  Any rank $N \leq P$ sublattice $\mathscr{L}'$ of $\mathscr{L}$ is  then a subgroup. The structure theorem for finitely generated Abelian groups   implies that there always exists a special basis $\mathbfcal B \equiv \{ \bs b_{1}, ..., \bs b_{P} \}$  for $\mathscr{L}$  with the property that $\{ a_{1} \bs b_{1}, ..., a_{N} \bs b_{N} \}$ is a basis for $\mathscr{L}'$, where $\{a_{1}, ... a_{N}\}$ are a set of integers with the property that each divides the next. However if $\mathscr{L}'$ is the intersection of a linear subspace $\Sigma$ with the lattice $\mathscr{L}$, then  $a_{i} \bs b_{i} \in \mathscr{L}'$ implies $\bs b_{i} \in \mathscr{L}'$.  Therefore  $\{\bs b_{1}, ..., \bs b_{N}\}$ must in fact be a basis for $\mathscr{L}'$ (because it generates $\{a_{1} \bs b_{1}, ..., a_{N} \bs b_{N}\}$). But this basis can trivially be extended to the basis $\mathbfcal B$ for $\mathscr{L}$ by  appending $\{\bs b_{N+1}, ... , \bs b_{P}\}$.  

To see that \emph{any} basis for $\mathscr{L}'$ can be extended to a basis for $\mathscr{L}$, note that any basis for $\mathscr{L}'$ is related to any other basis (for instance, $\{\bs b_{1}, ..., \bs b_{N}\}$) by some $N \times N$ unimodular matrix.  But any such $N \times N$ unimodular matrix can obviously be extended to a block-diagonal $P \times P$ unimodular matrix.  Acting with this matrix on $\mathbfcal B$  gives the extended basis.

\section{An explicit example}\label{coordinateexample}
In this appendix we illustrate the construction of the aligned lattice basis and the reduction of the relative phases in an explicit example with $P=N+1=3$. In particular, we consider a theory with charges and phases
\be
{\mathbfcal Q}=\left(\begin{matrix} 1&1\\2&-3\\-3&0\end{matrix}\right)\,,~~~\bs\delta=\left(\begin{matrix}2.04\\6.20\\4.16\end{matrix}\right)\,,
\ee
and take the non-perturbative scales $\Lambda_I^4$ to be identical for simplicity. The potential is therefore
\be\label{thetapota}
V(\bs \theta)=\Lambda^4 \left[3-\cos\left(\theta^1+\theta^2+2.04 \right)-\cos\left(2\theta^1-3\theta^2+6.20\right)-\cos\left(-3\theta^1 + 4.16\right)\right]\,.
\ee
The auxiliary coordinates $\bs\phi$ are constrained by the condition $\Pob \bs \phi=\bs \delta$ to reproduce (\ref{thetapota}) on-shell, where the orthogonal projector onto the orthogonal complement of the constraint surface $\Sigma$ is given by
\be
\Pob ={\mathbbold 1}- {\mathbfcal Q} ({\mathbfcal Q}^\top{\mathbfcal Q})^{-1}{\mathbfcal Q}^\top={1\over 115}\left(
\begin{array}{ccc}
 81 & 27 & 45 \\
 27 & 9 & 15 \\
 45 & 15 & 25 \\
\end{array}
\right)\,.
\ee

As expected, the rank of the orthogonal projector is $P-N=1$ in this example. Employing the LLL lattice reduction algorithm \cite{Lenstra1982,lllpackage}, we find the aligned basis vectors,
\be
\trafo{\omega}{\phi}=\left(
\begin{array}{ccc}
 -1 & -1 & 1 \\
 3 & -2 & -1 \\
 0 & 3 & -1 \\
\end{array}
\right)\,,~~\trafo{\omega_\parallel}{\phi}=\left(
\begin{array}{ccc}
 -1 & -1  \\
 3 & -2 \\
 0 & 3  \\
\end{array}
\right)\,,~~\trafo{\omega_\nparallel}{\phi}=\left(
\begin{array}{ccc}
 1 \\
 -1 \\
 -1 \\
\end{array}
\right)\,.
\ee
It is easy to verify that the first two basis vectors, $\trafo{\omega_\parallel}{\phi}$ are parallel to $\Sigma$, while the last basis vector $\trafo{\omega_\nparallel}{\phi}$ has a very small projection onto the orthogonal complement of $\Sigma$,
\be
\Pob\trafo{\omega_\nparallel}{\phi}={1\over 115}\left(
\begin{array}{ccc}
 9 \\
 3  \\
 5  \\
\end{array}
\right) \,.
\ee
Note that the length of the shortest lattice vector of the lattice generated by $\Pob$ agrees with $1/\sqrt{\det(\mathbfcal Q^\top\mathbfcal Q)}=1/\sqrt{115}$, as expected. The inverse transformation is given by
\be
\trafo{\phi}{\omega}=\left(
\begin{array}{ccc}
 5 & 2 & 3 \\
 3 & 1 & 2 \\
 9 & 3 & 5 \\
\end{array}
\right)\,,~~\trafo{\phi}{\omega_\parallel}=\left(
\begin{array}{ccc}
 5 & 2 & 3 \\
 3 & 1 & 2 \\
\end{array}
\right)\,,~~\trafo{\phi}{\omega_\nparallel}=\left(
\begin{array}{ccc}
 9 & 3 & 5 \\
\end{array}
\right)\,.
\ee

We now can express the potential in terms of the aligned coordinates $\bs \omega_\parallel=\trafo{\phi}{\omega_\parallel}\bs\phi$, 
\be\label{omegapota}
V(\bs \omega_\parallel)=\Lambda^4 \left[3-\cos\left(2.04-\omega_\parallel^1-\omega_\parallel^2\right)-\cos\left(3\omega_\parallel^1-2\omega_\parallel^2+6.20\right)-\cos\left(-3\omega_\parallel^2 + 4.16\right)\right]\,.
\ee
Finally, we note that considering the shift $\bs\phi\rightarrow\bs\phi+2\pi \trafo{\omega_\nparallel}{\phi} n_\delta$ the constraint equation can be rewritten as
\be
\Pob (\bs \phi+2\pi \trafo{\omega_\nparallel}{\phi} n_\delta-\Pob \bs \delta)=0\,,
\ee
which allows to reduce the phase via (\ref{ndelta}), 
\be\label{ndeltaa}
\bs n_{\bs\delta} = \left[ {1\over 2\pi} \trafo{\phi}{\omega_{\nparallel}}  \Pob \bs \delta\right]_{\text{n.i.}}=21\,,
\ee
which gives the potential
\be\label{omegapotaa}
V(\bs \omega_\parallel) = \Lambda^4 \left[3-\cos\left(-\omega_\parallel^1-\omega_\parallel^2 -0.09 \right)-\cos\left(3\omega_\parallel^1-2\omega_\parallel^2-0.03\right)-\cos\left(-3\omega_\parallel^2 -0.05 \right)\right]\,.
\ee
As expected, the phases are significantly reduced by employing the approximate shift symmetry.

\section{Derivation of \eqref{minimapnplus1}}\label{311deriv}
In this appendix we derive \eqref{minimapnplus1}, starting from \eqref{quadraticconstrained}.
For the case of equal $\Lambda_I$, one has $\bs \Delta^\perp = \bs P^\perp = \mathbfcal R^\top \left( \mathbfcal R \mathbfcal R^\top \right)^{-1} \mathbfcal R$. Using \eqref{quadraticconstrained}, it is easy to see that energies at the minima can be written
\be\label{minimapnplus2}
V(\bs\phi_{n_\omega})\approx{2\pi^2}\Lambda^4 \left[ \trafo{\omega_\nparallel}{\phi}^\top\Pob \trafo{\omega_\nparallel}{\phi} \right] n_\omega^2\approx{2\pi^2}\Lambda^4 \left({c \, n_\omega\over \sqrt{\det\mathbfcal Q^\top \mathbfcal Q}}\right)^2\,,~~~\forall |m|<{\cal N}_{\text{vac}}\,,
\ee
where the positive integer $c$ is 
\be \label{cdef}
c\equiv \sqrt{ \det \left[ \left( \Pob \trafo{\omega_\nparallel}{\phi} \right)^\top \left( \Pob \trafo{\omega_\nparallel}{\phi} \right) \right] \det\mathbfcal Q^\top \mathbfcal Q} = \sqrt{\det\mathbfcal Q^\top \mathbfcal Q \over \det \trafo{\omega_{\parallel}}{\phi}^\top \trafo{\omega_{\parallel}}{\phi}} \,.
\ee
 To derive \eqref{cdef} we first used that $\trafo{\omega_\nparallel}{\phi}^\top\Pob \trafo{\omega_\nparallel}{\phi} = \trafo{\omega_\nparallel}{\phi}^\top \left( \Pob \right)^2 \trafo{\omega_\nparallel}{\phi} = $ \\ $\left( \Pob \trafo{\omega_\nparallel}{\phi} \right)^\top \left( \Pob \trafo{\omega_\nparallel}{\phi} \right)$ is a number, because $P - N = 1$, and thus equal to the determinant of the matrices that form it. Then we use the identity 
\begin{equation} \label{detequality}
	\det \left[ \left( \Pob \trafo{\omega_\nparallel}{\phi} \right)^\top \left( \Pob \trafo{\omega_\nparallel}{\phi} \right) \right] = \left( \det \trafo{\omega_{\parallel}}{\phi}^\top \trafo{\omega_{\parallel}}{\phi} \right)^{-1}.
\end{equation}
To see this note that
\begin{align}
\det \trafo{\omega}{\phi} &= 1 \notag \\
&= \det \left( \trafo{\omega_\parallel}{\phi} ~~ \trafo{\omega_\nparallel}{\phi} \right) \notag \\
&= \det \left( \trafo{\omega_\parallel}{\phi} ~~ \Pob \trafo{\omega_\nparallel}{\phi} \right) \equiv \det \bs T'_{\omega \phi} \,,
\end{align}
where we used the invariance of the determinant under adding linear combinations of some columns to other columns. Then, by computing $\det \left( \bs T'_{\omega \phi} \right)^\top \bs T'_{\omega \phi}$ and using that the columns of $\trafo{\omega_\parallel}{\phi}$ and $\Pob \trafo{\omega_\nparallel}{\phi}$ are orthogonal, one obtains \eqref{detequality}.

\section{Derivation of \eqref{D0bounds}} \label{D0boundsappendix}
To derive the bounds \eqref{D0bounds} on the diameter of the tile containing the origin, note first the general inequality for $P$-vectors $\bs v$ : $\lVert \bs v \rVert_2 / \sqrt{P} \leq \lVert \bs v \rVert_\infty \leq \lVert \bs v \rVert_2$, or
\begin{equation} \label{infinityvs2norm}
	\frac{1}{\lVert \bs v \rVert_2} \leq \frac{1}{\lVert \bs v \rVert_\infty} \leq \frac{\sqrt{P}}{\lVert \bs v \rVert_2} \,.
\end{equation}
The field range inside $\mathcal{T}_{\bs 0}$ along any specific direction is a lower bound for (half of) the diameter (cf. \eqref{D0alt}). This holds in particular for the field range along the eigenvector $\hat{\bs \Psi}_{|\bold Q|}$ corresponding to the smallest eigenvalue of $| \bold Q | \equiv \sqrt{ \bold Q^\top \bold Q }$. Combining this with the left-most inequality in \eqref{infinityvs2norm} we have therefore 
\begin{equation}
	\mathcal{D}_{\bs 0} \geq {2 \pi \over \lVert \bold Q \hat{\bs \Psi}_{|\bold Q|} \rVert_\infty} \geq {2 \pi \over \lVert \bold Q \hat{\bs \Psi}_{|\bold Q|} \rVert_2} = {2 \pi \over \lambda_\text{min}(|\bold Q|)} \,.\footnote{It is not hard to see that this lower bound can never be saturated.}
\end{equation}
On the other hand, each $P$-vector $\bold Q \hat{\bs \Theta}$ is subject to the right-most inequality in \eqref{infinityvs2norm}. Therefore the same inequality holds between the maxima of both sides over all $\hat{\bs \Theta} \in S^{N-1}$. Using this relation in the expression \eqref{D0alt} for the diameter, we arrive at
\begin{equation}
	\mathcal{D}_{\bs 0} \leq \max \left\{ {2 \pi \sqrt{P} \over \lVert \bold Q \hat{\bs \Theta} \rVert_2} ~ \Big| ~ \hat{\bs \Theta} \in S^{N-1} \right\} = {2 \pi \sqrt{P} \over \lambda_\text{min}(|\bold Q|)} \,,
\end{equation}
where we used $\lambda_\text{min}(|\bold Q|) = \lVert \bold Q \hat{\bs \Psi}_{|\bold Q|} \rVert_2 \leq \lVert \bold Q \hat{\bs \Theta} \rVert_2$ for all $\hat{\bs \Theta} \in S^{N-1}$.

\section{Axion potentials and Gaussian random fields}\label{gaussianfield}
In the body of this work we developed tools that allow for a systematic approach to general  (multi-)axion theories. This analytic approach is most powerful for well-aligned axion theories. Unfortunately, when the number $P$ of non-trivial  non-perturbative terms becomes very large, alignment typically fails and all approximate shift symmetries are broken. We now turn to a complimentary description of the axion potential, in terms of a Gaussian random field, that is valid precisely when the theory ceases to be well-aligned, and again allows for a simple statistical description of the theory. In particular, we find that at large $P\gg N$ the potential statistics are typically well approximated by an isotropic Gaussian random field with Gaussian covariance function, henceforth referred to by the shorthand {\it Gaussian field}. The statistical properties and the distribution of minima in Gaussian fields is very well understood, and efficient numerical algorithms exist to numerically sample such fields \cite{Bachlechner:2014rqa}. Therefore, by providing an effective description of the axion potential in terms of Gaussian fields a host of tools become available to study axion theories.

\subsection{Gaussian fields}
Before discussing the connection to axion potentials, let us review some of the basic properties of a stationary, isotropic Gaussian field $V_{{\mathcal G}}(\bs \chi)$ in $N$ dimensions $\chi^i$. We will assume a Gaussian covariance function for simplicity, although more general results exist. The ensemble is specified fully by the mean and the two-point function,
\bea
\langle V_{{\mathcal G}}(\bs \chi)\rangle &=&\bar{V}_{{\mathcal G}}\,,\\
\langle (V_{{\mathcal G}}(\bs \chi)-\bar{V}_{{\mathcal G}})(V_{{\mathcal G}}(\bs \chi^\prime)-\bar{V}_{{\mathcal G}})\rangle&=&\Lambda^8_{{\mathcal G}}~ e^{-{\left\lVert \bs \chi-\bs \chi^\prime\right\rVert_2^2 / 2\Delta_{{\mathcal G}}^2}}\,.
\eea
where the typical length scale over which the potential varies significantly is called $\Delta_{{\mathcal G}}$ and the overall scale is set by $\Lambda_{{\mathcal G}}$. $\bar{V}_{{\mathcal G}}$ denotes the mean of the random function. In order to understand the distribution of minima in the potential defined above, we will be interested in the correlations between the potential, its gradient and the Hessian matrix $H_{ij}= \partial_i\partial_j  V_{{\mathcal G}}$. The correlations of the Hessian are give by \cite{Bachlechner:2014rqa} (see also \cite{Easther:2016ire})
\be\label{grfhessiancorrelations}
\langle H_{ab}(\bs \chi)H_{cd}(\bs \chi) \rangle = \left(\delta_{ab}\delta_{cd}+\delta_{ad}\delta_{bc}+\delta_{ac}\delta_{bd}\right){\Lambda_{{\mathcal G}}^8 \over \Delta_{{\mathcal G}}^4}\,.
\ee
Let us consider the ensemble of points at which the random function takes on a particular value, $V$, and denote the corresponding ensemble average as $\langle\dots\rangle_{V}$. The only non-vanishing correlations between the field and its derivatives are given by
\bea
\langle H_{ab}(\bs \chi)\rangle_{V}&=&-{V-\bar{V}_{{\mathcal G}}\over \Delta_{{\mathcal G}}^2}\delta_{ab}\,,\\
\langle H_{ab}(\bs \chi)H_{cd}(\bs \chi)\rangle_{V}&=&\left({(V-\bar{V}_{{\mathcal G}})^2\over \Lambda_{{\mathcal G}}^8}\delta_{ab}\delta_{cd}+\delta_{ad}\delta_{bc}+\delta_{ac}\delta_{bd}\right){\Lambda_{{\mathcal G}}^8\over \Delta_{{\mathcal G}}^4}\,.
\eea
Note that crucially the gradient is uncorrelated with the potential and the Hessian. These correlations can be cast into a simple random matrix model,
\be\label{grfhess}
H=M-{V-\bar{V}_{{\mathcal G}}\over \Delta_{{\mathcal G}}^2}\mathbbold{1} \,,
\ee
where the matrix $M$ is a real, symmetric random matrix in the Gaussian orthogonal ensemble (GOE), i.e. it can be written as
\be
M={1\over \sqrt{2}}(A+A^\top)\,,~~A_{ij}\sim \mathcal{N} \left(0,\sigma_{M}\right)\,,
\ee
where $\sigma_{M}={ \Lambda_{{\mathcal G}}^4 / \Delta_{{\mathcal G}}^2}$. In the large $N$-limit the eigenvalue spectrum of GOE matrices is given by the famous Wigner semicircle,
\be
\rho(\Lambda)={1\over 2\pi N\sigma_M^2}\sqrt{4 N\sigma_M^2-\lambda^2}\,.
\ee
By considering the case where the smallest eigenvalue is no longer negative, $\lambda_{\text{min}}>0$, we can therefore easily solve for the mean value of stable minima in the large $N$-limit. We find
\be
\langle V\rangle_{\text{minima}}\approx \bar{V}_{{\mathcal G}}- 2\sqrt{N} \Lambda_{{\mathcal G}}^4 \,.
\ee
The probability distribution function of energies at minima with barely positive definite Hessian matrix is simply obtained by considering the smallest eigenvalue of the Hessian in (\ref{grfhess}) and solving for $V$: it is approximated by the convolution of the Tracy-Widom distribution with a normal distribution.

\subsection{Axion theories at large $P$}
In the previous section we reviewed the statistical properties of Gaussian random fields with Gaussian covariance function. We are now in a position to consider the statistics of the axion potential in the large $N$ and $P\gg N$ limit, and compare those results to a Gaussian random field.

The non-perturbative potential for the axions with unbroken discrete shift symmetry is given in (\ref{potentialnophases}),
\be\label{nppotentialapp}
V=\sum_{I=1}^P \Lambda_I^4  \left[1-\cos\left(\mathbfcal Q \bs\theta \right)^I \right]\,,
\ee
where as above ${\mathbfcal Q}$ is a $P\times N$ integer charge matrix. In the following we will assume that the couplings $\Lambda_I^4$ are of similar magnitude and independent of the charges $\mathbfcal Q^I$. In the large $P$-limit the potential approaches a Gaussian random field. However, the potential (\ref{nppotentialapp}) clearly is not isotropic (it is periodic under $v_{\parallel}^i \rightarrow v_{\parallel}^i + 2\pi$ only for some directions $\bs v_\parallel$), nor does it have a Gaussian covariance function. Curiously, however, when sampling over random one-dimensional slices through the $N$-dimensional potential, the mean power spectrum is very well approximated by a Gaussian. It is therefore not very surprising that there are some similarities between the distribution of minima in Gaussian random landscapes and random axion landscapes, as we make precise below.

When sampling over ensembles of potentials defined by random $\mathbfcal Q$, containing i.i.d. random integers distributed uniformly in the interval $[-s,s]$ and random phases, the mean $\bar{V}_{\text{np}}$ and variance $\Lambda^8_{\text{np}}$ of the potential $V$  respectively are given by
\bea
\bar{V}_{\text{np}}&\equiv&\langle V(\bs \theta)\rangle=\sum_{I=1}^P\Lambda_I^4 \,,\nonumber\\
\Lambda^8_{\text{np}}&\equiv&\langle [V(\bs \theta)-\langle V(\bs \theta)\rangle]^2\rangle={1\over 2}\sum_{I=1}^P\Lambda_I^8\,.
\eea
Just as in the case of a Gaussian field, the gradient is not correlated with the potential, or the Hessian. The correlations of the Hessian $H_{ab}\equiv \partial_a\partial_b V(\bs\theta)$ of the non-perturbative potential are given by
\be\label{hessnpcorrelations}
\langle H_{ab}(\bs \theta)H_{cd}(\bs \theta)\rangle=\left(\delta_{ab}\delta_{cd}+\delta_{ad}\delta_{bc}+\delta_{ac}\delta_{bd}-{6\over 5}\delta_{ac}\delta_{bc}\delta_{cd}\right){\Lambda_{\text{np}}^8 \over \Delta_{\text{np}}^4}\,,
\ee
where we defined an effective correlation length of the axion potential,
\be
\Delta_{\text{np}}={\sqrt{3}\over s}\,.
\ee
The correlations (\ref{hessnpcorrelations}) are very similar to the Hessian correlations of a Gaussian fields in (\ref{grfhessiancorrelations}), and only deviate for the diagonal elements of the Hessian. Furthermore, we have for the correlation between the Hessian and the potential
\be
\langle H_{ab}(\bs \theta)\rangle_{V}=-{V-\bar{V}_{\text{np}}\over \Delta_{\text{np}}^2}\delta_{ab}\,.
\ee
\begin{figure}
  \centering
  \includegraphics[width=1\textwidth]{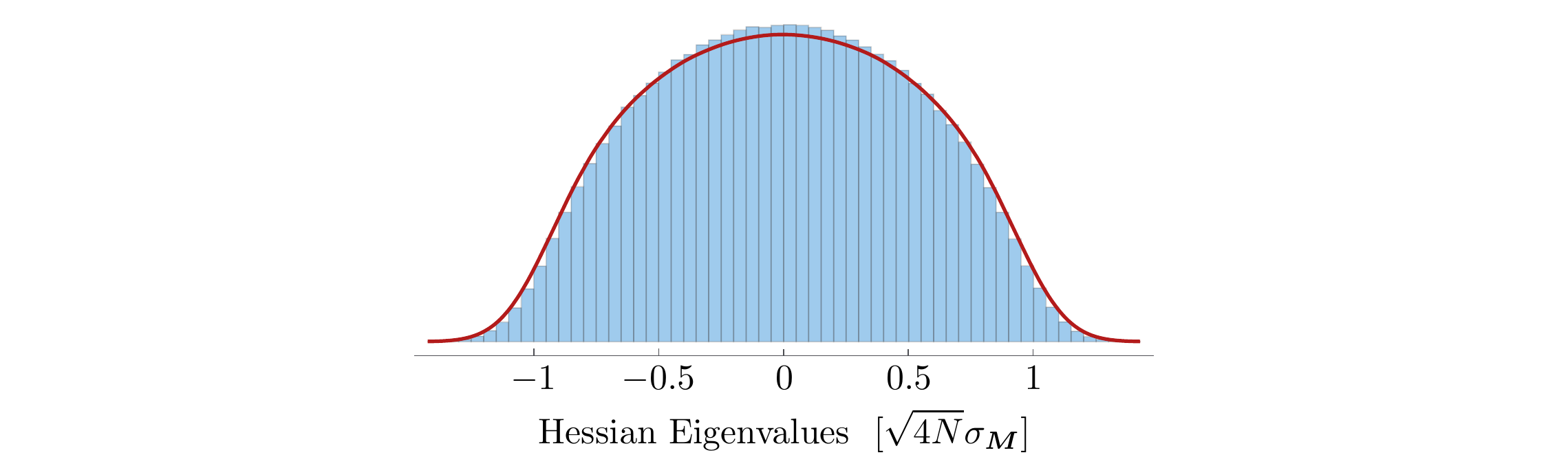}
  \caption{\small Normalized eigenvalue spectrum of Hessian of potential along with random matrix model for $N=20$, $P=2000$.}\label{rmtcompare} 
\end{figure}
We can therefore propose an approximate random matrix model for the Hessian, that reproduces the correct correlations, except for the variance of the diagonal terms,
\be\label{nphess}
{\bs H}\approx {\bs M}-{V-\bar{V}_{\text{np}}\over \Delta_{\text{np}}^2}\mathbbold 1\,,
\ee
where ${\bs M}$ is a GOE matrix with standard deviation $\sigma_{{\bs M}} = \Lambda_{\text{np}}^4 /\Delta_{\text{np}}^2$. We display the eigenvalue spectrum of the full Hessian matrix along with this simple random matrix model in the left part of Figure \ref{rmtcompare}. Using the main result of \cite{tracy1996}, this implies that in the large $N$-limit we expect most minima at
\be
\langle V\rangle_{\text{minima}}\approx V - 2\sqrt{N} \Lambda_{\text{np}}^4 \left( 1 - \frac{0.6}{N^{2/3}} \right) = \bar{V}_{\text{np}} - \sqrt{2N} \langle\Lambda_I^4 \rangle_{\text{r.m.s}} \left( 1 - \frac{0.6}{N^{2/3}} \right)\,,
\ee
so that the leading term behaves just like in the case of Gaussian fields. It is extremely hard to accurately sample the distribution of minima of the axion potential at large $P$. However, we can obtain a (not necessarily representative) sample of minima by numerically solving for local minima. We find a good agreement between the numerical results and the random matrix theory expectation.

\bibliographystyle{klebphys2.bst}
\bibliography{tunnelingrefs}

\providecommand{\href}[2]{#2}\begingroup\raggedright\begin{thebibliography}{10}

\bibitem{Peccei:1977hh}
{\sc R.~D. Peccei} and {\sc H.~R. Quinn}, ``{CP Conservation in the Presence of
  Instantons},''
\href{http://dx.doi.org/10.1103/PhysRevLett.38.1440}{{\em Phys. Rev. Lett.}
  {\bfseries 38} (1977) 1440--1443}.

\bibitem{Wen:1985jz}
{\sc X.~G. Wen} and {\sc E.~Witten}, ``{World Sheet Instantons and the
  {Peccei-Quinn} Symmetry},''
\href{http://dx.doi.org/10.1016/0370-2693(86)91587-X}{{\em Phys. Lett.}
  {\bfseries 166B} (1986) 397--401}.

\bibitem{Dine:1986vd}
{\sc M.~Dine} and {\sc N.~Seiberg}, ``{Nonrenormalization Theorems in
  Superstring Theory},''
\href{http://dx.doi.org/10.1103/PhysRevLett.57.2625}{{\em Phys. Rev. Lett.}
  {\bfseries 57} (1986) 2625}.

\bibitem{Dine:1986zy}
{\sc M.~Dine}, {\sc N.~Seiberg}, {\sc X.~G. Wen}, and {\sc E.~Witten},
  ``{Nonperturbative Effects on the String World Sheet},''
\href{http://dx.doi.org/10.1016/0550-3213(86)90418-9}{{\em Nucl. Phys.}
  {\bfseries B278} (1986) 769--789}.

\bibitem{Denef:2004cf}
{\sc F.~Denef} and {\sc M.~R. Douglas}, ``{Distributions of nonsupersymmetric
  flux vacua},'' \href{http://dx.doi.org/10.1088/1126-6708/2005/03/061}{{\em
  JHEP} {\bfseries 0503} (2005) 061},
\href{http://arxiv.org/abs/hep-th/0411183}{{\ttfamily arXiv:hep-th/0411183
  [hep-th]}}.

\bibitem{Svrcek:2006yi}
{\sc P.~Svrcek} and {\sc E.~Witten}, ``{Axions In String Theory},''
  \href{http://dx.doi.org/10.1088/1126-6708/2006/06/051}{{\em JHEP} {\bfseries
  0606} (2006) 051},
\href{http://arxiv.org/abs/hep-th/0605206}{{\ttfamily arXiv:hep-th/0605206
  [hep-th]}}.

\bibitem{Denef:2007pq}
{\sc F.~Denef}, {\sc M.~R. Douglas}, and {\sc S.~Kachru}, ``{Physics of String
  Flux Compactifications},''
  \href{http://dx.doi.org/10.1146/annurev.nucl.57.090506.123042}{{\em Ann. Rev.
  Nucl. Part. Sci.} {\bfseries 57} (2007) 119--144},
\href{http://arxiv.org/abs/hep-th/0701050}{{\ttfamily arXiv:hep-th/0701050
  [hep-th]}}.

\bibitem{Denef:2008wq}
{\sc F.~Denef}, ``{Les Houches Lectures on Constructing String Vacua},'' in
  {\em {String theory and the real world: From particle physics to
  astrophysics. Proceedings, Summer School in Theoretical Physics, 87th
  Session, Les Houches, France, July 2-27, 2007}}, pp.~483--610.
\newblock 2008.
\newblock \href{http://arxiv.org/abs/0803.1194}{{\ttfamily arXiv:0803.1194
  [hep-th]}}.
\newblock
\url{https://inspirehep.net/record/780946/files/arXiv:0803.1194.pdf}.
\newblock

\bibitem{Baumann:2014nda}
{\sc D.~Baumann} and {\sc L.~McAllister}, {\em {Inflation and String Theory}}.
\newblock Cambridge University Press, 2015.
\newblock \href{http://arxiv.org/abs/1404.2601}{{\ttfamily arXiv:1404.2601
  [hep-th]}}.
\newblock
\url{https://inspirehep.net/record/1289899/files/arXiv:1404.2601.pdf}.
\newblock

\bibitem{Long:2016jvd}
{\sc C.~Long}, {\sc L.~McAllister}, and {\sc J.~Stout}, ``{Systematics of Axion
  Inflation in Calabi-Yau Hypersurfaces},''
  \href{http://dx.doi.org/10.1007/JHEP02(2017)014}{{\em JHEP} {\bfseries 02}
  (2017) 014},
\href{http://arxiv.org/abs/1603.01259}{{\ttfamily arXiv:1603.01259 [hep-th]}}.

\bibitem{Halverson:2017deq}
{\sc J.~Halverson}, {\sc C.~Long}, and {\sc P.~Nath}, ``{An Ultralight Axion in
  Supersymmetry and Strings and Cosmology at Small Scales},''
\href{http://arxiv.org/abs/1703.07779}{{\ttfamily arXiv:1703.07779 [hep-ph]}}.

\bibitem{Halverson:2017ffz}
{\sc J.~Halverson}, {\sc C.~Long}, and {\sc B.~Sung}, ``{On Algorithmic
  Universality in F-theory Compactifications},''
\href{http://arxiv.org/abs/1706.02299}{{\ttfamily arXiv:1706.02299 [hep-th]}}.

\bibitem{Hu:2000ke}
{\sc W.~Hu}, {\sc R.~Barkana}, and {\sc A.~Gruzinov}, ``{Cold and fuzzy dark
  matter},'' \href{http://dx.doi.org/10.1103/PhysRevLett.85.1158}{{\em Phys.
  Rev. Lett.} {\bfseries 85} (2000) 1158--1161},
\href{http://arxiv.org/abs/astro-ph/0003365}{{\ttfamily arXiv:astro-ph/0003365
  [astro-ph]}}.

\bibitem{Arvanitaki:2009fg}
{\sc A.~Arvanitaki}, {\sc S.~Dimopoulos}, {\sc S.~Dubovsky}, {\sc N.~Kaloper},
  and {\sc J.~March-Russell}, ``{String Axiverse},''
  \href{http://dx.doi.org/10.1103/PhysRevD.81.123530}{{\em Phys. Rev.}
  {\bfseries D81} (2010) 123530},
\href{http://arxiv.org/abs/0905.4720}{{\ttfamily arXiv:0905.4720 [hep-th]}}.

\bibitem{Hui:2016ltb}
{\sc L.~Hui}, {\sc J.~P. Ostriker}, {\sc S.~Tremaine}, and {\sc E.~Witten},
  ``{On the hypothesis that cosmological dark matter is composed of ultra-light
  bosons},''
\href{http://arxiv.org/abs/1610.08297}{{\ttfamily arXiv:1610.08297
  [astro-ph.CO]}}.

\bibitem{Marsh:2015xka}
{\sc D.~J.~E. Marsh}, ``{Axion Cosmology},''
  \href{http://dx.doi.org/10.1016/j.physrep.2016.06.005}{{\em Phys. Rept.}
  {\bfseries 643} (2016) 1--79},
\href{http://arxiv.org/abs/1510.07633}{{\ttfamily arXiv:1510.07633
  [astro-ph.CO]}}.

\bibitem{Diez-Tejedor:2017ivd}
{\sc A.~Diez-Tejedor} and {\sc D.~J.~E. Marsh}, ``{Cosmological production of
  ultralight dark matter axions},''
\href{http://arxiv.org/abs/1702.02116}{{\ttfamily arXiv:1702.02116 [hep-ph]}}.

\bibitem{Natural}
{\sc K.~Freese}, {\sc J.~A. Frieman}, and {\sc A.~V. Olinto}, ``{Natural
  inflation with pseudo - Nambu-Goldstone bosons},''
\href{http://dx.doi.org/10.1103/PhysRevLett.65.3233}{{\em Phys.Rev.Lett.}
  {\bfseries 65} (1990) 3233--3236}.

\bibitem{Dimopoulos:2005ac}
{\sc S.~Dimopoulos}, {\sc S.~Kachru}, {\sc J.~McGreevy}, and {\sc J.~G.
  Wacker}, ``{N-flation},''
  \href{http://dx.doi.org/10.1088/1475-7516/2008/08/003}{{\em JCAP} {\bfseries
  0808} (2008) 003},
\href{http://arxiv.org/abs/hep-th/0507205}{{\ttfamily arXiv:hep-th/0507205
  [hep-th]}}.

\bibitem{Kim:2004rp}
{\sc J.~E. Kim}, {\sc H.~P. Nilles}, and {\sc M.~Peloso}, ``{Completing natural
  inflation},'' \href{http://dx.doi.org/10.1088/1475-7516/2005/01/005}{{\em
  JCAP} {\bfseries 0501} (2005) 005},
\href{http://arxiv.org/abs/hep-ph/0409138}{{\ttfamily arXiv:hep-ph/0409138
  [hep-ph]}}.

\bibitem{SW}
{\sc E.~Silverstein} and {\sc A.~Westphal}, ``{Monodromy in the CMB: Gravity
  Waves and String Inflation},''
  \href{http://dx.doi.org/10.1103/PhysRevD.78.106003}{{\em Phys.Rev.}
  {\bfseries D78} (2008) 106003},
\href{http://arxiv.org/abs/0803.3085}{{\ttfamily arXiv:0803.3085 [hep-th]}}.

\bibitem{McAllister:2008hb}
{\sc L.~McAllister}, {\sc E.~Silverstein}, and {\sc A.~Westphal}, ``{Gravity
  Waves and Linear Inflation from Axion Monodromy},''
  \href{http://dx.doi.org/10.1103/PhysRevD.82.046003}{{\em Phys. Rev.}
  {\bfseries D82} (2010) 046003},
\href{http://arxiv.org/abs/0808.0706}{{\ttfamily arXiv:0808.0706 [hep-th]}}.

\bibitem{Kaloper:2008fb}
{\sc N.~Kaloper} and {\sc L.~Sorbo}, ``{A Natural Framework for Chaotic
  Inflation},'' \href{http://dx.doi.org/10.1103/PhysRevLett.102.121301}{{\em
  Phys. Rev. Lett.} {\bfseries 102} (2009) 121301},
\href{http://arxiv.org/abs/0811.1989}{{\ttfamily arXiv:0811.1989 [hep-th]}}.

\bibitem{Kaloper:2011jz}
{\sc N.~Kaloper}, {\sc A.~Lawrence}, and {\sc L.~Sorbo}, ``{An Ignoble Approach
  to Large Field Inflation},''
  \href{http://dx.doi.org/10.1088/1475-7516/2011/03/023}{{\em JCAP} {\bfseries
  1103} (2011) 023},
\href{http://arxiv.org/abs/1101.0026}{{\ttfamily arXiv:1101.0026 [hep-th]}}.

\bibitem{Bousso:2000xa}
{\sc R.~Bousso} and {\sc J.~Polchinski}, ``{Quantization of four form fluxes
  and dynamical neutralization of the cosmological constant},'' {\em JHEP}
  {\bfseries 0006} (2000) 006,
\href{http://arxiv.org/abs/hep-th/0004134}{{\ttfamily arXiv:hep-th/0004134
  [hep-th]}}.

\bibitem{Bachlechner:2015gwa}
{\sc T.~C. Bachlechner}, ``{Axionic Band Structure of the Cosmological
  Constant},'' \href{http://dx.doi.org/10.1103/PhysRevD.93.023522}{{\em Phys.
  Rev.} {\bfseries D93} no.~2, (2016) 023522},
\href{http://arxiv.org/abs/1510.06388}{{\ttfamily arXiv:1510.06388 [hep-th]}}.

\bibitem{bejk2}
{\sc T.~Bachlechner}, {\sc O.~Janssen}, {\sc K.~Eckerle}, and {\sc M.~Kleban}
  {\em , to appear} .

\bibitem{Bachlechner:2017zpb}
{\sc T.~C. Bachlechner}, {\sc K.~Eckerle}, {\sc O.~Janssen}, and {\sc
  M.~Kleban}, ``{Axions of Evil},''
\href{http://arxiv.org/abs/1703.00453}{{\ttfamily arXiv:1703.00453 [hep-th]}}.

\bibitem{1981RSPSA.377..147D}
{\sc P.~C.~W. {Davies}} and {\sc S.~D. {Unwin}}, ``{Why is the cosmological
  constant so small},'' \href{http://dx.doi.org/10.1098/rspa.1981.0119}{{\em
  Proceedings of the Royal Society of London Series A} {\bfseries 377} (June,
  1981) 147--149}.

\bibitem{Sakharov:1984ir}
{\sc A.~D. Sakharov}, ``{Cosmological Transitions With a Change in Metric
  Signature},'' {\em Sov. Phys. JETP} {\bfseries 60} (1984) 214--218.
[Sov. Phys. Usp.34,409(1991)].

\bibitem{Banks:1984cw}
{\sc T.~Banks}, ``{T C P, Quantum Gravity, the Cosmological Constant and All
  That...},''
\href{http://dx.doi.org/10.1016/0550-3213(85)90020-3}{{\em Nucl. Phys.}
  {\bfseries B249} (1985) 332--360}.

\bibitem{Linde:1986dq}
{\sc A.~D. Linde}, ``{Inflation and Quantum Cosmology},''
{\em Submitted to: Newton Centenary Volume (Cambridge U. Press)} (1986) .

\bibitem{Weinberg:1987dv}
{\sc S.~Weinberg}, ``{Anthropic Bound on the Cosmological Constant},''
\href{http://dx.doi.org/10.1103/PhysRevLett.59.2607}{{\em Phys.Rev.Lett.}
  {\bfseries 59} (1987) 2607}.

\bibitem{Linde:2015edk}
{\sc A.~Linde}, ``{A brief history of the multiverse},''
  \href{http://dx.doi.org/10.1088/1361-6633/aa50e4}{{\em Rept. Prog. Phys.}
  {\bfseries 80} no.~2, (2017) 022001},
\href{http://arxiv.org/abs/1512.01203}{{\ttfamily arXiv:1512.01203 [hep-th]}}.

\bibitem{Bachlechner:2014rqa}
{\sc T.~C. Bachlechner}, ``{On Gaussian Random Supergravity},''
  \href{http://dx.doi.org/10.1007/JHEP04(2014)054}{{\em JHEP} {\bfseries 1404}
  (2014) 054},
\href{http://arxiv.org/abs/1401.6187}{{\ttfamily arXiv:1401.6187 [hep-th]}}.

\bibitem{mathematicademo}
{\sc T.~C. Bachlechner}, {\sc K.~Eckerle}, {\sc O.~Janssen}, and {\sc
  M.~Kleban}, ``{A Systematic Framework for Axion Theories}.''
  \url{http://cosmo.nyu.edu/kleban/}, 2017

\bibitem{Vafa:2005ui}
{\sc C.~Vafa}, ``{The string landscape and the swampland},''
\href{http://arxiv.org/abs/hep-th/0509212}{{\ttfamily arXiv:hep-th/0509212
  [hep-th]}}.

\bibitem{ArkaniHamed:2006dz}
{\sc N.~Arkani-Hamed}, {\sc L.~Motl}, {\sc A.~Nicolis}, and {\sc C.~Vafa},
  ``{The string landscape, black holes and gravity as the weakest force},''
  \href{http://dx.doi.org/10.1088/1126-6708/2007/06/060}{{\em JHEP} {\bfseries
  06} (2007) 060},
\href{http://arxiv.org/abs/hep-th/0601001}{{\ttfamily arXiv:hep-th/0601001
  [hep-th]}}.

\bibitem{Rudelius:2014wla}
{\sc T.~Rudelius}, ``{On the Possibility of Large Axion Moduli Spaces},''
  \href{http://dx.doi.org/10.1088/1475-7516/2015/04/049}{{\em JCAP} {\bfseries
  1504} no.~04, (2015) 049},
\href{http://arxiv.org/abs/1409.5793}{{\ttfamily arXiv:1409.5793 [hep-th]}}.

\bibitem{Cheung:2014vva}
{\sc C.~Cheung} and {\sc G.~N. Remmen}, ``{Naturalness and the Weak Gravity
  Conjecture},'' \href{http://dx.doi.org/10.1103/PhysRevLett.113.051601}{{\em
  Phys. Rev. Lett.} {\bfseries 113} (2014) 051601},
\href{http://arxiv.org/abs/1402.2287}{{\ttfamily arXiv:1402.2287 [hep-ph]}}.

\bibitem{Heidenreich:2015wga}
{\sc B.~Heidenreich}, {\sc M.~Reece}, and {\sc T.~Rudelius}, ``{Weak Gravity
  Strongly Constrains Large-Field Axion Inflation},''
  \href{http://dx.doi.org/10.1007/JHEP12(2015)108}{{\em JHEP} {\bfseries 12}
  (2015) 108},
\href{http://arxiv.org/abs/1506.03447}{{\ttfamily arXiv:1506.03447 [hep-th]}}.

\bibitem{Bachlechner:2015qja}
{\sc T.~C. Bachlechner}, {\sc C.~Long}, and {\sc L.~McAllister}, ``{Planckian
  Axions and the Weak Gravity Conjecture},''
  \href{http://dx.doi.org/10.1007/JHEP01(2016)091}{{\em JHEP} {\bfseries 01}
  (2016) 091},
\href{http://arxiv.org/abs/1503.07853}{{\ttfamily arXiv:1503.07853 [hep-th]}}.

\bibitem{Rudelius:2015xta}
{\sc T.~Rudelius}, ``{Constraints on Axion Inflation from the Weak Gravity
  Conjecture},'' \href{http://dx.doi.org/10.1088/1475-7516/2015/09/020,
  10.1088/1475-7516/2015/9/020}{{\em JCAP} {\bfseries 1509} no.~09, (2015)
  020},
\href{http://arxiv.org/abs/1503.00795}{{\ttfamily arXiv:1503.00795 [hep-th]}}.

\bibitem{Hebecker:2015rya}
{\sc A.~Hebecker}, {\sc P.~Mangat}, {\sc F.~Rompineve}, and {\sc L.~T.
  Witkowski}, ``{Winding out of the Swamp: Evading the Weak Gravity Conjecture
  with F-term Winding Inflation?},''
  \href{http://dx.doi.org/10.1016/j.physletb.2015.07.026}{{\em Phys. Lett.}
  {\bfseries B748} (2015) 455--462},
\href{http://arxiv.org/abs/1503.07912}{{\ttfamily arXiv:1503.07912 [hep-th]}}.

\bibitem{Ibanez:2015fcv}
{\sc L.~E. Ibanez}, {\sc M.~Montero}, {\sc A.~Uranga}, and {\sc I.~Valenzuela},
  ``{Relaxion Monodromy and the Weak Gravity Conjecture},''
  \href{http://dx.doi.org/10.1007/JHEP04(2016)020}{{\em JHEP} {\bfseries 04}
  (2016) 020},
\href{http://arxiv.org/abs/1512.00025}{{\ttfamily arXiv:1512.00025 [hep-th]}}.

\bibitem{Heidenreich:2016aqi}
{\sc B.~Heidenreich}, {\sc M.~Reece}, and {\sc T.~Rudelius}, ``{Evidence for a
  Lattice Weak Gravity Conjecture},''
\href{http://arxiv.org/abs/1606.08437}{{\ttfamily arXiv:1606.08437 [hep-th]}}.

\bibitem{Bachlechner:2014hsa}
{\sc T.~C. Bachlechner}, {\sc M.~Dias}, {\sc J.~Frazer}, and {\sc
  L.~McAllister}, ``{A New Angle on Chaotic Inflation},''
\href{http://arxiv.org/abs/1404.7496}{{\ttfamily arXiv:1404.7496 [hep-th]}}.

\bibitem{Choi:2014rja}
{\sc K.~Choi}, {\sc H.~Kim}, and {\sc S.~Yun}, ``{Natural inflation with
  multiple sub-Planckian axions},''
  \href{http://dx.doi.org/10.1103/PhysRevD.90.023545}{{\em Phys.Rev.}
  {\bfseries D90} (2014) 023545},
\href{http://arxiv.org/abs/1404.6209}{{\ttfamily arXiv:1404.6209 [hep-th]}}.

\bibitem{Bachlechner:2014gfa}
{\sc T.~C. Bachlechner}, {\sc C.~Long}, and {\sc L.~McAllister}, ``{Planckian
  Axions in String Theory},''
  \href{http://dx.doi.org/10.1007/JHEP12(2015)042}{{\em JHEP} {\bfseries 12}
  (2015) 042},
\href{http://arxiv.org/abs/1412.1093}{{\ttfamily arXiv:1412.1093 [hep-th]}}.

\bibitem{HT1}
{\sc T.~Higaki} and {\sc F.~Takahashi}, ``{Natural and Multi-Natural Inflation
  in Axion Landscape},'' \href{http://dx.doi.org/10.1007/JHEP07(2014)074}{{\em
  JHEP} {\bfseries 1407} (2014) 074},
\href{http://arxiv.org/abs/1404.6923}{{\ttfamily arXiv:1404.6923 [hep-th]}}.

\bibitem{HT2}
{\sc T.~Higaki} and {\sc F.~Takahashi}, ``{Axion Landscape and Natural
  Inflation},''
\href{http://arxiv.org/abs/1409.8409}{{\ttfamily arXiv:1409.8409 [hep-ph]}}.

\bibitem{Shiu:2015uva}
{\sc G.~Shiu}, {\sc W.~Staessens}, and {\sc F.~Ye}, ``{Widening the Axion
  Window via Kinetic and Stueckelberg Mixings},''
  \href{http://dx.doi.org/10.1103/PhysRevLett.115.181601}{{\em Phys. Rev.
  Lett.} {\bfseries 115} (2015) 181601},
\href{http://arxiv.org/abs/1503.01015}{{\ttfamily arXiv:1503.01015 [hep-th]}}.

\bibitem{Bao:2017thx}
{\sc N.~Bao}, {\sc R.~Bousso}, {\sc S.~Jordan}, and {\sc B.~Lackey}, ``{Fast
  optimization algorithms and the cosmological constant},''
\href{http://arxiv.org/abs/1706.08503}{{\ttfamily arXiv:1706.08503 [hep-th]}}.

\bibitem{lllpackage}
{\sc W.~van~der Kallen}, ``{Implementations of extended LLL}.''
  \url{http://www.staff.science.uu.nl/~kalle101/lllimplementations.html}, 1998

\bibitem{minkowski1911}
{\sc H.~Minkowski}, {\em Gesammelte Abhandlungen von Hermann Minkowski: Vol.:
  2}.
\newblock B.G. Teubner, 1911.
\newblock \url{http://books.google.com/books?id=yRiiQwAACAAJ}.

\bibitem{Lenstra1982}
{\sc A.~Lenstra}, {\sc H.~Lenstra}, and {\sc L.~Lov\'{a}sz}, ``Factoring
  polynomials with rational coefficients,''
  \href{http://dx.doi.org/10.1007/BF01457454}{{\em Mathematische Annalen}
  {\bfseries 261} no.~4, (1982) 515--534}.
  \url{http://dx.doi.org/10.1007/BF01457454}.

\bibitem{Nguyen2004}
{\sc P.~Q. Nguyen} and {\sc D.~Stehl{\'e}}, {\em Low-Dimensional Lattice Basis
  Reduction Revisited},
  \href{http://dx.doi.org/10.1007/978-3-540-24847-7_26}{pp.~338--357}.
\newblock Springer Berlin Heidelberg, Berlin, Heidelberg, 2004.
\newblock \url{https://doi.org/10.1007/978-3-540-24847-7_26}.

\bibitem{Gama2006}
{\sc N.~Gama}, {\sc N.~Howgrave-Graham}, {\sc H.~Koy}, and {\sc P.~Q. Nguyen},
  {\em Rankin's Constant and Blockwise Lattice Reduction},
  \href{http://dx.doi.org/10.1007/11818175_7}{pp.~112--130}.
\newblock Springer Berlin Heidelberg, Berlin, Heidelberg, 2006.
\newblock \url{https://doi.org/10.1007/11818175_7}.

\bibitem{Chen2011}
{\sc Y.~Chen} and {\sc P.~Q. Nguyen}, {\em BKZ 2.0: Better Lattice Security
  Estimates}, \href{http://dx.doi.org/10.1007/978-3-642-25385-0_1}{pp.~1--20}.
\newblock Springer Berlin Heidelberg, Berlin, Heidelberg, 2011.
\newblock \url{https://doi.org/10.1007/978-3-642-25385-0_1}.

\bibitem{Dadush:2013:ADS:2627817.2627896}
{\sc D.~Dadush} and {\sc D.~Micciancio}, ``Algorithms for the densest
  sub-lattice problem,'' in {\em Proceedings of the Twenty-fourth Annual
  ACM-SIAM Symposium on Discrete Algorithms}, SODA '13, pp.~1103--1122.
\newblock Society for Industrial and Applied Mathematics, Philadelphia, PA,
  USA, 2013.
\newblock \url{http://dl.acm.org/citation.cfm?id=2627817.2627896}.

\bibitem{Li:2014:ADS}
{\sc J.~Li} and {\sc P.~Q. Nguyen}, ``Approximating the densest sublattice from
  {Rankin}'s inequality,''
  \href{http://dx.doi.org/http://dx.doi.org/10.1112/S1461157014000333}{
  {\bfseries 17} no.~A, (2014) 92--111}.

\bibitem{KreinMilman}
{\sc M.~Krein} and {\sc D.~Milman}, ``On extreme points of regular convex
  sets,'' {\em Studia Math.} {\bfseries 9} (1940) 133--138.

\bibitem{CIS-115135}
{\sc L.~Lovasz}, {\sc L.~LovÃ¡sz}, and {\sc M.~Simonovits}, ``Random walks in
  a convex body and an improved volume algorithm,'' {\em Random Structures \&
  Algorithms} {\bfseries 4} no.~4, (1993) 359--412.

\bibitem{Cousins:2015:BKG:2746539.2746563}
{\sc B.~Cousins} and {\sc S.~Vempala},
  \href{http://dx.doi.org/10.1145/2746539.2746563}{``Bypassing kls: Gaussian
  cooling and an o*(n3) volume algorithm,''} in {\em Proceedings of the
  Forty-seventh Annual ACM Symposium on Theory of Computing}, STOC '15,
  pp.~539--548.
\newblock ACM, New York, NY, USA, 2015.
\newblock \url{http://doi.acm.org/10.1145/2746539.2746563}.

\bibitem{matlabsampling}
{\sc B.~Cousins}, ``Volume computation of convex bodies,'' June, 2015.
\newblock
  \url{https://www.mathworks.com/matlabcentral/fileexchange/43596-volume-computation-of-convex-bodies?requestedDomain=www.mathworks.com}

\bibitem{goodman1963}
{\sc N.~R. Goodman}, ``{The Distribution of the Determinant of a Complex
  Wishart Distributed Matrix},''
  \href{http://dx.doi.org/10.1214/aoms/1177704251}{{\em Ann. Math. Statist.}
  {\bfseries 34} no.~1, (03, 1963) 178--180}.
  \url{http://dx.doi.org/10.1214/aoms/1177704251}.

\bibitem{MP1967}
{\sc V.~A. Mar\u{c}enko} and {\sc L.~A. Pastur}, ``Distribution of eigenvalues
  for some sets of random matrices,'' {\em Mathematics of the USSR-Sbornik}
  {\bfseries 1} no.~4, (1967) 457.
  \url{http://stacks.iop.org/0025-5734/1/i=4/a=A01}.

\bibitem{Long:2014dta}
{\sc C.~Long}, {\sc L.~McAllister}, and {\sc P.~McGuirk}, ``{Aligned Natural
  Inflation in String Theory},''
  \href{http://dx.doi.org/10.1103/PhysRevD.90.023501}{{\em Phys.Rev.}
  {\bfseries D90} (2014) 023501},
\href{http://arxiv.org/abs/1404.7852}{{\ttfamily arXiv:1404.7852 [hep-th]}}.

\bibitem{Czerny:2014xja}
{\sc M.~Czerny}, {\sc T.~Higaki}, and {\sc F.~Takahashi}, ``{Multi-Natural
  Inflation in Supergravity},''
  \href{http://dx.doi.org/10.1007/JHEP05(2014)144}{{\em JHEP} {\bfseries 1405}
  (2014) 144},
\href{http://arxiv.org/abs/1403.0410}{{\ttfamily arXiv:1403.0410 [hep-ph]}}.

\bibitem{Tye:2014tja}
{\sc S.~H.~H. Tye} and {\sc S.~S.~C. Wong}, ``{Helical Inflation and Cosmic
  Strings},''
\href{http://arxiv.org/abs/1404.6988}{{\ttfamily arXiv:1404.6988
  [astro-ph.CO]}}.

\bibitem{Ali:2014mra}
{\sc T.~Ali}, {\sc S.~S. Haque}, and {\sc V.~Jejjala}, ``{Natural Inflation
  from Near Alignment in Heterotic String Theory},''
\href{http://arxiv.org/abs/1410.4660}{{\ttfamily arXiv:1410.4660 [hep-th]}}.

\bibitem{Burgess:2014oma}
{\sc C.~Burgess} and {\sc D.~Roest}, ``{Inflation by Alignment},''
  \href{http://dx.doi.org/10.1088/1475-7516/2015/06/012}{{\em JCAP} {\bfseries
  1506} no.~06, (2015) 012},
\href{http://arxiv.org/abs/1412.1614}{{\ttfamily arXiv:1412.1614 [hep-th]}}.

\bibitem{Madison}
{\sc J.~Brown}, {\sc W.~Cottrell}, {\sc G.~Shiu}, and {\sc P.~Soler},
  ``{Fencing in the Swampland: Quantum Gravity Constraints on Large Field
  Inflation},''
\href{http://arxiv.org/abs/1503.04783}{{\ttfamily arXiv:1503.04783 [hep-th]}}.

\bibitem{Madrid}
{\sc M.~Montero}, {\sc A.~M. Uranga}, and {\sc I.~Valenzuela},
  ``{Transplanckian axions !?},''
\href{http://arxiv.org/abs/1503.03886}{{\ttfamily arXiv:1503.03886 [hep-th]}}.

\bibitem{Shiu:2015xda}
{\sc G.~Shiu}, {\sc W.~Staessens}, and {\sc F.~Ye}, ``{Large Field Inflation
  from Axion Mixing},'' \href{http://dx.doi.org/10.1007/JHEP06(2015)026}{{\em
  JHEP} {\bfseries 06} (2015) 026},
\href{http://arxiv.org/abs/1503.02965}{{\ttfamily arXiv:1503.02965 [hep-th]}}.

\bibitem{Palti:2015xra}
{\sc E.~Palti}, ``{On Natural Inflation and Moduli Stabilisation in String
  Theory},'' \href{http://dx.doi.org/10.1007/JHEP10(2015)188}{{\em JHEP}
  {\bfseries 10} (2015) 188},
\href{http://arxiv.org/abs/1508.00009}{{\ttfamily arXiv:1508.00009 [hep-th]}}.

\bibitem{Kappl:2015esy}
{\sc R.~Kappl}, {\sc H.~P. Nilles}, and {\sc M.~W. Winkler}, ``{Modulated
  Natural Inflation},''
  \href{http://dx.doi.org/10.1016/j.physletb.2015.12.073}{{\em Phys. Lett.}
  {\bfseries B753} (2016) 653--659},
\href{http://arxiv.org/abs/1511.05560}{{\ttfamily arXiv:1511.05560 [hep-th]}}.

\bibitem{Liddle:1998jc}
{\sc A.~R. Liddle}, {\sc A.~Mazumdar}, and {\sc F.~E. Schunck}, ``{Assisted
  inflation},'' \href{http://dx.doi.org/10.1103/PhysRevD.58.061301}{{\em
  Phys.Rev.} {\bfseries D58} (1998) 061301},
\href{http://arxiv.org/abs/astro-ph/9804177}{{\ttfamily arXiv:astro-ph/9804177
  [astro-ph]}}.

\bibitem{silverstein1985}
{\sc J.~W. Silverstein}, ``{The Smallest Eigenvalue of a Large Dimensional
  Wishart Matrix},'' \href{http://dx.doi.org/10.1214/aop/1176992819}{{\em Ann.
  Probab.} {\bfseries 13} no.~4, (11, 1985) 1364--1368}.
  \url{http://dx.doi.org/10.1214/aop/1176992819}.

\bibitem{Rudelson1}
{\sc M.~{Rudelson}} and {\sc R.~{Vershynin}}, ``{Delocalization of eigenvectors
  of random matrices with independent entries},'' {\em ArXiv e-prints} (June,
  2013) , \href{http://arxiv.org/abs/1306.2887}{{\ttfamily arXiv:1306.2887
  [math.PR]}}.

\bibitem{edelman1988}
{\sc A.~Edelman}, ``Eigenvalues and condition numbers of random matrices,''
  {\em SIAM Journal on Matrix Analysis and Applications} {\bfseries 9} no.~4,
  (1988) 543--560.

\bibitem{edelman1991}
{\sc A.~Edelman}, ``The distribution and moments of the smallest eigenvalue of
  a random matrix of wishart type,''
  \href{http://dx.doi.org/http://dx.doi.org/10.1016/0024-3795(91)90076-9}{{\em
  Linear Algebra and its Applications} {\bfseries 159} (1991) 55 -- 80}.

\bibitem{Wirtz2015}
{\sc T.~Wirtz}, {\sc G.~Akemann}, {\sc T.~Guhr}, {\sc M.~Kieburg}, and {\sc
  R.~Wegner}, ``{The smallest eigenvalue distribution in the real
  Wishart-Laguerre ensemble with even topology},''
  \href{http://dx.doi.org/10.1088/1751-8113/48/24/245202}{{\em J. Phys.}
  {\bfseries A48} no.~24, (2015) 245202},
\href{http://arxiv.org/abs/1502.03685}{{\ttfamily arXiv:1502.03685 [math-ph]}}.

\bibitem{Easther:2016ire}
{\sc R.~Easther}, {\sc A.~H. Guth}, and {\sc A.~Masoumi}, ``{Counting Vacua in
  Random Landscapes},''
\href{http://arxiv.org/abs/1612.05224}{{\ttfamily arXiv:1612.05224 [hep-th]}}.

\bibitem{tracy1996}
{\sc C.~A. Tracy} and {\sc H.~Widom}, ``On orthogonal and symplectic matrix
  ensembles,'' {\em Comm. Math. Phys.} {\bfseries 177} no.~3, (1996) 727--754.
  \url{http://projecteuclid.org/euclid.cmp/1104286442}.

\end{thebibliography}\endgroup
\end{document}